\documentclass[11pt]{article}
\usepackage{amsfonts,amsmath,amsxtra}
\usepackage{amssymb,amscd}
\catcode`\@=11
\def\marginnote#1{}
\newcount\hour
\newcount\minute
\newtoks\amorpm
\hour=\time\divide\hour by60
\minute=\time{\multiply\hour by60 \global\advance\minute by-\hour}
\edef\standardtime{{\ifnum\hour<12 \global\amorpm={am}%
     \else\global\amorpm={pm}\advance\hour by-12 \fi
     \ifnum\hour=0 \hour=12 \fi
   \number\hour:\ifnum\minute<10 0\fi\number\minute\the\amorpm}}
\edef\militarytime{\number\hour:\ifnum\minute<10 0\fi\number\minute}
\def\draftlabel#1{{\@bsphack\if@filesw {\let\thepage\relax
\xdef\@gtempa{\write\@auxout{\string
   \newlabel{#1}{{\@currentlabel}{\thepage}}}}}\@gtempa
\if@nobreak \ifvmode\nobreak\fi\fi\fi\@esphack}
     \gdef\@eqnlabel{#1}}
\def\@eqnlabel{}
\def\@vacuum{}
\def\draftmarginnote#1{\marginpar{\raggedright\scriptsize\tt#1}}
\def\draft{\oddsidemargin -0.1truein
     \def\@oddfoot{\sl preliminary draft \hfil
     \rm\thepage\hfil\sl\today\quad\militarytime}
     \let\@evenfoot\@oddfoot \overfullrule 3pt
     \let\label=\draftlabel
     \let\marginnote=\draftmarginnote
\def\@eqnnum{{\rm (\theequation)}
\rlap{\kern\marginparsep\tt\@eqnlabel}%
\global\let\@eqnlabel\@vacuum}  }
\def\numberbysection{\@addtoreset{equation}{section}
     \def\theequation{\thesection.\arabic{equation}}}
\numberbysection

\renewcommand{\theequation}{\thesection.\arabic{equation}}
\makeatletter
\newdimen\normalarrayskip            % skip between lines
\newdimen\minarrayskip               % minimal skip between lines
\normalarrayskip\baselineskip
\minarrayskip\jot
\newif\ifold             \oldtrue            \def\new{\oldfalse}
\def\arraymode{\ifold\relax\else\displaystyle\fi}%mode of array enrties
\def\eqnumphantom{\phantom{(\theequation)}} % ight phantom in eqnarray
\def\@arrayskip{\ifold\baselineskip\z@\lineskip\z@
  \else
  \baselineskip\minarrayskip\lineskip1\baselineskip\fi}
\def\@arrayclassz{\ifcase \@lastchclass \@acolampacol \or
\@ampacol \or \or \or \@addamp \or
\@acolampacol \or \@firstampfalse \@acol \fi
\edef\@preamble{\@preamble
\ifcase \@chnum
  \hfil$\relax\arraymode\@sharp$\hfil
  \or $\relax\arraymode\@sharp$\hfil
  \or \hfil$\relax\arraymode\@sharp$\fi}}
\def\@array[#1]#2{\setbox\@arstrutbox=\hbox{\vrule
  height\arraystretch \ht\strutbox
  depth\arraystretch \dp\strutbox
width\z@}\@mkpream{#2}\edef\@preamble{\halign \noexpand\@halignto
\bgroup \tabskip\z@ \@arstrut \@preamble \tabskip\z@ \cr}%
\let\@startpbox\@@startpbox \let\@endpbox\@@endpbox
\if #1t\vtop \else \if#1b\vbox \else \vcenter \fi\fi
\bgroup \let\par\relax
\let\@sharp##\let\protect\relax
\@arrayskip\@preamble}
\def\eqnarray{\stepcounter{equation}%
           \let\@currentlabel=\theequation
           \global\@eqnswtrue
           \global\@eqcnt\z@
           \tabskip\@centering              %formulae  centering
           \let\\=\@eqncr
           $$%
         \halign to \displaywidth  \bgroup
          \eqnumphantom \@eqnsel
   \hskip\@centering                               %right tab%
 $\displaystyle  \tabskip\z@ {##}$%
 &\global\@eqcnt\@ne \hskip 2\arraycolsep
      $ \displaystyle  \arraymode{##}$\hfil
 &\global\@eqcnt\tw@ \hskip 2\arraycolsep
      $\displaystyle\tabskip\z@{##}$\hfil
      \tabskip\@centering
 &{##}\tabskip\z@\cr}
\makeatother
\newcounter{mo}

\newcounter{bk}

\newcommand{\Si}{\Sigma}
\newcommand{\tr}{{\rm tr}}

\newcommand{\Ad}{{\rm Ad}}
\newcommand{\ti}[1]{\tilde{#1}}
\newcommand{\om}{\omega}
\newcommand{\Om}{\Omega}
\newcommand{\de}{\delta}
\newcommand{\al}{\alpha}
\newcommand{\te}{\theta}

\newcommand{\be}{\beta}
\newcommand{\la}{\lambda}
\newcommand{\La}{\Lambda}

\newcommand{\ep}{\epsilon}

\newcommand{\G}{\Gamma}
\newcommand{\ka}{\kappa}

\newcommand{\ga}{\gamma}

\newcommand{\si}{\sigma}

\def\bea{\begin{eqnarray}\new\begin{array}{cc}}
\def\ee{\end{array}\end{eqnarray}}
\def\bel{\bea\label}
\newcommand{\beq}[1]{\begin{equation}\label{#1}}
\newcommand{\eq}{\end{equation}}
\newcommand{\beqn}[1]{\begin{small} \begin{eqnarray}\label{#1}}
\newcommand{\eqn}{\end{eqnarray} \end{small}}
\newcommand{\p}{\partial}
\def\sq2{\sqrt{2}}

\newcommand{\oh}{\frac{1}{2}}

\def\sl2{{\rm sl}(2, {\mathbb C})}
\def\SLN{{\rm SL}(N, {\mathbb C})}

\def\f1#1{\frac{1}{#1}}

\newcommand{\bp}{\bar{\partial}}
\newcommand{\bz}{\bar{z}}

\newcommand{\bA}{\bar{A}}

\def\mC{{\mathbb C}}

\def\mR{{\mathbb R}}

\def\frak{\mathfrak}
\def\gb{{\frak b}}
\def\gg{{\frak g}}
\def\gp{{\frak p}}
\def\gn{{\frak n}}
\def\gs{{\frak s}}

\def\gS{{\frak S}}
\def\gL{{\frak L}}

\def\gk{{\frak k}}

\def\gh{{\frak h}}
\def\gu{{\frak u}}

\def\gt{{\frak t}}

\def\gR{{\frak R}}

\def\bfe{{\bf e}}

\def\bfu{{\bf u}}
\def\bfv{{\bf v}}

\def\bfP{{\bf P}}

\def\bfS{{\bf S}}

\def\bfT{{\bf T}}

\def\clD{\mathcal{D}}

\def\clG{\mathcal{G}}
\def\clF{\mathcal{F}}
\def\clR{\mathcal{R}}
\def\clU{\mathcal{U}}
\def\clT{\mathcal{T}}
\def\clO{\mathcal{O}}
\def\clH{\mathcal{H}}
\def\clK{\mathcal{K}}

\def\clM{\mathcal{M}}
\def\clN{\mathcal{N}}
\def\clP{\mathcal{P}}
\def\clQ{\mathcal{Q}}
\def\clS{\mathcal{S}}
\def\clV{\mathcal{V}}
\def\clX{\mathcal{X}}

\def\clZ{\mathcal{Z}}

\def\bag2{{\bf g_2}}
\def\bas8{{\bf so(8)}}

\def\sr2{\sqrt{2}}
\newcommand{\ran}{\rangle}
\newcommand{\lan}{\langle}
\def\f1#1{\frac{1}{#1}}

\newtheorem{rem}{Remark}[section]

\def\Ad{{\rm Ad}}

\def\Bun{{\rm Bun}}
\def\rank{{\rm rank}}
\def\Hom{{\rm Hom}}
\def\Lie{{\rm Lie}}
\def\Res{{\rm Res}}
\def\Isom{{\rm Isom}}

\begin{document}
%\draft \vspace{-1cm}
 \begin{flushright}
 ITEP-TH-32/17\\
 IITP-TH-19/17
 \end{flushright}
\vspace{1.cm}

 \begin{center}
{\LARGE Quasi-compact Higgs bundles and Calogero-Sutherland systems with two types spins}\\
 \vspace{10mm}
 {{\large S. Kharchev}$^{\flat\,\S}$\ \ {{\large A. Levin}$^{\,\natural\,\,\flat}$\ \
  {\large M. Olshanetsky}$^{\,\flat\,\S\,\sharp}$\ \  {\large A. Zotov}$^{\,\diamondsuit\,\flat\, \natural\, \sharp}$}\\
   \vspace{7mm}
  \vspace{2mm}$^\flat$ -
%  {\sf Institute of Theoretical and Experimental Physics, Moscow, 117218, Russia}\\
  {\sf ITEP, B. Cheremushkinskaya, 25, Moscow, 117259, Russia}\\
   \vspace{2mm}$^\S$ - {\sf Institute for Information Transmission Problems RAS (Kharkevich Institute),
 \\  Bolshoy Karetny per. 19, Moscow, 127994,  Russia}\\
 \vspace{2mm}$^\natural$ - {\sf National Research University Higher School of Economics, Russian Federation,\\
%  Mathematics Department of
 NRU HSE,
 Usacheva str. 6,  Moscow, 119048, Russia}\\
% \vspace{2mm}$^\natural$ - {\sf International Laboratory of Representation Theory and Mathematical Physics,\\
%  Mathematics Department of NRU HSE,
% Usacheva str. 6,  Moscow, 119048, Russia}\\
 \vspace{2mm} $^\sharp$ - {\sf Moscow Institute of Physics and Technology,\\
 Inststitutskii per.  9,  Dolgoprudny, 141700 Moscow Region, Russia}\\
  \vspace{2mm} $^\diamondsuit$ -
 {\sf Steklov Mathematical Institute of Russian
Academy of Sciences,\\ Gubkina str. 8, Moscow, 119991,  Russia
 }\\
}
 \vspace{0.5cm}
 {\footnotesize Emails: kharchev@itep.ru, alevin2@hse.ru, olshanet@itep.ru,
 zotov@mi.ras.ru}\\
 \end{center}

%\today
 \begin{abstract}
We define the quasi-compact Higgs $G^{\mathbb C}$-bundles over
singular curves introduced in our previous paper for the Lie group
SL($N$). The quasi-compact structure means that the automorphism
groups of the bundles are reduced to the maximal compact subgroups
of $G^{\mathbb C}$ at marked points of the curves.
 We demonstrate that in  particular cases
this construction leads to the classical integrable systems of
Hitchin type. The examples of the systems are analogues of the
classical Calogero-Sutherland systems related to a simple complex
Lie group $G^{\mathbb C}$ with two types of interacting spin variables. These
type models were introduced previously by Feher and Pusztai. We
construct the Lax operators of the systems  as the Higgs fields
defined over a singular rational curve. We also construct
 hierarchy of independent integrals of motion.
 Then we pass to a fixed point set of real involution
related to one of the complex structures on the moduli space of the
Higgs bundles. We prove that the number of independent integrals of
motion is equal to the half of dimension of the fixed point set. The
latter is a phase space of a real completely integrable system. We
construct the classical $r$-matrix depending on the spectral
parameter on a real singular curve, and in this way  prove the
complete integrability of the system. We present three equivalent
descriptions of the system and establish their equivalence.
 \end{abstract}

\footnotesize \tableofcontents \normalsize

\section{Introduction and summary}

\setcounter{equation}{0}

In this paper we describe  \emph{the quasi-compact
Higgs bundles}
 (see the definition below).
The aim  is to construct  integrable systems of the Hitchin type
\cite{Hi} using the quasi-compact Higgs bundles. As an example we
consider  generalizations (GCS) of the spin Calogero-Sutherland (CS)
model and its extensions related to simple Lie algebras. These
systems were introduced in \cite{Feher,F} and for the Lie group $\SLN$
in our previous work \cite{KLOZ}, where  the
quasi-compact Higgs bundles were defined. The systems are similar
to the spin CS systems, where the starting point are \emph{the
quasi-parabolic Higgs bundles} \cite{K,Si} over a singular rational
curve \cite{Ne,TC}.
 The crucial distinction from the CS systems beyond the quasi-compactness, is that  we use
  a specific real involution on the
moduli space of the Higgs bundles considered for the classical
Hitchin systems in \cite{BS} and for some  quantum Hitchin systems in
\cite{NS,NW}. After this involution we come to the real completely integrable systems.

\paragraph{Description of systems.}
%We prove that
%More precisely, we generalize our previous results related to Lie
%algebras and $\rm{so}(N)$
%%
The CS model \cite{Ca,Su} describes one-dimensional system of
 pairwise interacting particles through long range potentials. We deal with
the classical model. It is an integrable system in the Liouville
sense, as well as its spin extension \cite{GH,W}. The latter model
can be written as the Euler-Arnold $\SLN$ top (its inertia tensor
depends on the positions of interacting particles).

Original CS system is related to the Lie group $\SLN$,
and it can be generalized for arbitrary simple complex Lie algebra
$\gg^\mC$ in the following way \cite{OP}. Let $R$ be the
corresponding root system and  $\gh^\mC$ is a Cartan subalgebra of
 Lie algebra $\gg^\mC$. Denote the coordinates of the particles
$\bfu=\sum_{i=1}^lu_ie_i\in\gh^\mC$, where $(e_1,\ldots,e_l)$ is a
canonical basis in $\gh^\mC$ and $\bfv$ is the momenta vector. Let
$\bfS\in\gg^\mC$ be an element of a (co)adjoint orbit of the group
$G^\mC$ in $(\gg^\mC)^*\sim\gg^\mC$. The coordinates $S_\al$ of
$\bfS$ in the root basis $E_\al$, ($\al\in R$) are called \emph{the
spin variables}.
 The Hamiltonian of the spin CS system
related to $\gg^\mC$ has the form
  \beq{i0}
H^{CS}=\oh\,(\bfv,\bfv)+\sum_{\al\in
R}\frac{1}{(\al,\al)}\,\frac{S_{\al}S_{-\al}}{\sinh^{2}(\bfu_\al)}\,.
   \eq
The Poisson brackets for $\bfv$, $\bfu$ are the Darboux brackets,
 while the Poisson structure for
$\{S_{\al},S_{\be}\}$ is obtained from the Lie-Poisson brackets on
the Lie co-algebra $(\gg^\mC)^*$ via Hamiltonian reduction with respect to the
coadjoint action of the Cartan subgroup of $G^\mC$ on the spin
variables.
 However, for $\bfS\in{\rm so}(N)$ the additional reduction is not
 needed \cite{BAB}. In our previous paper \cite{KLOZ} we described the ${\rm
so}(N)$ model  with two types of spins.
 Here we use this approach for an arbitrary simple Lie group $G^\mC$.

The extension of (\ref{i0}) is as follows.
 Let $G^\mR$ be the normal
real form of the group $G^\mC$. Such form exists for any simple
complex group. Consider the maximal compact subgroup $K\subset
G^\mC$. The maximal compact subgroup $U\subset G^\mR$ is the
intersection $K\cap G^\mR$. For example, if $G^\mC=\SLN$, then
$G^\mR={\rm SL}(N,\mR)$, $K={\rm SU}(N)$, $U={\rm SO}(N)$. Let
$E_\al$ be a root basis in the Lie algebra $\gg^\mC$. The Lie
algebra $\gu={\rm Lie}(U)$ has the basis $E_\al-E_{-\al}$, where
$\al\in R^+$ are the positive roots. Define the two type of spin
variables $\bfS, \bfT\in\gu$
 $$
\bfS=\sum_{\al\in R^+}S_\al(E_\al-E_{-\al})\,,~~\bfT=\sum_{\al\in R^+}T_\al(E_\al-E_{-\al})\,,
~~S_\al\,,T_\al\in\mR
 $$
belonging to the same coadjoint orbit $\clO\subset\gu^*$. Then the
Hamiltonian assumes the form \cite{Feher}\footnote{In the systems
described in \cite{Feher} different notations were used}:
 \beq{i1}
H=\oh(\bfv,\bfv)+\sum_{\al\in R^+}\frac{2}{(\al,\al)}\,\frac{
S_\al^2 +T_\al^2-2 S_{\al}T_{\al}\cosh(\bfu_\al)}
{\sinh^{2}(\bfu_\al)}\,,
 \eq
  where  $\bfv$ and $\bfu$ are real ($\bfv,\bfu\in\gh^\mR$ - Cartan subalgebra of $\gg^\mR$).
  In contrast with the spin CS system the Poisson brackets for the spin variables are the genuine
  Poisson-Lie brackets without additional constraints
\bel{i101}
\{T_\al,T_\be\}=N_{\al,\be}T_{\al+\be}-N_{\al,-\be}T_{\al-\be}, \\
\{S_\al,S_\be\}=-N_{\al,\be}S_{\al+\be}+N_{\al,-\be}S_{\al-\be}, \\
\{T_\al,S_\be\}=0,
\ee
with the structure constants $N_{\al,\be}$ defined in (\ref{na}). Equations of
motion\footnote{ In order to reproduce the results of \cite{KLOZ} in
the ${\rm so}(N)$ case one should redefine $S\rightarrow -S$ and
$T\rightarrow -T$. This also leads to changing the signs in the
r.h.s. of (\ref{i101}).} are generated by (\ref{i1}), (\ref{i101})
and the canonical brackets $\{v_j,\bfu_\al\}=\al(j)\equiv\al(e_j)$:
 \beq{i102}
 \dot S_\al= \sum_{\be+\ga=\al}C_{\be\ga}
S_\be\frac{S_\ga-T_\ga\cosh(\bfu_\ga)}{\sinh^2(\bfu_\ga)},
 \quad\quad
 \dot T_\al= \sum_{\be+\ga=\al}C_{\be\ga}
 T_\be\frac{S_\ga\cosh(\bfu_\ga)-T_\ga}{\sinh^2(\bfu_\ga)},
 \eq
 \beq{i103}
 {\dot v}_j=\sum_{\ga\in R^+}\frac{4\ga(j)}{(\ga,\ga)}
 \left( (S_\ga^2+T_\ga^2)\frac{\cosh(\bfu_\ga)}{\sinh^3(\bfu_\ga)}-S_\ga T_\ga\frac{1+\cosh^2(\bfu_\ga)}{\sinh^3(\bfu_\ga)}
 \right).
 \eq
 In this way the Hamiltonian (\ref{i1}) describes particles with real coordinates $\bfu$ and
 real momenta $\bfv$
 together with  two interacting tops %\cite{BM}
 on the compact Lie group $U$
 with the inertia tensors  and the interaction term depending on the dynamical coordinates
 of particles on $\gh^\mR$. For the group $\SLN$ these tops are defined on the group
 $U=$SO$(N)$. For other simple groups the corresponding subgroups are presented in
 Table 1 in the Appendix A.

  For $N=2$ the
algebra so(2) is commutative and one can fix values of the  spin
variables. In this case we obtain from (\ref{i1}) the Hamiltonian
with two constants
  \beq{i4}
 H=\frac{v^2}{2}+\frac{
m_1^2 +m_2^2-2 m_1m_2\cosh(2u)} {\sinh^{2}(2u)}\,.
  \eq
This Hamiltonian reproduces the CS model of  the BC$_1$ type
\cite{OP}. In the quantum case this Hamiltonian coincides with the
Casimir operator acting on a special matrix element of the principal
series representations of SU$(1,1)$ written in the spherical
coordinates \cite{V}.

  We prove that the Hamiltonian (\ref{i1}) is the first nontrivial one in the family
  of Poisson commuting real independent integrals of motion.

 % that makes the system completely integrable  in the Liouville sense.

  The hierarchy is described by the Lax operator. In the Chevalley basis of
  the algebra $\gg^\mR$ it  assume the form
 \beq{i104a}
 L(x)=\sum\limits_j v_je_j+\sum_{\be\in R}
 \frac{T_\be\exp(-\bfu_\be)-S_\be}{\sinh(\bfu_\be)}E_\be\,,
 \eq
  The equations (\ref{i102}), (\ref{i103}) are equivalent to the Lax
 equations ${\dot L}(x)=[L(x),M]$  with the $M$ operator
  \beq{i105}
 M=\sum_{\be\in
 R}\frac{S_\be\cosh(\bfu_\be)-T_\be}{\sinh^2(\bfu_\be)}E_\be\,.
 \eq

 To derive the system and the Lax operator we use the symplectic reduction
 staring from three different finite-dimensional symplectic spaces.
 We call the corresponding systems the Model I, the Model II and the Model III.
 Literally, the equations (\ref{i102}), (\ref{i103}) and the Lax operator
 corresponds to the Model I and the Model II. There are the  changes of variables that leads to the
 Model III. In next paragraph we clarify interrelations between the
 models coming from comparison of different Higgs bundles.

 While the spin variables
in the Models I and II are elements of a coadjoint orbit of the
compact subgroup $U$ ($U=G^\mR\cap K$), in the model III they are
elements of the cotangent bundle $T^*(\clX^\mR)$ to the symmetric
space $\clX^\mR=G^\mR/U$.
%For example, for $G^\mC=\SLN$, $G^\mR$=SL$(N,\mR)$, $K$=SU$(N)$ and $U$=SO$(N)$.
In this way the GCS models  are\emph{ real integrable systems}.
To derive them we start with \emph{complex symplectic manifolds}
for the following reason. In the second part of paper we relate GCS systems
with a special class of Hitchin systems
\cite{Hi}. The latter are based on considerations of holomorphic bundles over
complex curves.

In the first part of paper we use the finite-dimensional approach similar
to \cite{OP}.
 For the Model I and III the unreduced symplectic spaces are  products of complex and
 real spaces. This makes the definition of Hamiltonian systems obscure. For the Model I
 we come to the well-defined \emph{real Hamiltonian systems} after the symplectic reduction with respect
 to the real group. For the Model III the real systems arise after passing to the
 real subgroup $G^\mR\subset G^\mC$.  The unreduced symplectic space of the Model II
 is complex and the real Hamiltonian system is result of the symplectic reduction with respect
 to the real group.

% In our case the part of coordinates in $\clM(G^\mC)$ are real and the definition
%
%To pass to Hamiltonian systems  we consider a specific real involution on the

 %\vskip2mm

  \paragraph{Quasi-compact Higgs bundles.}
 In order to construct the complete family of integrals
 of motion we define the Lax pair depending on the spectral parameter
 $x\in\mR$. For example, the Lax operator corresponding to the Model I has the form
 \beq{i104}
 L(x)=\sum\limits_j v_je_j+\sum_{\be\in R}\left( \frac{T_\be\exp(-\bfu_\be)-S_\be}{\sinh(\bfu_\be)}
 + (1+\coth(x))T_\be \right)E_\be\,.
 \eq
 % ZAMENA
 % VMESTO PREDLOZHENIYA POSLE ETOY FORMULY NADO VSTAVIT' ABZAC - NAPISANNYY NIZHE
 %
It was obtained in \cite{Feher} in a slightly different form. We
will construct the classical $r$-matrix related to $L(x)$. The
classical $r$-matrix without spectral parameter was previously
obtained in \cite{Feher2}.

The Lax operators play the role of the Higgs fields in the Hitchin
approach to integrable systems \cite{Hi}. To construct the Lax
matrix (\ref{i104}) we  consider the $G^\mC$ Higgs bundle over
singular curves as in \cite{Ne,TC}. To construct the defined above
systems we endow the Higgs bundles  with the so-called
\emph{quasi-compact structure}. It means that the gauge group at the
marked points on the base spectral curve is reduced to the maximal
compact subgroups $K$. In the standard approach to the Hitchin
systems the gauge group may have the quasi-parabolic structure, i.e.
the gauge group is reduced at marked points to parabolic subgroups.
In the latter case the moduli space of the Higgs bundles
$\clM(G^\mC)$ are the phase space of \emph{complex integrable
systems} \cite{Hi}.

In our case the part of coordinates in the local description of the Higgs bundles are real.
As it was mentioned above in this case the definition
of Hamiltonian systems is  obscure.
To pass to real Hamiltonian systems  we consider a specific real involution on the
moduli space of the Higgs bundles defined in  \cite{BS}.
The fixed points set of the corresponding involution
becomes \emph{real completely integrable system}, since the number of independent
integrals of motion is exactly equal to the half dimension of the
moduli space.

 Thus, we break the
complex structure in two stages. First, we do it locally by
introducing the quasi-compact structure. Next we pass to the real
involution of the Higgs bundle. The last step allows one to consider
the spin variables as genuine elements of  the coadjoint
orbits in the Lie coalgebra $\gu^*$, while for the quasi-parabolic bundles they were elements
of  symplectic quotient of the orbits with respect to the Cartan group action.

But first, before  using the real involution we define the moduli
space of the Higgs
 bundles $\clM_A(G^\mC)$ $\,(A= I,\, II,\, III)$
that lead eventually to phase spaces of Models I, II and III.
The Higgs bundle corresponding to the Model I is define over the  curve
$\Si^I$. It is the normalization of two rational curves glued in two points.
The result of the symplectic reduction with respect to the automorphism group of the bundle
(the gauge group) is the moduli space $\clM_I(G^\mC)$.
For the Models II and III the base curve is a rational curve glued in two points.
It coincides with the singular curve for the CS system
\footnote{The $\SLN$ Higgs bundle I
constructed in  \cite{KLOZ} is unstable. It is one of the reasons to
construct the Higgs bundles II and III.}.
We prove that the moduli spaces
of the Higgs bundles $\clM_A(G^\mC)$ $\,(A= I,\, II\, III)$
after the involution coincide with the phase spaces of
the integrable system corresponding to the Models I, II and III.

There is a correspondence between the Higgs bundles I and II,
provided by the Universal Higgs Bundle and its projections on the
Higgs bundles I and II:
%Here we construct another Higgs bundles
%over different base spectral curve that however  after symplectic
%reduction lead to the same integrable systems.
\beq{0}
\begin{array}{ccccc}
     &  & \fbox{Universal~Higgs~Bundle} &  &  \\
   & \pi_1\,\swarrow&  & \searrow\,\pi_2 & \\
    \fbox{Higgs~Bundle~I} &  &  &  & \fbox{Higgs~Bundle~II} \\
     &SR_1 \searrow &  & \swarrow SR_2 &  \\
     &  & \fbox{GCS ~Phase~Space} &  &
  \end{array}
  \eq
where SR means a symplectic reduction and the passing to the fixed point sets
of the involution.

%{\rm Reduction~I}{\rm Reduction~II}
%he Higgs bundles are endowed with the so-called \emph{quasi-compact
%structure}.
%It means that the gauge group at the singular points on the
%base spectral curve is reduced to the maximal compact subgroups $K$
%\footnote{In the standard approach to the Hitchin systems the gauge
%group may have the quasi-parabolic structure, i.e. the gauge group
%is reduced at singular points to a parabolic subgroup.}. This definition leads to
%the Higgs bundles of type II.

The quasi-compact bundles can be equivalently defined in a different way.
Instead of restriction of the gauge group at the marked points
 to the maximal compact subgroups $K$, we attach to the points the symmetric spaces
  $\clX^\mC=K\setminus G^\mC$. For the corresponding Higgs bundle it amounts to
  attaching to the marked point the cotangent bundle $T^*\clX^\mC$.
   This definition leads to the moduli space  $\clM_{III}(G^\mC)$.
  The equivalence of these descriptions defines the isomorphism
%Thus, in addition to the diagram (\ref{0}) we have
\beq{011}
\clM_{II}(G^\mC)\sim\clM_{III}(G^\mC)\,.
\eq
One of the aims of the paper is the derivation of the moduli spaces $\clM_{A}(G^\mC)$ and the
description of the interrelations between them.
It turns out that the isomorphism (\ref{011}) is symplectomorphism. We do not prove this fact here.
In a separate publication we will construct general quasi-compact Higgs bundles and
prove that (\ref{011}) is symplectomorphism.

%e are no sufficient independent integrals of motion. To

% considered for the general classical
%Hitchin systems in \cite{BS} and for some quantum Hitchin systems in
%\cite{NS,NW}.

\bigskip

In the first Section we develop the finite-dimensional approach to
the system as in \cite{OP} for the CS system. We start with  free
systems related to the complex simple Lie groups. As we explained
above
 there are three models of these systems. For the Model I it
 is a free system defined on the product of two cotangent bundles
 $T^*G^\mC\times T^*K$ (see (\ref{sss})). It was considered in  \cite{KLOZ}
 for $G^\mC=\SLN$. By symplectic reduction with respect to the action of the
 group $K\times K$ we come to the reduced phase space $(T^*G^\mC\times T^*K)//(K\times K)$.
 We obtain the Lax pair corresponding to the Hamiltonian (\ref{i1}).
 For the Model II we start with the phase space $T^*G^\mC$ and the symmetry group $K$.
We prove that the reduced phase space $T^*G^\mC//K$ is
symplectomorphic to the previous one and leads to the same
integrable family. For the Model III we start with the phase space
$T^*G^\mC\times T^*\clX^\mC$ and the symmetry $G^\mC$.
To come to integrable situation we pass to the subgroups $G^\mR$, $U=G^\mR\cap K$,
and real symmetric space $\clX^\mR=U\setminus G^\mR$.
 The resulting
model is a system of particles similar to the spin CS system, but
the spin variables are elements of the cotangent bundle
$T^*\clX^\mR$.

 The drawback of this approach is that the Lax operators derived in this
way are independent of the spectral parameter. To bring it in the
game %in the  next four sections
we derive   the systems by replacing the cotangent bundles
$T^*G^\mC$ with the Higgs bundles over the singular curves endowed
with the quasi-compact structures. In particular, we construct the
Universal Higgs bundle that allows one to establish the
correspondence presented in the diagram (\ref{0}). We also discuss
the interrelations between the Higgs bundles of type II and III
(\ref{011}).

It is also instructive to compare our construction with the Higgs
bundle description of the CS system. The standard approach to the CS
system \cite{Ne,TC} is similar to the Model III.
 The Model I type description is specific for the GCS system and has not
analog for the CS model. The Model II type description is implicit
and  interrelation between the Model II and Model III for the CS systems is
complicated.

 In  Appendix B we  consider of a gyrostat related
 to a simple Lie group.
It is the Euler-Arnold top with additional rotator moment.
We derive equations of motion for this system.
We analyze the Lax equations in Appendix C.

% ZAMENA - MOZHNO POMENYAT' BLAGODARNOSTI CELIKOM

%%%%%%%%%%%%%%%%%%%%%%%%%%%%%%%%%%%%%%%%%%%%%%%%%%%%%%%%%%%%%%%%%%%%%%%%%%%%%%%%%%
\noindent \paragraph{Acknowledgements} We are grateful to L. Feh\'er
for useful comments and remarks. This work was supported by RFBR
grants 15-01-99504, RFBR 18-01-00460 (S. Kharchev) and 15-02-04175,
18-02-01081 (A.L., M.O. and A.Z.). A. Levin was partially supported
by Laboratory of Mirror Symmetry NRU HSE, RF Government grant, ag.
14.641.31.0001. The work of A. Zotov was funded by the Russian
Academic Excellence Project `5-100'.
%%%%%%%%%%%%%%%%%%%%%%%%%%%%%%%%%%%%%%%%%%%%%%%%%%%%%%%%%%%%%%%%%%%%%%%%%%%%%%%%%%%

\section{Finite-dimensional description}

\setcounter{equation}{0}

\subsection{Model I}
Let $G^\mC$ be a simple complex Lie group and $K$ -- its maximal
compact subgroup. Consider the symplectic space
 \beq{sss}
\clR_I(G^\mC)=T^*G^\mC\times T^*K\,.
\footnote{This model is similar to the model used in \cite{Feher}, except that
 $T^*K$ is replaced by the two coadjoint $K$-orbits}
 \eq
For example, $G=\SLN$, $K=$SU$(N)$. Then
 \beq{R}
\dim_\mR\,\clR=2\dim_\mR\,G^\mC+2\dim_\mR\,K.
 \eq
%For $G=\SLN$ $\,\dim\,(\clR)=2(N^2-1)+N(N-1)$.
In the left trivialization $T^*G^\mC \sim G^\mC\times (\gg^\mC)^*$
is described by the coordinates   $(g,\eta)$  $(g\in
G^\mC,\eta\in(\gg^\mC)^*)$. Similarly, let $(h,\nu)$ be the
coordinates on $T^*K=K\times\gk^\ast$, $(h\in K,\nu\in\gk^*)$.
 Define the symplectic form $\om=d\vartheta$, where 1-form $\vartheta$
 is the sum
 \bel{8}
\vartheta=\vartheta^G+\vartheta^K,\\
\vartheta^G=\lan\eta,\Omega^L(g)\ran\,,\ \
\vartheta^K=\lan\nu,\Omega^L(h)\ran
 \ee
  and $\Omega^L$ is the
left-invariant Maurer-Cartan form.
\begin{rem}
The symplectic space $\clR_I(G^\mC)$ (\ref{sss}) is the product of complex and real
symplectic spaces. Correspondingly,
the one-form $\vartheta$ (\ref{8}) is the sum of complex and real one-forms.
As we mentioned above the definition of dynamical systems on $\clR_I(G^\mC)$
is ambiguous. Thus, defined below Hamiltonians and integrals are formal, until
passing to the real symplectic space. We will see that the same is true for the considered below
Model III.
\end{rem}

Define  the quadratic Hamiltonian
 \beq{ha}
H=\oh(\eta,\eta)\,.
 \eq
where $(\,,\,)$ comes from (\ref{kk}), (\ref{are6}), and the Poisson
brackets are defined by (\ref{8}).

Consider the symplectic quotient of this system with respect the following
 gauge symmetry. The gauge group
  $G^{gauge}=K_1\times K_2$, $K_1,K_2\sim K$ acts as
   \beq{9}
K_1\,:\,g\to g f_1^{-1}  \,,\eta\to \Ad_{f_1}^*\eta  \,;h\to h f_1^{-1}
\,, \nu\to \Ad_{f_1}^*\nu\,.
 \eq
 \beq{10}
K_2\,:\,g\to f_2g\,,\eta\to \eta \,; \, h\to f_2h\,,\,\nu\to \nu.
 \eq
The Hamiltonian vector fields $V_{1,2}$  corresponding to
(\ref{9}) (\ref{10}), are given by $V_1=\{F_1,~\}_K$ and
$V_2=\{F_2,~\}_K$, where  the Poisson brackets are defined on the co-algebra $\gk^*$ and
$F_1$, $F_2$ are defined as
 \bel{9'}
  i_{V_{1,2}}\om=-dF_{1,2},\\
F_{1}=\lan\mu_1,\ep_1\ran, \ \ \
F_{2}=\lan\mu_2,\ep_2\ran\,,~~(\ep_{1,2}\in\gk),
 \ee
and corresponding moment maps $\mu_1$,
$\mu_2:\,\clR_I(G^\mC)\to\gk^*$ take the form
 \bel{m1}
 \mu_1=\eta|_{\gk^*}+ \nu,\\
 \mu_2=\Ad_g^\ast\eta|_{\gk^*}+ \Ad_h^\ast \nu\,.
\ee
Introduce notations
 \beq{ov}
\bfT\doteq\nu\in \gk^*\,,~~ \bfS\doteq\Ad^*_h\nu\in\gk^*\, .
 \eq
The coordinates on $\bfT$ and $\bfS$ are functions on the co-algebra
$\gk^*$. They have the linear Lie-Poisson structure. We write
tentative
   $\{\bfT,\bf T\}\sim \bf  T$, $\{\bf S,\bf S\}\sim \bf S$ and
$\{\bf T,\bf S\}=0$ (see (\ref{20a}), (\ref{30a}) (\ref{30b})).
\footnote{In fact, these brackets are defined on the subalgebra
$\gu^*\subset\gk^*$ (\ref{gu}) and will be used below. But this
structure holds on $\gk^*$ as well.}
 \begin{rem}
One can fix the Casimir functions in the Poisson algebra on $\gk^*$.
In this way  (\ref{ov}) is the map of the cotangent bundle $T^*K$ to
the pair of the coadjoint $K$-orbits $\clO_K$ corresponding to these
Casimir functions. These two orbits are results of two ways
describing coadjoint orbits as symplectic quotients from the
cotangent bundle by either left or right shifts.
 \end{rem}
Using this construction we replace $\clR_I(G^\mC)$ (\ref{sss}) with
 \beq{sss1}
\ti\clR_I(G^\mC)=T^*G^\mC\times (\clO_K\times\clO_K).
 \eq
It follows from (\ref{dp}) that  the gauge actions (\ref{10}) and (\ref{9}) transform $g$  to the Cartan
subgroup
 \beq{10a}
f_1gf_2=e(\bfu)=\exp\,\bfu\,,~~\bfu\in\gh^\mR\,,~
\bfu=(u_1,\ldots,u_l)\,,~(l=\rank\,\gg)\,.
 \eq
Using the invariant form on $\gk$ we identify $\gk$ and $\gk^*$ and
 decompose $\bfT$ and $\bfS$
 in the root basis (\ref{4.2e})
 \bel{11}
 \bfT=\bfT_{\gh_K}+\sum_{\al\in R^+}(T_\al E_\al-\bar T_\al E_{-\al})\,,~~
 \bfS=\bfS_{\gh_K}+\sum_{\al\in R^+}(S_\al E_\al-\bar S_\al E_{-\al})\,,\\
 \bfT_{\gh_K}=\imath \sum_{j=1}^lT_je_j\,,~~\bfS_{\gh_K}=\imath\sum_{j=1}^l S_je_j
\ee
We put the moment constraints (\ref{m1}) $\mu_L=0$, $\mu_R=0$.
Plugging $g=e(\bfu)$ (\ref{10a}) into these equations we come to the
linear system
  \begin{subequations}\label{me}
  \begin{align}
 &\eta|_{\gk}=-\bfT\,,\\
 &\Ad_{e(\bfu)}\eta|_{\gk}=-\bfS\,.
 \end{align}
  \end{subequations}
Consider the a  solution of (\ref{me})
 \bel{eta}
\eta=P+X\,,~~P\in\gh^\mC\,,\quad X=\sum_{\al\in R}X_\al E_\al\,.
  \ee
   Then (see (\ref{11}) and
(\ref{bk}))
  \bel{ogr} \eta|_{\gk}=\imath\Im m\, P+ \oh\sum_{\al\in
R^+}(X_\al-\bar X_{-\al})E_\al-(\bar X_\al- X_{-\al})E_{-\al}\,.
 \ee
  From (\ref{me})  we obtain
 \bel{me1}
 \oh(X_\al-\bar X_{-\al})=-T_\al, \\
      \oh(X_\al e(\bfu_\al)-\bar X_{-\al}e(-\bfu_\al))=- S_\al
 \ee
and
  \beq{me2}
\imath\Im m\, P=\bfT_{\gh_K}\,,~~ \imath\Im m\, P=\bfS_{\gh_K}\,.
  \eq
% We come again to the last conditions below (\ref{me3}).
 The system (\ref{me1}) has a unique solution
  \bel{13}
 X=\sum_{\al\in R^+}(X_\al E_\al+ X_{-\al}E_{-\al})\,,
   \ee
   $$
X_\al=\frac{T_\al e(-\bfu_\al) -S_\al}{\sinh(\bfu_\al)}\,,~~
  X_{-\al}=\frac{\bar T_\al
  e(\bfu_\al)-\bar{S}_\al}{\sinh(\bfu_\al)}\,.
   $$
The general solution of the linear system (\ref{me}) has the form
  \beq{12}
 \eta=P+X\,,~~P=\bfv+S_{\gh_K},
% T_{\gh_K}=\imath(T_1,\ldots,T_l)\in \gh_K\,.
 \eq
where
$\bfv=(v_1,\ldots,v_l)\in \gh^\mR\,,~~ S_{\gh_K}=\imath(S_1,\ldots,S_l)$.
 In what follows $\eta$ plays the role of the Lax operator (without spectral parameter).
 Thus, the reduced space
$$
 \ti\clR_I^{red}(G^\mC)=\ti\clR_I(G^\mC)//G^{gauge}\doteq(\mu_L^{-1}(0)\times\mu_R^{-1}(0))/K_L\times K_R
$$
  is defined by the coordinates
  \beq{vu}
 \bfv,\bfu\in \gh^\mR\,,~~\exp(\bfu)\in\clH^\mR\,,~~ \bfS=\Ad_h\bfT\,,\,\bfT\in
 \gk\,,
  \eq
so that
 \beq{rs1}
 \ti\clR_I^{red}(G^\mC)=T^* \clH^\mR\times T^*K\,.
 \eq
It has the dimension $\dim_\mR\,\ti\clR_I^{red}(G^\mC)=2\dim\,K+2l$.

The gauge fixing (\ref{10a}) is not complete. Multiplication by an
element $s\in\clT$ from the Cartan torus (from $K$) $f_2\to f_2s$,
$f_1\to f_1s$ preserves the gauge.
%One
%There are residual gauge transformations preserving this condition.
%They are generated by the
% semi-direct product $\clT$, where $W$ is the Weyl group and $\clT$
%is the Cartan torus in $K$.
 The torus acts on the  variables as
 \beq{rg}
 h\to shs^{-1}\,,~~\bfT \to \Ad_s\bfT \,,~~
 \bfS\to \Ad_s\bfS \,,~~s\in\clT\,.
  \eq
 The moment map equation generating this action (see (\ref{10}) and (\ref{9})) is
  \beq{me3}
 \bfT_{\gh_K}-\bfS_{\gh_K}=0\,.
  \eq
Notice that this condition is consistent with (\ref{me2}).
%Then from (\ref{me1}) $P|_{\gk^*} =S_{\gh_K}=0$. It means that
% $$
%P=\bfv\in\gh^\mR\,.
% $$
After fixing the gauge action we come
 to the symplectic quotient $(\clO\times\clO)//\clT$.
The elements of the reduced space $\clR^{red}$ are particles momenta and coordinates $(\bfv,\bfu)$
and the reduced spin variables. Thus
  \beq{ps2}
\clR_I^{red}(G^\mC)= T^* \clH^\mR\times ((\clO_K\times\clO_K)//\clT)\,.
 \eq
Due to the action of the Weyl group  the coordinates belong to some
Weyl chamber $\bfu\in\La\subset\gh^\mR$. Using (\ref{dk}) we find
the dimension of $\clR^{red}$
 \beq{reds}
 \dim\,\clR_I^{red}(G^\mC)=2\dim_{\mR}\,\clO_K=4\sum_{j=1}^l(d_j-1)\,.
 \eq
Here we assume that the orbits $\clO_K$ are generic. For
$G^\mC=\SLN$ (and $K=$SU$(N)$)
 \beq{reds1}
 \dim\,\clR_{\SLN}^{red}=2N(N-1)\,.
 \eq

At this stage we come to the real symplectic space (\ref{ps2}).
The Poisson structure on  $\clR_I^{red}(G^\mC)$ is defined by the canonical brackets
for the coordinates and momenta
of particles $(\bfv,\bfu)$ and the Dirac brackets for the spin variables $(\bfS,\bfT)$.
They come from
the Lie-Poisson brackets on the Lie algebra $\gk$ upon imposing the constraints (\ref{me3})
and the gauge fixing with respect to the $\clT$ action.

From (\ref{ha}) and (\ref{are6}) we find the Hamiltonian
 \beq{ha4}
H=\oh\sum_{j=1}^l(v_j^2-S_j^2)+U(\bfu,\bfS,\bfT)\,,
 \eq
 \beq{h1}
U(\bfu,\bfS,\bfT)=\sum_{\al\in R^+}\frac{2 X_\al X_{-\al}}{(\al,\al)}=
2\sum_{\al\in R^+}\frac{|S_\al|^2+|T_\al|^2-S_\al \bar T_\al e(\bfu_\al)
-\bar S_\al  T_\al e(-\bfu_\al)}{(\al,\al)\sinh^2(\bfu_\al)}\,.
 \eq
The symplectic form (\ref{8}) on the space $\ti\clR_I(G^\mC)$
assumes the form
 \beq{sf1}
 (D\bfv,D\bfu) +D( \bfT,h^{-1}Dh))\,.
  \eq
Summarizing, we came to the Hamiltonian (\ref{h1}) defined on the phase space (\ref{ps2}) by the two-step
symplectic reduction
 \beq{tss}
\Big\{\ti\clR_I(G^\mC)\,,~H\,\,(\ref{ha})\Big\}\stackrel{K_L\times
K_R}{\longrightarrow} \Big\{
\ti\clR_{G^\mC}^{red}\,\,(\ref{rs1})\Big\}\stackrel{\clT}{\rightarrow}
\Big\{ \clR_{G^\mC}^{red}\,\,(\ref{ps2})\,,\, H\,\,(\ref{ha4})\Big\}
 \eq

\subsubsection{Integrals of motion I}

%We construct the invariant integrals using the Lax operator $\clL$ (\ref{12}).
Let $d_j$ $(j=1,\ldots,l)$ be the order of the invariants of the
algebra $\gg^\mC$ (\ref{dj}).
All independent integrals of motion $I_{jk}$
$(k=0,\ldots,d_j)$, $(j=1,\ldots,r)$
can be described in the following way.
 Consider the polynomials $(\bfT+\eta)^{d_j}$ on the algebra $\gg^\mC$
 (this choice will be justify in Section 3.3.)
  with $\bfT$  (\ref{11}) and $\eta$  (\ref{12}).
The integrals of motion $I_{jk}$ are monomials
 \beq{im1}
I_{jk}=
({\bf T}^{d_j-k}\eta^k)\,, ~~k=0,\ldots,d_j\,.
 \eq
 coming from the expansion
 $
({\bf T}+\eta)^{d_j}= ({\bf T}^{d_j})+d_j({\bf
T}^{d_j-1}\eta)+\ldots+(\eta^{d_j})\,.
 $
In particular, the Hamiltonian $H$ (\ref{ha}) is  $I_{1,2}$. Note
that the integrals $I_{j,0}=({\bf T})^{d_j}$ are the Casimir
functions of the Lie-Poisson algebra
 on $(\gk)^*$. They are the Casimir functions on the whole Lie-Poisson algebra
 on $(\gg^\mC)^*$, since $T_\al$ and $\bfT|_{\gh_K}$ Poisson commute with
 other variables (see  (\ref{30b})).

Let $m_j$ be the number of integrals of order $d_j$. The total number of integrals $\clN_G$
is $\sum_{j=1}^lm_j$.
It coincides with the dimension of
the space of homogeneous polynomials of two variables of order $d_j$.
Then $m_j=d_j+1$. Excluding the  $l$ Casimir functions we find the number of integrals
 \beq{ni}
\clN_G=\sum_{j=1}^l(d_j+1)-l=\sum_{j=1}^ld_j=\oh(\dim\,K+l)\,.
 \eq
 Evidently, all integrals are functionally independent.
 We prove their involutivity below using the classical $r$-matrix.
For the complete integrability it is necessary to have
$\clN_G=\oh\dim_{\mR}\,(\clR_{G^\mC}^{red})=\dim\,(\clO_K)$ (see
(\ref{reds})). So, we have a  total deficiency of the integrals (see
(\ref{dk}))
 \beq{dc}
\de_{G^\mC}=\oh\dim\,\clR_{G^\mC}^{red}-\clN_G=\sum_{j=1}^ld_j-2l\,.
 \eq
Notice that $\de_G\geq 0$, where the equality runs up for the rank
one systems. In particular, for $G^\mC=\SLN$ ($A_{N-1}$ type
systems) $d_j=j+1$, $j=1,\dots,N-1\,,~$ $m_j=j+2$
 $$
\clN_{\SLN}=\sum_{j=1}^lm_j-(N-1)=\sum_{j=1}^{N-1}(j+2)-(N-1)=\oh(N-1)(N+2)\,.
 $$
From (\ref{reds1}) we find $\de_{\SLN}=\oh(N-1)(N-2)$.

%%%%%%%%%%%%%%%%%%%%%%%%%%%%%%%%%%%%%%%%%%%%%%%%%%%%%%%%%%%%%%%%%%%%%%%%%%%%%%%%%

%%%%%%%%%%%%%%%%%%%%%%%%%%%%%%%%%%%%%%%%%%%%%%%%%%%%%%%%%%%%%%%%%%%%%%%%%%%%%%%

\subsubsection{Real forms}

 To come to the completely integrable systems we pass
from the complex group $G^\mC$ to its \emph{normal real form}
$G^\mR$ (see Appendix A). We describe this construction in detail for
the corresponding Higgs bundles in Section 3.4.  The phase space of the real form of the model I
is
\beq{tu}
\clR^{red}_{I}(G^\mR)=T^*\clH^\mR\times (\clO_U\times\clO_U)\,.
   \eq
 is described by the variables $(\xi,r=h^{-1} \exp(\bfu))$, $(h\in U$, $\exp(\bfu))\in\clH^\mR$.
 The variables $\bfT,\bfS $ are
elements of the compact subalgebra $\gu=\gg^\mR\cap \gk$ (\ref{gu}).
In the root basis they have the expansion (compare with (\ref{11}))
 \beq{ST}
\bfT=\sum_{\al\in R^+}T_\al (E_\al- E_{-\al})\,,~~
 \bfS=\sum_{\al\in R^+}S_\al (E_\al- E_{-\al})\,,
 \eq
where  $T_\al$, $S_\al$ are real.
\beq{tud}
\dim_\mR\,\clR^{red}_{I}(G^\mR)=2\sum_{j=1}^ld_j-2\rank\,U\,.
\eq
As in (\ref{eta}), (\ref{13})
 \beq{231}
\eta=P+X\,,~~P\in\gh^\mR\,,~~~
P=\bfv=(v_1,\ldots,v_l)\,,
 \eq
 \beq{131}
X=\sum_{\al\in R}X_\al E_\al\,,~~
X_\al=\frac{T_\al e(-\bfu_\al)-S_\al}{\sinh(\bfu_\al)}\,,~~
  X_{-\al}=\frac{T_\al e(\bfu_\al)-S_\al}{\sinh(\bfu_\al)}\,.
   \eq
%
%parameterized by $\bfv,\bfu\in\gh^\mR$, $\bfS,\bfT\in\gu$.
%
 The Poisson structure on $\clR_{I}^{red}(G^\mR)$ takes the form
 \beq{pls}
\{v_j,u_k\}=\de_{jk}\,,~~\{\bfT,\bfT\}=\bfT\,\,(\ref{20a})\,,~\{\bfS,\bfS\}=-\bfS\,\,(\ref{30a})\,,~
\{\bfS,\bfT\}=0\,\,(\ref{30b})\,.
 \eq
 % As above we pass from the variables
% $h\in U$ and $\bfT\in\gu$ on $T^*U$ to the product of two orbits $\clO_U\times\clO_U$
% because the brackets for $\bfS$, $\bfT$ are degenerated. Then instead (\ref{tu}) we have
%  \beq{ps1}
%\clR_{II}^{red}(G^\mR)=T^*\clH^\mR\times (\clO_U\times\clO_U)\,.
%   \eq
% Thus, from (\ref{du}) we obtain
%\beq{ss4}
%\dim_\mR\,\clR_{II}^{red}(G^\mR)//U=2\,\rank\, (G^\mR)+2\dim\,(\clO_U)= 2\sum_{j=1}^ld_j-2\rank\,U\,.
% \eq
%
%
From (\ref{ha4}) and (\ref{h1}) we obtain  the Hamiltonian $H$
 \beq{ha2}
H=\oh(\bfv,\bfv)+U(\bfu,\bfS,\bfT)\,,
 \eq
 \beq{ha3}
U(\bfu,\bfS,\bfT)=\sum_{\al\in R^+}\frac{2 X_\al X_{-\al}}{(\al,\al)}=
2\sum_{\al\in R^+}\frac{S_\al^2+T_\al^2-
2S_\al  T_\al \cosh(\bfu_\al)}{(\al,\al)\sinh^2(\bfu_\al)}\,.
 \eq
 %Summarizing, we come to symplectic quotient
%\beq{tss1}
%\fbox{$T^*G^\mR\,,~H\,\,(\ref{ha})$}
%\stackrel{U_L\times U_R}{\longrightarrow}
%\fbox{$ \clR_{G^\mR}^{red}\,\,(\ref{ps1})\,,\, H\,\,(\ref{ha2})$}
%\eq

%%%%%%%%%%%%%%%%%%%%%%%%%%%%%%%%%%%%%%%%%%%%%%%%%%%%%%%%%%%%%%%%%%%%%%%%%%%%%%%%%%%

\subsubsection{Integrals of motion II}
From (\ref{ni}) and (\ref{tud}) we have the redundant number of integrals
\beq{dc1}
\de_{G^\mR}=\oh\dim\,\clR^{red}_I(G^\mC)-\clN_G=-\rank\,U\,.
\eq
For example,  $\de_{SL(N,\mR)}=-[N/2]$. However,
it turns out that the redundant integrals  are Casimir functions
on $(\gg^\mR)^*\supset\gu^*$ and thereby they are constants on the whole phase space
$\clR^{red}_I(G^\mR)$ (\ref{tu}). Namely, the following statement holds.

\noindent
\emph{$\bullet$ The integrals (\ref{im1}) for $d_j$ even
\beq{inv}
 I_{j,1}=(\bfT^{d_j-1}\eta)\,,~~(d_j\,-\,{\rm even})
\eq
are the Casimir functions on  $(\gg^\mR)^*$.\\
$\bullet\bullet $
The number of even invariants $d_j$ of $\gg^\mR$ is equal to $\rank\, \gu$.}

%\bigskip
It follows from these two statements that (\ref{inv}) provides the
necessary number of the redundant integrals of motion.
Let us first prove that (\ref{inv}) assumes the form
 \beq{fp}
 I_{j,1}=(\bfT^{d_j-1}\eta)=(\bfT^{d_j})\,,~(d_j - {\rm even})\,.
 \eq
  Consider the action of the involutive automorphism $\rho$
(\ref{rho}) on $I_{j,1}$. Let us choose a basis in the adjoint
representation of the algebra $\gg^\mR$ such that $\gh^\mR$ acts as
diagonal matrices, $E_{\pm\al}$ for  $\al\in R^+$ as upper (for +)
or lower (for -) triangular matrices. Then   the operator
$\ti\rho=-\rho$ acts as the matrix  transposition (see (\ref{rho})).
The Killing form is invariant with respect to this action. Then
$(\ti\rho (\bfT^{d_j-1}),\ti\rho (\eta))=(\bfT^{d_j-1},\eta)$.
%(\ti\rho (\bfT^{d_j-1}),\ti\rho (\eta))=(\ti\rho(\eta),\rho(\bfT^{d_j-1}))\,.
%$$
 Let $\eta=\eta^++\eta^-$, where $\ti\rho(\eta^\pm)=\pm\eta^\pm$. Since $\ti\rho(\bfT)=-\bfT$
we have
$$
(\ti\rho(\bfT^{d_j-1}),\ti\rho(\eta)))=
((\eta^+-\eta^-),\ti\rho(\bfT^{d_j-1}))=(-1)^{d_j-1}((\eta^+,\bfT^{d_j-1})
-(\eta^-,\bfT^{d_j-1}))\,.
$$
But the l.h.s. is equal to $(\eta^+\bfT^{d_j-1})+(\eta^-\bfT^{d_j-1})$.
 Therefore, for even $d_j$ we have $(\bfT^{d_j-1}\eta^+)=0$ and $(\bfT^{d_j-1}\eta)=(\bfT^{d_j-1}\eta^-)$.
   From (\ref{131}) we find that
$\eta^-=\sum_{\al\in R^+}T_\al(E_\al-E_{-\al})$. In this way for
$d_j$ even $I_{j,1}=(\bfT^{d_j-1}\eta^-)=(\bfT^{d_j})$. Thus
(\ref{fp}) holds. These expressions are the Casimir functions on
$\gu^*$.

Now let us prove that we have exactly rank$(U)$ Casimir functions.
It follows from Table 1 in Appendix A that for  the groups
SO$(2N+1)$, Sp$(N)$, SO$(2N)$ ($N$ is even),
 G$_2$, F$_4$, E$_7$, $E_8$ 1. $\rank\,G^\mR=\rank\,U$, 2. all $d_j$ are even.
Thus, the number of the Casimir functions is equal to the rank$(G^\mR)=$rank$(K)$.
%We find also in this Table that for the corresponding algebras rank$(K)=$rank$(U)$.
It means that the redundant integrals are the Casimir functions.

For  $G^\mR=$SL$(N,\mR)$   the number of even invariants is equal $[N/2]=$rank(SO$(N)=U$).
For SO$(2N)$, ($N$ is odd) only $d_N=N$ is odd and we have $N-1$ Casimir functions in (\ref{inv}).
 On the other hand,  rank$(U=$SO$(N)\times $SO$(N))=N-1$.
For E$_6$ there are two odd orders $5$ and $9$ of the invariants and thereby we have
four Casimir functions. But the  rank$(U=Sp(4))=4=6-2$. $\Box$

%%%%%%%%%%%%%%%%%%%%%%%%%%%%%%%%%%%%%%%%%%%%%%%%%%%%%%%%%%%%%%%%%%%%%%%%%%%%%%%%%%%%

\subsubsection{Equations of motion and Lax equation}

Equations of motion follow from the Hamiltonian
(\ref{ha2})-(\ref{ha3}) and the Poisson brackets (\ref{pls}). For
$\bfv=\sum_{j=1}^lv_je_j$ and $\bfu=\sum_{j=1}^lu_je_j$ we have
 \beq{16}
\dot u_j=v_j\,.
 \eq
and
 \beq{16a}
\dot v_j=-\p_{u_j}U(\bfu,\bfS,\bfT)=%-\sum_{\al\in R^+}\frac{2\al(j)Y_\al(X_\al+X_{-\al})}{(\al,\al)}=
 \eq
 $$
-4\sum_{\al\in R^+}\frac{\al(j)}{\al^2}\left(\frac{S_\al
T_\al}{\sinh(\bfu_\al)} + (S_\al^2+T_\al^2-2\cosh(\bfu_\al)S_\al
T_\al)\frac{\cosh(\bfu_\al)}{\sinh^3(\bfu_\al)}\right)\,,~
(\al(j)=\al(e_j))\,.
 $$
The equations for $\bfS$ and $\bfT$ have form (\ref{fe}), (\ref{fe1}) with the Hamiltonian (\ref{qh})
 $$
H^{top}(\bfS,\bfT)=\oh\sum_{\nu\in R}\left(\oh T_\nu^2f(\nu)+  T_\nu
g(\nu)\right)\,,
 $$
where from (\ref{ha3}) we find
 \beq{gf}
f(\nu)=\frac{4}{\nu^2\sinh^2(\bfu_\nu)}\,,~~
g_S(\nu)=-\frac{4T_{\nu}\cosh(\bfu_\nu)}{\nu^2\sinh^2(\bfu_\nu)}\,,~~
g_T(\nu)=-\frac{4S_{\nu}\cosh(\bfu_\nu)}{\nu^2\sinh^2(\bfu_\nu)}\,.
 \eq
Then using (\ref{fe}) and (\ref{fe1}) we obtain
 \beq{17}
\dot S_\ga=\frac{1}{2} \sum_{\al+\be=\ga}C_{\al\be} \Bigl(S_\al
S_\be(\f1{\sinh^2(\bfu_\be)}-\f1{\sinh^2(\bfu_\al)})
+\frac{S_{\be}T_{\al}\cosh(\bfu_\al)}{\sinh^2(\bfu_\al)}-
\frac{S_{\al}T_{\be}\cosh(\bfu_\be)}{\sinh^2(\bfu_\be)}\Bigr)\,,
 \eq
 \beq{18}
\dot T_\ga=\frac{1}{2}\sum_{\al+\be=\ga}C_{\al\be} \Bigl(T_\al
T_\be(\f1{\sinh^2(\bfu_\al)}-\f1{\sinh^2(\bfu_\be)})
-\frac{T_{\be}S_{\al}\cosh(\bfu_\al)}{\sinh^2(\bfu_\al)}+
\frac{T_{\al}S_{\be}\cosh(\bfu_\be)}{\sinh^2(\bfu_\be)}\Bigr)\,.
 \eq
We will prove in Appendix C that the equations of motion (\ref{16}), (\ref{16a})  (\ref{17}), (\ref{18}) have the Lax form
  \beq{15}
 \p_t\eta=[\eta,M]\,,
  \eq
where the $M$ operator is defined as
 \beq{14}
M=\sum_{\al\in R^+}Y_\al
U_\al\,,~~Y_\al(\bfu_\al)=\p_{t}X_\al(t)|_{t=\bfu_\al}\stackrel{(\ref{131})}{=}\frac{S_\al\cosh(\bfu_\al)-T_\al
}{\sinh^2(\bfu_\al)}\,.
 \eq
%The existence of the $M$ operator is not sufficient for the complete integrability of these systems
% Section 3.

%%%%%%%%%%%%%%%%%%%%%%%%%%%%%%%%%%%%%%%%%%%%%%%%%%%%%%%%%%%%%%%%%%%%%%%%%%%%%%%%

%%%%%%%%%%%%%%%%%%%%%%%%%%%%%%%%%%%%%%%%%%%%%%%%%%%%%%%%%%%%%%%%%%%%%%%%%%%%%%%%
\subsection{Model II}

In this model we consider (instead of $\clR_I(G^\mC)$ (\ref{sss}))
the unreduced phase space
 \beq{ss1}
\clR_{II}(G^\mC)=T^*G^\mC\,.
 \eq
 Then
$$
\dim_\mR\,\clR_{II}(G^\mC)=4\dim_\mR\,G^\mC.
$$
%For $G=\SLN$ $\,\dim\,(\clR)=2(N^2-1)+N(N-1)$.
In the left trivialization $T^*G^\mC \sim G^\mC\times (\gg^\mC)^*$.
It is described by the coordinates   $(r,\xi)$, where
 $r\in G^\mC,\xi\in(\gg^\mC)^*$.
 The symplectic form $\om=D\vartheta$, is similar to (\ref{8}), where 1-form $\vartheta$ is
 \beq{81}
\vartheta^{G^\mC}_{II}=(\xi,r^{-1}Dr))\,.
 \eq
Define the quadratic Hamiltonian
 \beq{ha1}
H=\oh(\xi,\xi)\,.
 \eq

Consider the symplectic quotient $\clR_{II}(G^\mC)$ with respect the
following gauge symmetry. The gauge group
  $ G^{gauge}_{II}=K$ acts as
 \beq{101}
K:\,r\to fr f^{-1}\,,~~\xi\to\Ad_f\xi\,.
 \eq
 The corresponding moment map is
 \beq{m11}
 \mu_{II}:=(\Ad_r \xi -\xi)|_{\gk}=0\,,~~\gk=\Lie\,(K)\,.
 \eq
 In accordance with (\ref{dp}) represent $r\in G^\mC$ as $r=k_1\exp(\bfu)k^{-1}_2$, where $k_{1,2}\in K$. Then
\beq{gac}
frf^{-1}=fk_1\exp(\bfu)k_2^{-1}f^{-1}\,,~~~\bfu\in\gh^\mR\,. \eq
 Let us fix the gauge assuming that
 \beq{fg2}
 f=k_2^{-1}\,.
 \eq
  Then
\beq{fg3}
frf^{-1}= h^{-1} \exp(\bfu)\,,~~~h=k_1^{-1}k_2\,.
\eq
 Note in (\ref{gac}) $k_1$, $k_2$ are defined up to
the left multiplication by an element from the Cartan torus $\clT$. It means that
 $h$ is defined up to the conjugations   $\clT$ (compare with (\ref{rg}))
 \beq{rg7}
h\to shs^{-1}\,.
\eq
Therefore, $h$ is an element of the quotient of $K$ under the $\clT$ action
 \beq{or}
h\in \ti K=\{h\sim shs^{-1}\,|\,s\in\clT\}\,.
 \eq
In these terms the moment map equation (\ref{m11}) assumes the form
\beq{m12} (\xi- \Ad_{h^{-1} \exp(\bfu)}\xi )|_{\gk}=0\,.
 \eq
 Let \beq{cde}
\xi=p+\xi_{\gk}\,,~~p\in\gp^\mC\,,~~\xi_{\gk}\in\gk
\eq
 is the Cartan decomposition (\ref{cd1}). For the Cartan
part $\xi|_{\gh^\mC}$ the corresponding decomposition is
 \beq{xd}
\xi_{\gh^\mC}=\bfv+\xi_{\gt}\,,
\eq
 where $\gt$ is the Cartan
subalgebra of $\gk$, and $\bfv\in\gh^\mR\subset \gp^\mC$. Since
$\Ad_K(\gp^\mC)\in\gp^\mC$ we have
 $$
\Ad_{h^{-1} \exp(\bfu)}\bfv)|_{\gk}=0\,.
 $$
Thus, solutions $\xi$ of (\ref{m12}) are defined up to the adding
$\bfv\in\gh^\mR$. The moment map constraints (\ref{m11}) along with
the gauge fixing define the reduced phase space
 $$
\ti\clR^{red}_{II}(G^\mC)=\clR_{II}(G^\mC)//K=\mu_{II}^{-1}(0)/K\,.
 $$
%Coordinates on this space are $(h\in K\,,\xi\in\gp^\mC\,,\bfv\,,\bfu)$.
%

The variables $(\xi,r)$ being defined on $\ti\clR^{red}_{II}$
satisfy the moment map equation (\ref{m12}), where
 \beq{rr}
r=h^{-1}\exp(\bfu)\,,~~h\in K\,,~~\exp(\bfu)\in\clH^\mR\,,\bfv\in\gh^\mR\,.
 \eq
 Thus
 \beq{ifo}
 \ti\clR^{red}_{II}(G^\mC)\sim T^*\clH^\mR\times T^*K=\{(\bfv,\exp\bfu),(\xi_\gk,h)\}\,.
 \eq

\begin{rem}
For this model upon the symplectic reduction we deal with the complex symplectic
space (\ref{ss1}) with the complex one form (\ref{81}) and the Hamiltonian
(\ref{ha1}). The real symplectic space (\ref{ifo}) arises after the symplectic reduction
with respect to the action of the real group (\ref{101}).
\end{rem}

  By comparing
$\ti\clR^{red}_{II}(G^\mC)$ with $\ti\clR^{red}_{I}(G^\mC)$
(\ref{rs1}).
% with
%variables $\eta$ (\ref{12}) and $(\bfv,\bfu)$ (\ref{vu}).
%%Let $\xi=p+\xi|_{\gk}$ is the Cartan decomposition (\ref{cd1}).
 define
the variables on $\ti\clR^{red}_{II}$ as
 \begin{subequations}\label{TS}
  \begin{align}
 &\xi=\eta \,,\\
 &\xi|_\gk=-\bfT\,,\\
 &\bfS=\Ad_h\bfT \,.
 \end{align}
  \end{subequations}
  In these terms the moment map equation (\ref{m12}) assumes the form
  $\bfS+\Ad_{\exp(\bfu)}\eta|_\gk=0$.
  It coincides with (\ref{me}b), while (\ref{TS}b) coincides with (\ref{me}a).
   The last equation (\ref{TS}c) is just the definition of $\bfS$ (\ref{ov}).

    The map (\ref{TS}a) (\ref{TS}b) along with the identification of $\exp(\bfu)$ in the both
 models is symplectomorphism. In fact,
 using (\ref{m12}), (\ref{xd}) and (\ref{TS}) one can prove that
$$
D(\xi,r^{-1}Dr)=(D\bfv,D\bfu)+D(\bfT,h^{-1}Dh)\,.
$$
It coincides with the form (\ref{sf1}). Note that
the Hamiltonian (\ref{ha1}) coincides with (\ref{ha}).

   %Our goal is to prove the isomorphism of the space $\clR^{red}_{II}$ and $\clR^{red}_{I}$ (\ref{ps2}).

In this way we described the passage from $\clR^{red}_{II}$ to
$\clR^{red}_{I}$. The inverse procedure is straightforward. In the
Model I we defined the gauge  action $g\to f_1gf_2^{-1}$ (\ref{10}),
(\ref{9}). Let $g=k_1\exp(\bfu)k_2^{-1}$ (\ref{dp}). We fix the
gauge $f_2=k_2^{-1}$ as in (\ref{gac}), (\ref{fg2}). Then we come to
the Model II with $r=g$ and the residual gauge action by $f_1\in K$.
 Thereby,
$\ti\clR^{red}_{II}$ is symplectomorphic to
 $\ti\clR^{red}_{I}$.
 Now take into account the
 residual gauge symmetry (\ref{rg7}). It is the same as in the model I (\ref{rg}).
 Thus,  we come to the symplectic quotient (see (\ref{ps2})
 \beq{dred}
 \ti\clR^{red}_{II}(G^\mC)//\clT=T^*\clH^\mR\times (T^*K//\clT))\,,
 \eq
 where $T^*K//\clT=\{(\bfT,h)\}$ is defined by the condition (\ref{or}) and
 the moment map equation with respect to the torus $\clT$ action $(\Ad_h\bfT-\bfT)|_\gt=0$.
 The dimension of the reduced space is
 \beq{dre}
 \dim_\mR\,( \ti\clR^{red}_{II}(G^\mC)//\clT )=2\dim\,T^*K=2\sum_{j=1}^l(2d_j-1)\,.
 \eq
%
%
 %Thenand (\ref{m12}) coincides with the moment equations (\ref{mc}a) and
% while (\ref{TS}c) is the definition of $\bfS$ (\ref{ov}).
%%\beq{m2v}
%%\xi|_{\gk}=\bfT\,,~~\xi=-\eta\,, ~~\bfS=h\bfT h^{-1}\,,~~h\,\,{\rm from\,}\,(\ref{rr})\,.
%%\eq
%%In these terms the moment constraint equations (\ref{m12}) coincides with (\ref{mc}a), (\ref{mc}b).
%
%
In fact,  one can choose
 $(\bfS\in\clO_K\,,\bfT\in\clO_K)$ as  coordinates on $\ti\clR^{red}_{II}$ as in
(\ref{ps2})
$$
  T^*\clH^\mR\times (\clO_K\times\clO_K)=\{(\bfv,\exp\bfu),(\bfT,\bfS)\}\,.
$$
 Passing to the symplectic quotient with respect to the $\clT$ action we come to the reduced
 space
\beq{grs1}
 \clR^{red}_{II}(G^\mC)\sim T^*\clH^\mR\times (\clO_K\times\clO_K)//\clT\,.
\eq
It has dimension (see (\ref{dre}))
\beq{dre1}
 \dim_\mR\, \clR^{red}_{II}(G^\mC)=2\dim\,(\clO_K\times\clO_K)=4\sum_{j=1}^l(d_j-1)\,.
 \eq
The dimension of $\clR^{red}_{II}$ coincides with the dimension of $\clR^{red}_{I}$ (\ref{reds}).

 The Model II as the Model I becomes integrable in the real form.
First, consider the real form $\ti\clR^{red}_{II}(G^\mR)$
 of the phase space $ \ti\clR^{red}_{II}(G^\mC)$  (\ref{ifo})
%\beq{ifor}
%\ti\clR^{red}_{II}(G^\mR)=T^*\clH^\mR\times (T^* U)=
%\{(\bfv,\exp\bfu),(\xi_\gu,h\in U=K\cap G^\mR)\}\,.
%\eq
\beq{inte}
\ti\clR^{red}_{II}(G^\mR)=T^*\clH^\mR\times (T^* U)\sim\{\xi=\eta\,,\,(\ref{231})\,,\,(\ref{131})\,, \,\bfT\in\gu\,,\,h\in U\}\,.
\eq
\beq{m2xa}
\xi=\bfv+\sum_{\al\in R}X_\al E_\al\,,~~
X_\al=\frac{T_\al e(-\bfu_\al)-S_\al}{\sinh(\bfu_\al)}\,,~~
  X_{-\al}=\frac{T_\al e(\bfu_\al)-S_\al}{\sinh(\bfu_\al)}\,.
   \eq
From (\ref{du}) we find its dimension
\beq{d2}
\dim\,\ti\clR^{red}_{II}(G^\mR)=2\sum_{j=1}^ld_j\,.
\eq
Passing from $T^* U$ to the coadjoint orbits $\clO_U\times\clO_U$ we come to
the similar to (\ref{tu}) phase space
\beq{grs}
 \clR^{red}_{II}(G^\mR)\sim T^*\clH^\mR\times (\clO_U\times\clO_U)\,.
\eq It has dimension $\dim\,
\clR^{red}_{II}(G^\mR)=2\sum_{j=1}^ld_j-2\rank\,U$ (\ref{tud}).
After identifications (\ref{TS}) of the models one can use the same
set of integrals of motion (\ref{im1}). Following Section 2.1.3 we find the needed
 for integrability number of integrals.

%%%%%%%%%%%%%%%%%%%%%%%%%%%%%%%%%%%%%%%%%%%%%%%%%%%%%%%%%%%%%%%%%%%%%%%%%%%%%%%%%%%%

%%%%%%%%%%%%%%%%%%%%%%%%%%%%%%%%%%%%%%%%%%%%%%%%%%%%%%%%%%%%%%%%%%%%%%%%%%%%%%%%
\subsection{Model III}
\subsubsection{Symplectic reduction}
Let $\clX^\mC=G^\mC/K$ be some Riemannian symmetric space. As the
staring phase space consider
 \beq{ss5}
\clR^{G^\mC}_{III}=T^*G^\mC\times T^*\clX^\mC\,,
 \eq
 where as above $T^*G^\mC$ is described by the coordinates $\rho\in G^\mC$ and
 $\varsigma\in \gg^\mC$.

Let us define the coordinates on the cotangent bundle $T^*\clX^\mC$.
It can be represented as the symplectic quotient $T^*G^\mC//K$. We
may choose the coordinates on $T^*G^\mC$
 as $(\zeta\in\gg^\mC\,,g\in G^\mC)$.
The one-form
\beq{oct}
\vartheta=(\zeta,Dgg^{-1})
\eq
 is invariant under the $K$-action
\beq{kak} g\to kg\,, ~~\zeta\to\Ad_k\zeta\,. \eq
 The symplectic
quotient $T^*G^\mC//K\sim T^*\clX^\mC$ is defined by the moment map
constraints $\zeta|_\gk=0$ and the gauge-invariant variables
$\Ad_{g^{-1}}\zeta$
 \beq{css}
T^*\clX^\mC=\{\bfP(g,\zeta)=\Ad_{g^{-1}}\zeta\,|\,g\sim
kg\,,\,\zeta|_\gk=0\}\,. \eq
 In fact, from (\ref{kak}) we have
$\bfP(kg,\Ad_k(\zeta))=\bfP(g,\zeta)$.
 The form on $\clR^{G^\mC}_{III}$ (\ref{ss5}) is the sum of two one-form
\beq{23}
\vartheta^{G^\mC}_{III}=(\varsigma,\rho^{-1}D\rho))+(\zeta,Dgg^{-1})\,,~~(\zeta|_\gk=0)
\eq
Consider the quadratic Hamiltonian
 \beq{ha5}
H=\oh(\varsigma,\varsigma)\,.
 \eq
The symplectic symmetry group $ G^{gauge}_{III}=G^\mC$ acts as
 \beq{301}
G^\mC\,:\,\rho\to f\rho f^{-1}\,,~~\varsigma\to\Ad_f\varsigma\,,~~g\to gf^{-1}\,,~~\zeta\to\zeta\,.
 \eq
 The corresponding moment map is
 \beq{m31}
 \mu_{III}:=-\varsigma+ \Ad_\rho \varsigma -\bfP\,.
 \eq
By the gauge action one can transform $r$ to the Cartan subgroup
\beq{31}
 \rho= f\exp\bfu^\mC f^{-1}\,,~~\exp\bfu^\mC\in\clH^\mC\,.
\eq
 \beq{32}
\Ad_{g^{-1}}\zeta=\bfP=\bfP|_{\gh^\mC}+\sum_{\al\in R}\bfP_\al E_\al
\,,~~ \bfP_\al=\bfP_\al^1+\imath\bfP_\al^2\,.
\eq
Then solution of
the moment map constraints $\mu_{III}=0$ takes the form
\beq{kc3}
\varsigma=\bfv^\mC+\bfP|_{\gh^\mC}+\sum_{\al\in
R}\frac{\bfP_\al}{1-\exp\bfu^\mC} E_\al\,, ~~\bfv^\mC\in\gh^\mC\,.
\eq
 The residual symmetry is the Cartan subgroup $\clH^\mC$. Its
adjoint action does not change the gauge (\ref{31}). It generates
the moment map $\mu=\bfP|_{\gh^\mC}$. Thus, finally taking
$\bfP|_{\gh^\mC}=0$ we obtain \footnote{In this expression we do not
fix the gauge action of $\Ad_{\clH^\mC}$ on $\bfP$.}
 \beq{33}
\varsigma=\bfv^\mC+\sum_{\al\in R}\frac{\bfP_\al}{1-\exp\bfu_\al^\mC} E_\al\,.
\eq
In this way the reduced phase space is
\beq{34}
\clR^{red}_{III}(G^\mC)\sim T^*\clH^\mC\times (T^*\clX^\mC//\clH^\mC)=
\{(\bfv^\mC,\exp\bfu^\mC),\bfP//\clH^\mC\}\,.
\eq
From (\ref{dgx}) we have
$$
\dim_\mR\,\clR^{red}_{III}(G^\mC)=\dim_\mR\,T^*\clX^\mC=2\sum_{j=1}^l(2d_j-1)\,.
$$
It coincides with $ \dim_\mR\,( \ti\clR^{red}_{II}(G^\mC)//\clT)$ (\ref{dre}).

%%%%%%%%%%%%%%%%%%%%%%%%%%%%%%%%%%%%%%%%%%%%%%%%%%%%%%%%%%%%%%%%%%%%%%%%%%%%%%%%

\subsubsection{Real form and integrals of motion}
Consider the real form of the model III. The symmetric space
$\clX=G^\mC/K$ is replaced by $\clX^\mR=G^\mR/U$ and $T^*\clH^\mC$
is replaced by $T^*\clH^\mR$:
 \beq{clr3} \clR^{red}_{III}(G^\mR)\sim
T^*\clH^\mR\times (T^*\clX^\mR//\clH^\mR)=
\{(\ti\bfv,\exp\ti\bfu),\ti\bfP//\clH^\mR\}\,,~~\ti\bfv\,,\ti\bfu\,,\ti\bfP_\al\in\mR\,.
 \eq
   From (\ref{dgxr}) we obtain
   \beq{clr10}
\dim_\mR\,\clR^{red}_{III}(G^\mR)=\dim_\mR\,T^*\clX^\mR=2\sum_{j=1}^ld_j\,.
 \eq
 Instead of (\ref{33}) we have
 \beq{333} \varsigma=\ti\bfv+\sum_{\al\in
R}\frac{\ti\bfP_\al}{1-\exp\ti\bfu_\al} E_\al\,,
 \eq where
$\ti\bfP_\al\in\mR$. It is defined up to the equivalence
$\ti\bfP_\al\sim\ti\bfP_\al\exp (x)$, $x\in\mR$. The corresponding
Hamiltonian (\ref{ha5}) is
 $$
H=\oh(\ti\bfv,\ti\bfv)-\sum_{\al\in R^+}\frac{\ti\bfP_\al\ti\bfP_{-\al}}{(\al,\al)\sinh^2(\ti\bfu_\al/2)}\,.
 $$

Now we deal with the well-defined real integrable system. We want to find the independent integrals
of motion needed for the integrability.
We will demonstrate below that the integrals of motion take the
form (compare with (\ref{im1}))
\beq{in3}
I_{jk}=({\bfP}^{d_j-k}\varsigma^k)\,, ~~k=0,\ldots,d_j\,.
 \eq
 Since $\mu_{III}=0$, we replace $\bfP$ in this expression with
 $\Ad_{\exp(\bfu)}\varsigma-\varsigma$. Therefore,
 $$
 I_{jk}=
((\Ad_{\exp(\bfu)}\varsigma-\varsigma)^{d_j-k}\varsigma^k)\,.
 $$
The number of integrals $I_{jk}$ for the fixed $j$ is $d_j$.
%Two terms $I_{j,0}=(\bfP^{d_j})$  and
%$I_{j,d_j}=(\varsigma^{d_j})$ coincide. Thereby, the number In this way we have
The total the number of independent integrals of motion is equal to
  $\sum_{j=1}^ld_j=\oh\dim_\mR\,\clR^{red}_{III}(G^\mR)$ (\ref{clr10})
  needed for integrability.

%%%%%%%%%%%%%%%%%%%%%%%%%%%%%%%%%%%%%%%%%%%%%%%%%%%%%%%%%%%%%%%%%%%%%%

\subsubsection{Model II and Model III}

Here we establish the equivalence of Model II and Model III.
First prove the symplectomorphism of the phase spaces (\ref{dred}) and (\ref{34})
\beq{i23}
 \ti\clR^{red}_{II}(G^\mC)//\clT=T^*\clH^\mR\times (T^*K//\clT))
 \sim
\clR^{red}_{III}(G^\mC)= T^*\clH^\mC\times (T^*\clX^\mC//\clH^\mC)
\eq
Consider first "the coordinate part". In the Model III one can take
$\rho=\exp(\bfu^\mC)$ (\ref{31}). Let $g=kp$ be the polar
decomposition of $g\in K\setminus G$. Assume that the gauge
transformation $f$ in (\ref{301}) is equal to $p$. Then by
(\ref{301}) transform $\exp(\bfu^\mC)$ to $\ti r_{II}$
$$
\exp(\bfu^\mC)=p\ti r_{II}p^{-1}\,.
$$
%=\ti k_1\exp(\ti\bfu)\ti k_2^{-1}

 After this transformation we are left with the gauge group $K$ as in the Model II.
Acting as in (\ref{fg3}) by $f\in K$ transform $\ti r_{II}$ to the form
$\ti r_{II}\to \ti h^{-1}\exp(\ti\bfu)$. In this way we come to the equality
\beq{2t3}
k^{-1}p^{-1}(g)\exp(\bfu^\mC)p(g)k=\ti h^{-1}\exp(\ti\bfu)\,,
\eq
where $\ti h$ is defined up to the conjugation and $p(g)$ up to the left multiplication
$$
h\sim tht^{-1}\,,~t\in\clT\,,~~\,p(g)\sim sp(g)\,,~s\in\clH^\mC\,.
$$
Thus, $\exp(\bfu^\mC)$ and $h^{-1}\exp(\bfu)$ are conjugated under the $G^\mC$ action.
The equality (\ref{2t3}) establishes the isomorphism
$$
\clH^\mC\times(\clH^\mC\setminus\clX^\mC)\sim\clH^\mR\times (\clT\setminus K/\clT)\,,
$$
$$
(\exp(\bfu^\mC),p(g))\sim(\exp(\ti\bfu),\ti h)\,.
 $$
Compare $\mu_{II}$ (\ref{m11}) and $\mu_{III}$ (\ref{m31}). After
the gauge transformation by $f=p$ the transformed $g$ belongs to
$K$. For this reason $\bfP=\Ad^{-1}_g(\zeta)\in\gp^\mC$ (\ref{32}).
Therefore, its projection on the subalgebra $\gk$ vanishes
$Pr\,\bfP|_{\gk}=0$.
% \mu_{III}:=-\varsigma+ \Ad_r \varsigma
 Then the moment map  equation
  $\mu_{III}=0$ becomes
$$
(\varsigma- \Ad_{\ti h^{-1}\exp(\ti\bfu)} \varsigma)|_{\gk}=0\,.
$$
It coincides with the moment map equation $\mu_{II}=0$ (\ref{m12})
in the Model II. As in (\ref{ifo}) we come to the space
$T^*\clH^\mR\times T^*K$
 with parameters $\{(\ti\bfv,\exp\ti\bfu),(\varsigma_\gk,\ti h)\}$.

Now compare $\clR^{red}_{III}(G^\mR)$ (\ref{clr3}) with
$\ti\clR^{red}_{II}(G^\mR)$  (\ref{inte}). Notice that their
dimensions (\ref{d2}) and (\ref{clr10}) coincide. The isomorphism is
given by the map
 \beq{i230}
\bfv=\ti\bfv\,,~~\bfu=\ti\bfu\,,~~\bfP_\al=(1-\exp\bfu_\al) X_\al\,,
 \eq
  where $X_\al$ is defined in (\ref{m2xa}). Then we have
%
%X=\sum_{\al\in R}X_\al E_\al\,,~~
%X_\al=\frac{T_\al e(-\bfu_\al)-S_\al}{\sinh(\bfu_\al)}\,,~~
%  X_{-\al}=\frac{T_\al e(\bfu_\al)-S_\al}{\sinh(\bfu_\al)}\
 $$
\left\{
\begin{array}{ll}
\bfP_{\al}=(1-\exp(\bfu_\al))\frac{\bfT_\al e(-\bfu_\al)-\bfS_\al}{\sinh(\bfu_\al)}\,, & \al\in R^+\,, \\
\bfP_{\al}=  (1-\exp(-\bfu_\al))\frac{\bfT_\al e(\bfu_\al)-\bfS_\al}{\sinh(\bfu_\al)}\,, &\al\in R^-\,.
\end{array}
\right.
 $$
On the other hand, $\bfS_\al$ and $\bfT_\al$ can be expressed
as linear combinations of $\bfP_\al$ and $\bfP_{-\al}$. In this way we come to the diffeomorphism
\beq{is23}
\clR^{red}_{III}(G^\mR)\,(\ref{clr3}) \,\sim\,
\ti\clR^{red}_{II}(G^\mR)\, (\ref{inte})\,.
\eq
It can be proved that
this map is the symplectomorphism.
%We come to this point later.

%%%%%%%%%%%%%%%%%%%%%%%%%%%%%%%%%%%%%%%%%%%%%%%%%%%%%%%%%%%%%%%%%%%%%%%%%%%%%%%%%%%%%%%%%%%%%%%%
%%%%%%%%%%%%%%%%%%%%%%%%%%%%%%%%%%%%%%%%%%%%%%%%%%%%%%%%%%%%%%%%%%%

%\newpage

\section{Higgs bundles. Model I}

\setcounter{equation}{0}

The described system is the Hitchin system over a singular curve.
Here we define the corresponding holomorphic bundles.
%\subsection{Model of type I}
\subsection{Holomorphic bundles}

\paragraph{Base spectral curve.}
The base spectral curve is a singular curve $\Si^I$. It is a
collection of two rational curves $\Si_\al\sim\mC P^1$, $(\al=1,2)$
with corresponding holomorphic coordinates $z_1\,,z_2\in\mC_{1,2}$.
The disjoint union of $\Si_\al$ is mapped to $\Si^I$ by the
normalization
 $$
\pi\,:\,\Si_1\cup\Si_2\to\Si^I\,,~~ \left\{
\begin{array}{l}
  \pi(z_1=0)=\pi(z_2=0)\,, \\
   \pi(z_1=\infty)=\pi(z_2=\infty)\,.
\end{array}
\right.
 $$

%\vspace{-1.5cm}

\unitlength 1mm % = 2.845pt
\linethickness{0.4pt}
\ifx\plotpoint\undefined\newsavebox{\plotpoint}\fi % GNUPLOT compatibility
\begin{picture}(108.75,120.25)(0,0)
\put(86.75,97.75){\oval(16,36.5)[]}
%\circle(86.75,97.5){36.031}
\put(104.766,97.5){\line(0,1){.857}}
\put(104.745,98.357){\line(0,1){.8551}}
\put(104.684,99.212){\line(0,1){.8512}}
\multiput(104.582,100.063)(-.028425,.169079){5}{\line(0,1){.169079}}
\multiput(104.44,100.909)(-.030363,.139613){6}{\line(0,1){.139613}}
\multiput(104.258,101.746)(-.031689,.118294){7}{\line(0,1){.118294}}
\multiput(104.036,102.574)(-.03262,.102071){8}{\line(0,1){.102071}}
\multiput(103.775,103.391)(-.033279,.089248){9}{\line(0,1){.089248}}
\multiput(103.476,104.194)(-.033738,.078808){10}{\line(0,1){.078808}}
\multiput(103.138,104.982)(-.031208,.064261){12}{\line(0,1){.064261}}
\multiput(102.764,105.753)(-.0315962,.0578804){13}{\line(0,1){.0578804}}
\multiput(102.353,106.506)(-.0318629,.0522896){14}{\line(0,1){.0522896}}
\multiput(101.907,107.238)(-.0320266,.0473337){15}{\line(0,1){.0473337}}
\multiput(101.427,107.948)(-.0321019,.0428968){16}{\line(0,1){.0428968}}
\multiput(100.913,108.634)(-.0321,.0388904){17}{\line(0,1){.0388904}}
\multiput(100.367,109.295)(-.0320296,.0352461){18}{\line(0,1){.0352461}}
\multiput(99.791,109.93)(-.03367,.0336825){18}{\line(0,1){.0336825}}
\multiput(99.185,110.536)(-.0352342,.0320427){18}{\line(-1,0){.0352342}}
\multiput(98.55,111.113)(-.0388785,.0321144){17}{\line(-1,0){.0388785}}
\multiput(97.89,111.659)(-.0428848,.0321179){16}{\line(-1,0){.0428848}}
\multiput(97.203,112.173)(-.0473218,.0320442){15}{\line(-1,0){.0473218}}
\multiput(96.494,112.653)(-.0522777,.0318823){14}{\line(-1,0){.0522777}}
\multiput(95.762,113.1)(-.0578687,.0316177){13}{\line(-1,0){.0578687}}
\multiput(95.009,113.511)(-.06425,.031232){12}{\line(-1,0){.06425}}
\multiput(94.238,113.886)(-.071632,.030698){11}{\line(-1,0){.071632}}
\multiput(93.45,114.223)(-.089236,.033312){9}{\line(-1,0){.089236}}
\multiput(92.647,114.523)(-.102059,.032658){8}{\line(-1,0){.102059}}
\multiput(91.831,114.784)(-.118283,.031733){7}{\line(-1,0){.118283}}
\multiput(91.003,115.006)(-.139601,.030415){6}{\line(-1,0){.139601}}
\multiput(90.165,115.189)(-.169068,.028488){5}{\line(-1,0){.169068}}
\put(89.32,115.331){\line(-1,0){.8512}}
\put(88.469,115.433){\line(-1,0){.8551}}
\put(87.614,115.495){\line(-1,0){1.714}}
\put(85.9,115.496){\line(-1,0){.8551}}
\put(85.045,115.435){\line(-1,0){.8512}}
\multiput(84.193,115.333)(-.169089,-.028362){5}{\line(-1,0){.169089}}
\multiput(83.348,115.191)(-.139624,-.030312){6}{\line(-1,0){.139624}}
\multiput(82.51,115.01)(-.118306,-.031645){7}{\line(-1,0){.118306}}
\multiput(81.682,114.788)(-.102084,-.032583){8}{\line(-1,0){.102084}}
\multiput(80.865,114.527)(-.08926,-.033246){9}{\line(-1,0){.08926}}
\multiput(80.062,114.228)(-.07882,-.033709){10}{\line(-1,0){.07882}}
\multiput(79.274,113.891)(-.064273,-.031184){12}{\line(-1,0){.064273}}
\multiput(78.503,113.517)(-.0578922,-.0315747){13}{\line(-1,0){.0578922}}
\multiput(77.75,113.106)(-.0523014,-.0318434){14}{\line(-1,0){.0523014}}
\multiput(77.018,112.661)(-.0473456,-.032009){15}{\line(-1,0){.0473456}}
\multiput(76.308,112.181)(-.0429087,-.032086){16}{\line(-1,0){.0429087}}
\multiput(75.621,111.667)(-.0389024,-.0320855){17}{\line(-1,0){.0389024}}
\multiput(74.96,111.122)(-.035258,-.0320165){18}{\line(-1,0){.035258}}
\multiput(74.325,110.545)(-.033695,-.0336575){18}{\line(-1,0){.033695}}
\multiput(73.718,109.94)(-.0320558,-.0352223){18}{\line(0,-1){.0352223}}
\multiput(73.141,109.306)(-.0321289,-.0388666){17}{\line(0,-1){.0388666}}
\multiput(72.595,108.645)(-.0321338,-.0428729){16}{\line(0,-1){.0428729}}
\multiput(72.081,107.959)(-.0320618,-.0473098){15}{\line(0,-1){.0473098}}
\multiput(71.6,107.249)(-.0319017,-.0522659){14}{\line(0,-1){.0522659}}
\multiput(71.154,106.517)(-.0316392,-.0578569){13}{\line(0,-1){.0578569}}
\multiput(70.742,105.765)(-.031255,-.064238){12}{\line(0,-1){.064238}}
\multiput(70.367,104.994)(-.030725,-.07162){11}{\line(0,-1){.07162}}
\multiput(70.029,104.207)(-.033346,-.089223){9}{\line(0,-1){.089223}}
\multiput(69.729,103.404)(-.032696,-.102047){8}{\line(0,-1){.102047}}
\multiput(69.468,102.587)(-.031777,-.118271){7}{\line(0,-1){.118271}}
\multiput(69.245,101.759)(-.030467,-.13959){6}{\line(0,-1){.13959}}
\multiput(69.062,100.922)(-.028551,-.169058){5}{\line(0,-1){.169058}}
\put(68.92,100.077){\line(0,-1){.8511}}
\put(68.817,99.225){\line(0,-1){.855}}
\put(68.755,98.37){\line(0,-1){2.5691}}
\put(68.815,95.801){\line(0,-1){.8513}}
\multiput(68.916,94.95)(.028299,-.1691){5}{\line(0,-1){.1691}}
\multiput(69.057,94.104)(.03026,-.139635){6}{\line(0,-1){.139635}}
\multiput(69.239,93.267)(.031601,-.118318){7}{\line(0,-1){.118318}}
\multiput(69.46,92.438)(.032545,-.102096){8}{\line(0,-1){.102096}}
\multiput(69.72,91.622)(.033213,-.089273){9}{\line(0,-1){.089273}}
\multiput(70.019,90.818)(.03368,-.078833){10}{\line(0,-1){.078833}}
\multiput(70.356,90.03)(.03116,-.064284){12}{\line(0,-1){.064284}}
\multiput(70.73,89.258)(.0315532,-.0579039){13}{\line(0,-1){.0579039}}
\multiput(71.14,88.506)(.031824,-.0523132){14}{\line(0,-1){.0523132}}
\multiput(71.586,87.773)(.0319914,-.0473575){15}{\line(0,-1){.0473575}}
\multiput(72.066,87.063)(.03207,-.0429206){16}{\line(0,-1){.0429206}}
\multiput(72.579,86.376)(.0320711,-.0389143){17}{\line(0,-1){.0389143}}
\multiput(73.124,85.715)(.0320034,-.0352699){18}{\line(0,-1){.0352699}}
\multiput(73.7,85.08)(.033645,-.0337075){18}{\line(0,-1){.0337075}}
\multiput(74.306,84.473)(.0352104,-.0320689){18}{\line(1,0){.0352104}}
\multiput(74.939,83.896)(.0388546,-.0321433){17}{\line(1,0){.0388546}}
\multiput(75.6,83.349)(.0428609,-.0321497){16}{\line(1,0){.0428609}}
\multiput(76.286,82.835)(.0472979,-.0320794){15}{\line(1,0){.0472979}}
\multiput(76.995,82.354)(.052254,-.0319211){14}{\line(1,0){.052254}}
\multiput(77.727,81.907)(.0578452,-.0316607){13}{\line(1,0){.0578452}}
\multiput(78.479,81.495)(.064226,-.031279){12}{\line(1,0){.064226}}
\multiput(79.249,81.12)(.071609,-.030751){11}{\line(1,0){.071609}}
\multiput(80.037,80.782)(.089211,-.033379){9}{\line(1,0){.089211}}
\multiput(80.84,80.481)(.102035,-.032734){8}{\line(1,0){.102035}}
\multiput(81.656,80.219)(.118259,-.031821){7}{\line(1,0){.118259}}
\multiput(82.484,79.997)(.139579,-.030519){6}{\line(1,0){.139579}}
\multiput(83.322,79.814)(.169047,-.028614){5}{\line(1,0){.169047}}
\put(84.167,79.671){\line(1,0){.8511}}
\put(85.018,79.568){\line(1,0){.855}}
\put(85.873,79.506){\line(1,0){1.714}}
\put(87.587,79.504){\line(1,0){.8551}}
\put(88.442,79.564){\line(1,0){.8513}}
\multiput(89.293,79.665)(.16911,.028237){5}{\line(1,0){.16911}}
\multiput(90.139,79.806)(.139646,.030208){6}{\line(1,0){.139646}}
\multiput(90.977,79.987)(.11833,.031557){7}{\line(1,0){.11833}}
\multiput(91.805,80.208)(.102108,.032507){8}{\line(1,0){.102108}}
\multiput(92.622,80.468)(.089285,.03318){9}{\line(1,0){.089285}}
\multiput(93.426,80.767)(.078845,.033651){10}{\line(1,0){.078845}}
\multiput(94.214,81.103)(.064296,.031136){12}{\line(1,0){.064296}}
\multiput(94.986,81.477)(.0579156,.0315317){13}{\line(1,0){.0579156}}
\multiput(95.738,81.887)(.0523251,.0318046){14}{\line(1,0){.0523251}}
\multiput(96.471,82.332)(.0473693,.0319738){15}{\line(1,0){.0473693}}
\multiput(97.182,82.812)(.0429325,.0320541){16}{\line(1,0){.0429325}}
\multiput(97.868,83.325)(.0389262,.0320566){17}{\line(1,0){.0389262}}
\multiput(98.53,83.87)(.0352818,.0319903){18}{\line(1,0){.0352818}}
\multiput(99.165,84.445)(.03372,.0336324){18}{\line(1,0){.03372}}
\multiput(99.772,85.051)(.0320819,.0351984){18}{\line(0,1){.0351984}}
\multiput(100.35,85.684)(.0321577,.0388427){17}{\line(0,1){.0388427}}
\multiput(100.896,86.345)(.0321657,.042849){16}{\line(0,1){.042849}}
\multiput(101.411,87.03)(.032097,.047286){15}{\line(0,1){.047286}}
\multiput(101.893,87.74)(.0319406,.0522422){14}{\line(0,1){.0522422}}
\multiput(102.34,88.471)(.0316822,.0578334){13}{\line(0,1){.0578334}}
\multiput(102.752,89.223)(.031303,.064215){12}{\line(0,1){.064215}}
\multiput(103.127,89.993)(.030778,.071598){11}{\line(0,1){.071598}}
\multiput(103.466,90.781)(.033412,.089199){9}{\line(0,1){.089199}}
\multiput(103.766,91.584)(.032772,.102023){8}{\line(0,1){.102023}}
\multiput(104.029,92.4)(.031865,.118247){7}{\line(0,1){.118247}}
\multiput(104.252,93.228)(.030571,.139567){6}{\line(0,1){.139567}}
\multiput(104.435,94.065)(.028676,.169036){5}{\line(0,1){.169036}}
\put(104.579,94.91){\line(0,1){.851}}
\put(104.682,95.761){\line(0,1){.855}}
\put(104.744,96.616){\line(0,1){.8838}}
%\end
\put(86.75,116){\circle*{1.118}}
\put(87,79.75){\circle*{1}}
\put(99.5,97.75){\circle*{1.118}}
\put(74.25,97.25){\circle*{.707}}
\put(81,120.25){$z_1=z_2=\infty$}
\put(81.5,75.5){$z_1=z_2=0$}
\put(61.5,98){$z_1=1$}
\put(108.75,98.75){$z_2=1$}
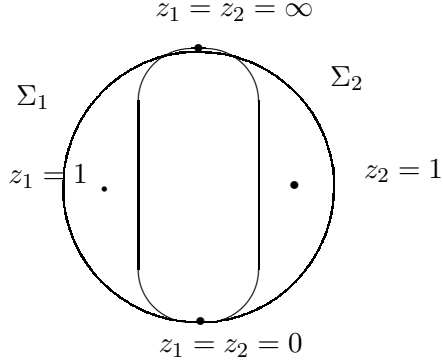
\put(89.5,70.5){\makebox(0,0)[cc]{Fig.1 The base spectral curve }}
\put(62.5,108.5){$\Sigma_1$}
\put(104.25,111){$\Sigma_2$}
\end{picture}

%%%%%%%%%%%%%%%%%%%%%%%%%%%%%%%%%%%%%%%%%%%%%%%%%%%%%%%%%%%%%%%%%%%%%%

\vspace{-70mm}

%%%%%%%%%%%%%%%%%%%%%%%%%%%%%%%%%%%%%%%%%%%%%%%%%%%%%%%

\paragraph{Holomorphic bundles.} Let $\clP(G^\mC)$ be the
principle $G^\mC$ - bundle over $\Si^I$, $V$ is a $G^\mC$ module and
$E(G^\mC)=\clP\times_{G^\mC}V$  is the associated vector bundle. It
is constructed by means of two vector bundles
$E_\al(G^\mC)=\clP_\al\times_{G^\mC}V$
 over $\Si_\al$.
 At the glued points we define the maps of the corresponding sections
\beq{hd1}
r_\infty\,:\gs|_{z_1=\infty}\in\G(E_1)\to \gs|_{z_2=\infty}\in\G(E_2)\,,~~
r_0\,:\gs|_{z_1=0}\in\G(E_1)\to \gs|_{z_2=0}\in\G(E_2)\,,
\eq
where we assume that $r_\infty\in G^\mC$ and $r_0\in K$ is a maximal compact subgroup
$K\subset G^\mC$. Let
 $\bp_\al+\bA_\al$ $(\bp_\al=\p_{\bz_\al})$ be the antiholomorphic connections
acting on the sections $\G(E_\al)$.
 Then
\beq{data}
\clD=\{(\bp_\al+\bA_\al)\,,~\al=1,2\,,~r_\infty\in G^\mC\,,~r_0\in K\}
\eq
 defines the  vector bundle $E^I(G^\mC)$ on the singular curve  $\Si^I$.
%The main difference with \cite{Ne} is that we fix a real  $K$-structure at the fibers
%over  the points $z_\al=0$. For $K=$SU$(N)$ it is an Hermitian structure.

\paragraph{Gauge group.} The group of  automorphisms $\clG$ of
the bundle $E(G^\mC)$ (the gauge group) is  the pair of smooth maps
\beq{320}
 f_\al\in\Si_\al\in C^\infty(\Si_\al)\to G^\mC\,.
  \eq
   Let
$\si$ be the involutive automorphism of $G^\mC$, such that its fixed
point set  is the maximal compact subgroup $K\subset G^\mC$.
 At the  point $z_\al=1$ we replace in (\ref{320}) $G^\mC$ with the maximal compact subgroup $K$
\beq{3.2}
f_\al(z_\al,\bz_\al)|_{z_\al=1}\in K\,.
\eq
% At the  point $z_2=1$ we assume that
%\beq{3.2a}
%f_2(z_\al,\bz_\al)|_{z_1=1}Id\,.
%\eq
%
We will refer to such vector bundle as the vector bundle with
\emph{the quasi-compact structure}. In particular, for $G^\mC=\SLN$
and $K=$SU$(N)$ the quasi-compact structure means that the gauge
group preserves a positive definite hermitian structures at the
marked points $z_\al=1$.

The gauge action on the data (\ref{dat}) has the form
 \beq{ggt2}
\bp_\al+\bA_\al\to f_\al(\bp_{\al}+\bA_\al) f^{-1}_\al\,,
 \eq
 \beq{ggt1} r_\infty\to f_2(\infty)r_\infty
f_1^{-1}(\infty)\,,~f_\al(\infty)=f_\al(z_\al,\bz_\al)|_{z_\al=\infty}\,,
\eq \beq{ggt3} r_0\to f_2(0) r_0
f_1^{-1}(0)\,,~f_\al(0)=f_\al(z_\al,\bz_\al)|_{z_\al=0} \,.
 \eq

\paragraph{The moduli space $\Bun_{I}(G^\mC)$.} The quotient
space $\Bun_{I}(G^\mC)=\clD/\clG$ of
 (\ref{data}), by (\ref{ggt2}), (\ref{ggt1}) , (\ref{ggt3})  is the moduli
 of vector $G^\mC$ bundles with the quasi-compact structure over the singular curve $\Si^I$.

It is proved below in Section 4.1 that  by the gauge group action
(\ref{ggt2}) of $\clG$ (\ref{320}), (\ref{3.2}) on the generic
configurations the connection forms  $\bA_\al$ on each component
$\Si_\al$ can be
 transformed to the trivial form
 \beq{cs1}
\bA_\al=0\,.
 \eq
Thereafter the residual gauge transformations are constant maps to $K$
 \beq{cgt}
\clG^{res}_\al =\{f_\al\in K\}\,.
 \eq
Put it otherwise, the almost all connections $\bA_\al$ have the pure
gauge form $\bA_\al=f_\al\bp f^{-1}_\al$ $f_\al\in\clG$. The
residual gauge transformations $\clG^{res}_\al$ preserving (\ref{cs1})
are constant maps to $G^\mC$. In fact, it is reduced to (\ref{cgt}).
 %$K$ the residual gauge transformations are
% \beq
%\clG^{res}_1 =\{f_\al\in K\}\,.
% \eq

For generic $r_\infty$ the transformation (\ref{ggt1}) allows one to take it to
the Cartan subgroup\\
 $\clH^\mR\subset G^\mR$ (see (\ref{dp3}))
\beq{ms1}
f_2(\infty)r_\infty f_1^{-1}(\infty)=\exp(\bfu)\,,~~\bfu\in\gh^\mR\,.
\eq
As in previous Section
these transformations are defined up to the action from the right
 by the final residual gauge transformation $\clG^{res}_2$
$$
f_\al(\infty)\to f_\al(\infty)s^{-1}\,, ~~s\in \clG^{res}_2=\clT\,,
$$
where  $\clT$ is the Cartan torus in $K$.
Denote the result of these actions on $r_0$ as
\beq{rh}
f_2 r_0 f_1^{-1}=h\in K\,.
\eq
%The transformation (\ref{ggt3}) is the action of the group $K\times K$ on the
%group $K$. The group $K=(K\times K)/K$  can be considered as the compact symmetric space.
In this way the element $h$  is defined up to the conjugation
 $$
h\to shs^{-1}\,,~~s\in \clG^{res}_2=\clT\,.
 $$
Therefore, $h$ is an element of the quotient  $\ti K=K/\clT$
(\ref{or}). Thus, a big cell of $\Bun^0_I(G^\mC)$ in the moduli
space is
 \beq{msh} \Bun^0_{I}(G^\mC)=\clD/\clG=(\clH^\mR,\ti K)\,.
  \eq
From (\ref{dk}) we find
 \beq{dms} \dim_\mR\,
\Bun_{I}(G^\mC)=\rank(G^\mR)+\dim\,K-\rank\,K=\dim\,K=
2\sum_{j=1}^ld_j-l\,.
 \eq
%%%%%%%%%%%%%%%%%%%%%%%%%%%%%%%%%%%%%%%%%%%%%%%%%%%%%%%%%%%%%%%%%%%

\subsection{Higgs bundles}

The quasi-compact Higgs bundle $\clH_{qc}(G^\mC)$ over $\Si$ is set of pairs
$$
\{(\p_{\bA_\al}=\bp_\al+\bA_\al,\Phi_\al)\,,~(r_\infty\,,\la_\infty)\,,~(r_0,\la_0)\}\,.
$$
Here
$$
\Phi_\al=\Phi_\al(z_\al,\bz_\al)dz_\al\in End(E_\al)\otimes\Om^{(1,0)}(\Si_\al)
$$
are the components of the Higgs fields.
The  pairs are
$$
(r_\infty,\xi_\infty)\in T^*(\Hom(E|_{z_1=\infty}\to E|_{z_2=\infty}))\sim T^*G\,,
$$
$$
(r_0,\xi_0)\in T^*(\Hom(E|_{z_1=0}\to E|_{z_2=0})) \sim T^*K\,,
$$
The Higgs bundle $\clH_G$ is a symplectic manifold equipped with the form
$$
\om=\sum_{\al=1,2}\int_{\Si_\al}( D\Phi_\al,D\bA_\al)
-\int_{\Si_1}\sum_{i=0,\infty}\de(z_i,\bz_i)|_{z_i}D(\la_i, r_i^{-1}Dr_i)+
$$
$$
\int_{\Si_2}\sum_{i=0,\infty}\de(z_i,\bz_i)|_{z_i}D(\la_i, r_i^{-1}Dr_i)\,.
$$
 The group of the automorphisms $\clG$ of the bundle $E$ (\ref{ggt2}), (\ref{ggt1})
 (\ref{ggt3}) can be lifted to
 symplectic automorphisms of the Higgs bundle $\clG$
$$
\Phi_\al\to f_\al\Phi_\al f^{-1}_\al\,,~~\la_0\to \Ad^{-1}_{f_1}|_{z_1=0}(\la_0)\,,
~~\la_\infty\to \Ad^{-1}_{f_1}|_{z_1=\infty}\la_\infty\,.
$$
Taking into account the gauge fixing (\ref{cs1}), (\ref{ms1}),
(\ref{rh}) we find the moment maps are
 \beq{m10}
 \mu_1=-\p_{\bz_1}\Phi_1+\de(z_1,\bz_1)|_{z_1=\infty}\la_\infty+
 \de(z_1,\bz_1)|_{z_1=0}\la_0\,,
 \eq
 \beq{m20}
\mu_2=-\p_{\bz_2}\Phi_2+\de(z_2,\bz_2)|_{z_2=\infty}(\Ad_{\exp(\bfu)}\la_\infty +
\de(z_2,\bz_2)|_{z_2=0}(\Ad_{h}\la_0)\,.
 \eq
%where
%$$
%p_\al=y_\al^{-1}p_\al y_\al- y_\al p_\al y_\al^{-1}\in\gk
%$$
% are the component of the moment maps coming from the forms
%$\om^{ss}_\al(y_\al,p_\al)$ (\ref{oss}).
%Note that due to (\ref{cd1}) the coordinates on the cotangent bundles $T^*\clX_\al$
%do not contribute to the moment maps.

Impose the moment map constraints $\mu_1=\mu_2=0$. They mean that
the Higgs fields $\Phi_\al (z_\al,\bz_\al)$ are holomorphic on
$\Si_\al$ and have simple poles at $z_\al=0,\infty$ with the
definite residues. Moreover, due to the quasi-compactness
(\ref{3.2}) the Higgs fields have the poles at $z_\al=1$ with
residues in the compact subalgebra $\gk$. We summarize this
structure in the Table:
 $$
\begin{tabular}{|c|c|c|c|}
\hline
  &$z_\al=0 $ & $z_\al=1$ & $z_\al=\infty$ \\
 \hline
\Res\,$\Phi_1$ & $\la_0 $ & $\clK_1\in\gk$ & $\la_\infty$\\
 \Res\,$\Phi_2$ & $\Ad_{r_0}\la_0 $ & $\clK_2\in\gk$ & $\Ad_{r_\infty}\la_\infty $ \\
  \hline
 \end{tabular}
$$
The fields $\Phi_\al$ satisfying these conditions take the form
\footnote{We omit here and in what follows for brevity the differentials $dz_\al$.}
$$
\Phi_1=\frac{\la_0}{z_1}+\frac{\la_0+\la_\infty+\clK_1}{1-z_1}\,,~~
\Phi_2=\frac{\Ad_{r_0}\la_0}{z_2}+\frac{\Ad_{r_0}\la_0+\Ad_{r_\infty}\la_\infty+\clK_2}{1-z_2}\,.
$$
Since the sums of residues vanishes
$$
\la_0+\la_\infty+\clK_1=0\,,~~\Ad_{r_0}\la_0+\Ad_{r_\infty}\la_\infty+\clK_2=0\,.
$$
%we have
%
%\beq{la1}
%\Phi_1=\frac{\la_0}{z_1}+\frac{\la_0+\la_\infty}{1-z_1}\,,~~
%\Phi_2=\frac{\Ad_{r_0}\la_0}{z_2}+\frac{\Ad_{r_0}\la_0+\Ad_{r_\infty}\la_\infty }{1-z_2}\,.
%\eq
and $\clK_\al\in\gk$ we have $\Res\,\Phi_\al(z_\al=1)|_{\gk}=0$, or
\beq{res5}
\la_0+\la_{\infty}|_{\gk}=0\,,~~\Ad_{r_0}\la_0+\Ad_{r_\infty}\la_\infty|_{\gk}=0\,.
\eq
The role of the Lax operators, depending on the spectral parameters
$z_\al$ will play the gauge transformed Higgs fields $\Phi_\al=
f_\al L_\al f^{-1}_\al$, where
 $f_\al(z_\al)|_{z_\al=\infty}$ transform $g$ to
 $e(\bfu)$ (\ref{ms1}) and $r_0$ to $h$.
Then
 \beq{la1}
L_1(z_1)=\frac{\la_0}{z_1}+\frac{\la_0+\la_\infty}{1-z_1}\,,~~
L_2(z_2)=\frac{\Ad_{h}\la_0}{z_2}+\frac{\Ad_{h}\la_0+\Ad_{\exp(\bfu)}\la_\infty }{1-z_2}\,.
 \eq
To come to notations of the previous Section we put
 $$
\la_0=\bfT:=\nu\,,~~\Ad_{h}\la_0=\bfS\,,~~\la_\infty=\eta\,.
$$
%
%where we replace $\nu=\bfT$ and $ h\nu h^{-1}=\bfS$.
%
In these notations (\ref{la1}) assumes the form
 \beq{la}
L_1(z_1)=\frac{\bfT}{z_1}+\frac{\bfT+\eta}{1-z_1}\,,~~
L_2(z_2)=\frac{\bfS}{z_2}+\frac{\bfS+\Ad_{\exp(\bfu)}\eta
}{1-z_2}\,.
 \eq
 Notice that  (\ref{res5})
 $$
 (\bfT+\eta)|_\gk=0\,,~~(\bfS+\Ad_{\exp(\bfu)}\eta)|_\gk=0
 $$
   becomes the moment map equations $\mu_l=\mu_R=0$ (\ref{m1}).

The moduli space of the Higgs bundle with the quasi-compact structure
 are parameterized  by $\bfS,\bfT\in\clO_K$ and
$\eta$ satisfying (\ref{m1}):
 \beq{hi1}
\clM_I(G^\mC)=\clH(G^\mC)//\clG=T^*H^\mR\times(\clO_K\times\clO_K)//\clT=
\{(\bfv,\exp\,\bfu,\,\bfT\,,\,\bfS=\Ad_h\bfT)\} \,.
\eq
Moreover,
\beq{dI}
\dim\,\clM_I(G^\mC)=\dim\,(\clO_K\times\clO_K)=4\sum_{j=1}^l(d_j-1)\,.
 \eq
From (\ref{ps2}) we then find that
$\clM_I(G^\mC)\sim\clR^{red}_{G^\mC}$ (\ref{ps2}). Thus, the moduli space $\clM_I(G^\mC)$
is the real symplectic manifold with the form (\ref{sf1}).

%%%%%%%%%%%%%%%%%%%%%%%%%%%%%%%%%%%%%%%%%%%%%%%%%%%%%%%%%%%%%%%%%%%%%%%%%%%%%%%%

\subsection{Integrals of motion}
The integrals of motion are obtained by means of the basis of
meromorphic differentials on $\Si_\al$
 \beq{dbe}
\mu^0_{k_j,d_j}=(z_\al)^{-k_j}(dz_\al)^{d_j}\,,~~
\mu^1_{k_j,d_j}=(1-z_\al)^{-k_j}(dz_\al)^{d_j}\,,~~(0<k_j\leq
d_j)\,.
 \eq
By expanding the gauge invariant quantities
 \beq{30}
( L_\al(z_\al)^{d_j})\,,~~k=1,\ldots,l\,,~ ~(d_j~{\rm are~ the~ invariants ~of}~ \gg)
\eq
 in this basis we obtain the invariant integrals.
Consider, for example, $L_1(z)$:
 \beq{il1}
( L_1(z)^{d_j})=
\sum_{k=0}^{d_j}C_{d_j}^kz^{k-d_j}(1-z)^{-k}(\bfT^{d_j-k}(\bfT+\eta)^k)=
\eq
$$
\sum_{k=0}^{d_j}C_{d_j}^kz^{k-d_j}(1-z)^{-k}\sum^k_{i=0}C_k^i(\bfT^{d_j-i}\eta^i)\,.
$$
Expanding $z^{k-d_j}(1-z)^{-k}$ as
$$
\f1{z^j(1-z)^k}=\sum_{i=0}^{j-1}c_iz^{i-j}+\sum_{n=0}^{k-1}b_n(1-z)^{n-k}\,,
$$
$$
c_i=\frac{k(k+1)\cdots(k+i-1)}{i!}\,,~~b_n=(-1)^n\frac{j(j+1)\cdots(j+n-1)}{n!}
 $$
we find the considered above integrals of motion (\ref{im1})
\beq{il2}
I_{jk}=(\bfT^{d_j-k}\eta^k)\,,~~k=0,\ldots,d_j\,.
 \eq
The number of  integrals is $\clN_G$ (\ref{ni}). Again, for the
integrability we have the deficit $\de_{G^\mC}$ (\ref{dc}).

%%%%%%%%%%%%%%%%%%%%%%%%%%%%%%%%%%%%%%%%%%%%%%%%%%%%%%%%%%%%%%%%%%%%%%%%%%%%%%%%%%%%%

\subsection{Real forms}

Following the general scheme \cite{BS} we  consider real points of some
complex structure  on the  moduli space of the Higgs bundles. We use a special involution $i$ of
the Higgs bundle.
 The operator is defined in the following way.
 \footnote{In \cite{BS} it is denoted $i_3$.}
Let $\si$ be the antiholomorphic involutive automorphism (\ref{gr}) acting on
the complex Lie algebra $\gg^\mC$
in such a way that the fixed point set $\gg^\mR$ is the normal real form of $\gg^\mC$.
Since the marked points $z_\al=0,1,\infty$ are invariant under
conjugation we can accompany it with the antiholomorphic
automorphism of $\Si$ $z_\al\to\bz_\al$.
 In general case the operator $i$ is defined as
 \footnote{The construction presented in \cite{BS} deals with bundles
over curves without marked points (\ref{01}). We extend it to
bundles with marked points (\ref{02}).}:
 \beq{01}
i(\p_{\bA_\al}(z_\al)\,,\Phi_\al(z_\al)= (\si\p_{\bA_\al}(\bar
z_\al)\,,\si\Phi_\al(\bar z_\al))\,, \eq

%%%%%%%%%%%%%%%%%%%%%%%%%%%%%%%%%%%%%%%%%%%%%%%%%%%%%%%%%%%%%%%%%%%%%%%%%%
\vspace{5mm}
%TeXCAD Picture [k12.tex]. Options:
%\grade{\on}
%\emlines{\off}
%\epic{\off}
%\beziermacro{\on}
%\reduce{\on}
%\snapping{\off}
%\pvinsert{% Your \input, \def, etc. here}
%\quality{8.000}
%\graddiff{0.005}
%\snapasp{1}
%\zoom{4.0000}
\unitlength 1mm % = 2.845pt
\linethickness{0.4pt}
\ifx\plotpoint\undefined\newsavebox{\plotpoint}\fi % GNUPLOT compatibility
\begin{picture}(125.75,101.75)(0,0)
%\circle(87.75,70.25){51.848}
\put(113.674,70.25){\line(0,1){1.1014}}
\put(113.651,71.351){\line(0,1){1.0994}}
\put(113.581,72.451){\line(0,1){1.0954}}
\multiput(113.464,73.546)(-.032651,.217893){5}{\line(0,1){.217893}}
\multiput(113.301,74.636)(-.029913,.154507){7}{\line(0,1){.154507}}
\multiput(113.091,75.717)(-.031894,.133959){8}{\line(0,1){.133959}}
\multiput(112.836,76.789)(-.033384,.117763){9}{\line(0,1){.117763}}
\multiput(112.536,77.849)(-.031383,.095104){11}{\line(0,1){.095104}}
\multiput(112.19,78.895)(-.032445,.085878){12}{\line(0,1){.085878}}
\multiput(111.801,79.925)(-.0332902,.0779279){13}{\line(0,1){.0779279}}
\multiput(111.368,80.938)(-.0316948,.0662508){15}{\line(0,1){.0662508}}
\multiput(110.893,81.932)(-.0323258,.0607916){16}{\line(0,1){.0607916}}
\multiput(110.376,82.905)(-.0328276,.0558714){17}{\line(0,1){.0558714}}
\multiput(109.817,83.855)(-.0332176,.0514026){18}{\line(0,1){.0514026}}
\multiput(109.22,84.78)(-.0335098,.0473163){19}{\line(0,1){.0473163}}
\multiput(108.583,85.679)(-.0337153,.0435574){20}{\line(0,1){.0435574}}
\multiput(107.909,86.55)(-.0323049,.0382597){22}{\line(0,1){.0382597}}
\multiput(107.198,87.392)(-.0324272,.0352504){23}{\line(0,1){.0352504}}
\multiput(106.452,88.203)(-.0324832,.0324309){24}{\line(-1,0){.0324832}}
\multiput(105.672,88.981)(-.0353026,.0323704){23}{\line(-1,0){.0353026}}
\multiput(104.86,89.725)(-.0383117,.0322432){22}{\line(-1,0){.0383117}}
\multiput(104.018,90.435)(-.0436117,.033645){20}{\line(-1,0){.0436117}}
\multiput(103.145,91.108)(-.0473702,.0334335){19}{\line(-1,0){.0473702}}
\multiput(102.245,91.743)(-.0514561,.0331348){18}{\line(-1,0){.0514561}}
\multiput(101.319,92.339)(-.0559243,.0327375){17}{\line(-1,0){.0559243}}
\multiput(100.368,92.896)(-.0608436,.0322278){16}{\line(-1,0){.0608436}}
\multiput(99.395,93.412)(-.0663018,.031588){15}{\line(-1,0){.0663018}}
\multiput(98.4,93.885)(-.0779814,.0331646){13}{\line(-1,0){.0779814}}
\multiput(97.387,94.317)(-.08593,.032307){12}{\line(-1,0){.08593}}
\multiput(96.355,94.704)(-.095154,.031229){11}{\line(-1,0){.095154}}
\multiput(95.309,95.048)(-.117817,.033194){9}{\line(-1,0){.117817}}
\multiput(94.248,95.346)(-.13401,.031678){8}{\line(-1,0){.13401}}
\multiput(93.176,95.6)(-.154555,.029664){7}{\line(-1,0){.154555}}
\multiput(92.094,95.808)(-.217945,.0323){5}{\line(-1,0){.217945}}
\put(91.005,95.969){\line(-1,0){1.0956}}
\put(89.909,96.084){\line(-1,0){1.0995}}
\put(88.81,96.153){\line(-1,0){2.2028}}
\put(86.607,96.149){\line(-1,0){1.0993}}
\put(85.508,96.077){\line(-1,0){1.0952}}
\multiput(84.412,95.958)(-.21784,-.033002){5}{\line(-1,0){.21784}}
\multiput(83.323,95.793)(-.154458,-.030162){7}{\line(-1,0){.154458}}
\multiput(82.242,95.582)(-.133908,-.03211){8}{\line(-1,0){.133908}}
\multiput(81.171,95.325)(-.117709,-.033573){9}{\line(-1,0){.117709}}
\multiput(80.111,95.023)(-.095053,-.031536){11}{\line(-1,0){.095053}}
\multiput(79.066,94.676)(-.085825,-.032584){12}{\line(-1,0){.085825}}
\multiput(78.036,94.285)(-.0778741,-.0334157){13}{\line(-1,0){.0778741}}
\multiput(77.023,93.851)(-.0661996,-.0318015){15}{\line(-1,0){.0661996}}
\multiput(76.03,93.374)(-.0607395,-.0324237){16}{\line(-1,0){.0607395}}
\multiput(75.059,92.855)(-.0558185,-.0329175){17}{\line(-1,0){.0558185}}
\multiput(74.11,92.296)(-.051349,-.0333004){18}{\line(-1,0){.051349}}
\multiput(73.185,91.696)(-.0472622,-.033586){19}{\line(-1,0){.0472622}}
\multiput(72.287,91.058)(-.0414315,-.0321766){21}{\line(-1,0){.0414315}}
\multiput(71.417,90.382)(-.0382076,-.0323665){22}{\line(-1,0){.0382076}}
\multiput(70.577,89.67)(-.0351981,-.032484){23}{\line(-1,0){.0351981}}
\multiput(69.767,88.923)(-.0323785,-.0325354){24}{\line(0,-1){.0325354}}
\multiput(68.99,88.142)(-.0323134,-.0353548){23}{\line(0,-1){.0353548}}
\multiput(68.247,87.329)(-.0337139,-.0401905){21}{\line(0,-1){.0401905}}
\multiput(67.539,86.485)(-.0335747,-.0436659){20}{\line(0,-1){.0436659}}
\multiput(66.868,85.612)(-.0333571,-.047424){19}{\line(0,-1){.047424}}
\multiput(66.234,84.711)(-.0330518,-.0515094){18}{\line(0,-1){.0515094}}
\multiput(65.639,83.784)(-.0326473,-.0559769){17}{\line(0,-1){.0559769}}
\multiput(65.084,82.832)(-.0321297,-.0608955){16}{\line(0,-1){.0608955}}
\multiput(64.57,81.858)(-.0337298,-.071092){14}{\line(0,-1){.071092}}
\multiput(64.097,80.862)(-.0330389,-.0780348){13}{\line(0,-1){.0780348}}
\multiput(63.668,79.848)(-.032168,-.085982){12}{\line(0,-1){.085982}}
\multiput(63.282,78.816)(-.031076,-.095205){11}{\line(0,-1){.095205}}
\multiput(62.94,77.769)(-.033004,-.11787){9}{\line(0,-1){.11787}}
\multiput(62.643,76.708)(-.031462,-.134061){8}{\line(0,-1){.134061}}
\multiput(62.391,75.636)(-.029415,-.154602){7}{\line(0,-1){.154602}}
\multiput(62.185,74.553)(-.031948,-.217997){5}{\line(0,-1){.217997}}
\put(62.026,73.463){\line(0,-1){1.0958}}
\put(61.912,72.368){\line(0,-1){1.0996}}
\put(61.846,71.268){\line(0,-1){3.3019}}
\multiput(61.927,67.966)(.03009,-.27376){4}{\line(0,-1){.27376}}
\multiput(62.047,66.871)(.033353,-.217787){5}{\line(0,-1){.217787}}
\multiput(62.214,65.782)(.030411,-.154409){7}{\line(0,-1){.154409}}
\multiput(62.427,64.701)(.032326,-.133856){8}{\line(0,-1){.133856}}
\multiput(62.685,63.63)(.030387,-.105889){10}{\line(0,-1){.105889}}
\multiput(62.989,62.571)(.031689,-.095002){11}{\line(0,-1){.095002}}
\multiput(63.338,61.526)(.032722,-.085773){12}{\line(0,-1){.085773}}
\multiput(63.73,60.497)(.0335412,-.0778202){13}{\line(0,-1){.0778202}}
\multiput(64.166,59.485)(.0319081,-.0661483){15}{\line(0,-1){.0661483}}
\multiput(64.645,58.493)(.0325215,-.0606871){16}{\line(0,-1){.0606871}}
\multiput(65.165,57.522)(.0330074,-.0557653){17}{\line(0,-1){.0557653}}
\multiput(65.726,56.574)(.0333831,-.0512953){18}{\line(0,-1){.0512953}}
\multiput(66.327,55.651)(.0336621,-.0472081){19}{\line(0,-1){.0472081}}
\multiput(66.967,54.754)(.0322433,-.0413796){21}{\line(0,-1){.0413796}}
\multiput(67.644,53.885)(.032428,-.0381554){22}{\line(0,-1){.0381554}}
\multiput(68.357,53.046)(.0325406,-.0351458){23}{\line(0,-1){.0351458}}
\multiput(69.106,52.237)(.0340044,-.0337315){23}{\line(1,0){.0340044}}
\multiput(69.888,51.461)(.0370162,-.0337226){22}{\line(1,0){.0370162}}
\multiput(70.702,50.719)(.0402448,-.0336491){21}{\line(1,0){.0402448}}
\multiput(71.548,50.013)(.0437199,-.0335043){20}{\line(1,0){.0437199}}
\multiput(72.422,49.343)(.0474777,-.0332807){19}{\line(1,0){.0474777}}
\multiput(73.324,48.71)(.0515626,-.0329688){18}{\line(1,0){.0515626}}
\multiput(74.252,48.117)(.0560295,-.0325571){17}{\line(1,0){.0560295}}
\multiput(75.205,47.564)(.0609472,-.0320315){16}{\line(1,0){.0609472}}
\multiput(76.18,47.051)(.0711463,-.0336152){14}{\line(1,0){.0711463}}
\multiput(77.176,46.58)(.0780879,-.0329131){13}{\line(1,0){.0780879}}
\multiput(78.191,46.153)(.086034,-.03203){12}{\line(1,0){.086034}}
\multiput(79.223,45.768)(.095255,-.030923){11}{\line(1,0){.095255}}
\multiput(80.271,45.428)(.117923,-.032814){9}{\line(1,0){.117923}}
\multiput(81.332,45.133)(.134112,-.031246){8}{\line(1,0){.134112}}
\multiput(82.405,44.883)(.154649,-.029166){7}{\line(1,0){.154649}}
\multiput(83.488,44.679)(.218048,-.031597){5}{\line(1,0){.218048}}
\put(84.578,44.521){\line(1,0){1.096}}
\put(85.674,44.409){\line(1,0){1.0997}}
\put(86.774,44.344){\line(1,0){1.1015}}
\put(87.875,44.326){\line(1,0){1.1013}}
\put(88.977,44.355){\line(1,0){1.099}}
\multiput(90.076,44.43)(.27371,.03053){4}{\line(1,0){.27371}}
\multiput(91.17,44.552)(.217733,.033704){5}{\line(1,0){.217733}}
\multiput(92.259,44.721)(.15436,.03066){7}{\line(1,0){.15436}}
\multiput(93.34,44.936)(.133803,.032541){8}{\line(1,0){.133803}}
\multiput(94.41,45.196)(.10584,.030557){10}{\line(1,0){.10584}}
\multiput(95.468,45.502)(.094951,.031842){11}{\line(1,0){.094951}}
\multiput(96.513,45.852)(.08572,.03286){12}{\line(1,0){.08572}}
\multiput(97.542,46.246)(.077766,.0336665){13}{\line(1,0){.077766}}
\multiput(98.553,46.684)(.0660968,.0320147){15}{\line(1,0){.0660968}}
\multiput(99.544,47.164)(.0606347,.0326192){16}{\line(1,0){.0606347}}
\multiput(100.514,47.686)(.0557121,.0330973){17}{\line(1,0){.0557121}}
\multiput(101.461,48.249)(.0512415,.0334657){18}{\line(1,0){.0512415}}
\multiput(102.384,48.851)(.0471538,.0337381){19}{\line(1,0){.0471538}}
\multiput(103.28,49.492)(.0413276,.0323099){21}{\line(1,0){.0413276}}
\multiput(104.147,50.17)(.0381031,.0324894){22}{\line(1,0){.0381031}}
\multiput(104.986,50.885)(.0350933,.0325972){23}{\line(1,0){.0350933}}
\multiput(105.793,51.635)(.0336767,.0340587){23}{\line(0,1){.0340587}}
\multiput(106.567,52.418)(.0336629,.0370705){22}{\line(0,1){.0370705}}
\multiput(107.308,53.234)(.0335842,.0402989){21}{\line(0,1){.0402989}}
\multiput(108.013,54.08)(.0334338,.0437738){20}{\line(0,1){.0437738}}
\multiput(108.682,54.956)(.0332041,.0475313){19}{\line(0,1){.0475313}}
\multiput(109.313,55.859)(.0328856,.0516157){18}{\line(0,1){.0516157}}
\multiput(109.905,56.788)(.0324668,.0560819){17}{\line(0,1){.0560819}}
\multiput(110.457,57.741)(.0319333,.0609987){16}{\line(0,1){.0609987}}
\multiput(110.968,58.717)(.0335005,.0712004){14}{\line(0,1){.0712004}}
\multiput(111.437,59.714)(.0327873,.0781408){13}{\line(0,1){.0781408}}
\multiput(111.863,60.73)(.031891,.086085){12}{\line(0,1){.086085}}
\multiput(112.246,61.763)(.030769,.095304){11}{\line(0,1){.095304}}
\multiput(112.584,62.811)(.032624,.117976){9}{\line(0,1){.117976}}
\multiput(112.878,63.873)(.03103,.134162){8}{\line(0,1){.134162}}
\multiput(113.126,64.946)(.033736,.180479){6}{\line(0,1){.180479}}
\multiput(113.328,66.029)(.031246,.218099){5}{\line(0,1){.218099}}
\put(113.484,67.12){\line(0,1){1.0961}}
\put(113.594,68.216){\line(0,1){2.0342}}
%\end
\put(87,96){\circle*{1.5}}
\put(88.25,44.75){\circle*{1.118}}
\put(61.5,69.75){\circle*{.707}}
\put(113.5,70.75){\circle*{.5}}
\put(80,101.75){$x_1=x_2=\infty$}
\put(80.25,39){$x_1=x_2=0$}
\put(49,70){$x_1=1$}
\put(125.75,70.75){$x_2=1$}
\put(113.25,70.75){\circle*{.5}}
\put(52.5,53.5){$\Sigma_1^0$}
\put(117.25,54.25){$\Sigma_2^0$}
\end{picture}
\vspace{-35mm}
\begin{center}
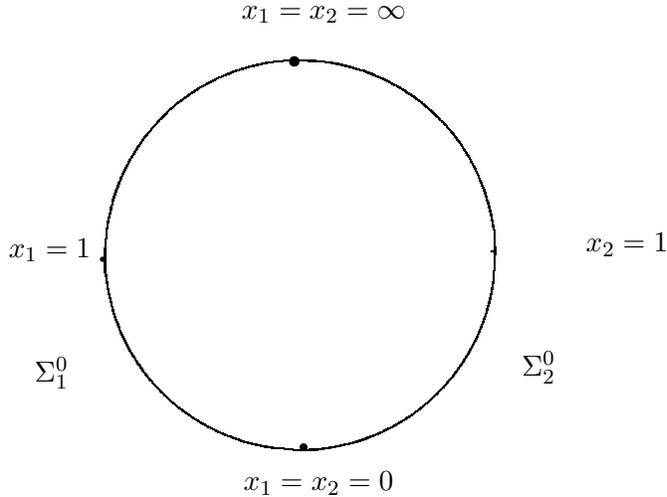
Fig 2. Real curve $\Si_0$
\end{center}

%%%%%%%%%%%%%%%%%%%%%%%%%%%%%%%%%%%%%%%%%%%%%%%%%%%%%%%%%%%%%%%%%%%%%
%%%%%%%%%%%%%%%%%%%%%%%%%%%%%%%%%%%%%%%%%%%%%%%%%%%%%%%%%%%%%%%%%%%%%%%%%%%%%%%%
 The moduli spaces of the Higgs
bundles are hyperk\"{a}hlerian. The operator $i$ is holomorphic in a
one complex structure and anti-holomorphic in two others, and its
fixed point set corresponds to a special configurations of branes in
terms of \cite{KW}.
In our case it acts  on  $\clM_I(G^\mC)$ (\ref{hi1}) as
 \beq{02}
i(L_\al(z_\al))=(\si L_\al(\bz_\al)\,.
 \eq The fixed point set
$\clM_I(G^\mR)$ of $i$ is described by the Lax operators (\ref{la})
 \beq{la2} L_1(x_1)=\frac{\bfT}{x_1}+\frac{\bfT+\eta}{1-x_1}\,,~~
L_2(x_2)=\frac{\bfS}{x_2}+\frac{\bfS+Ad_{\exp(\bfu)}(\eta)}{1-x_2}\,,
\eq
 where $\bfS$, $\bfT\in\gu$, $\bfu_\al\in\mR$, $\eta$ has the
form (\ref{231}), (\ref{131}) and $x_1$, $x_2$ are coordinates on
the real curve $\Si_0$ Fig.2.
$$
\clM_I(G^\mR)=T^*H^\mR\times(\clO_U\times\clO_U)\,.
$$
They coincide with $\clR^{red}_{G^\mR}$ (\ref{tu}) and
$$
\dim\,\clM_I(G^\mR)=2\sum_{j=1}^ld_j-2\rank\,U\,.
$$
%
%
%
%%%%%%%%%%%%%%%%%%%%%%%%%%%%%%%%%%%%%%%%%%%%%%%%%%%%%%%%%%%%%%%%%%%%%%%%%%%%
%
%\subsection{Lax equation}

There is an involution $\Upsilon$ on the moduli space $\clM_I(G^\mC)$. It acts on the coordinates as
$$
\Upsilon\left(\begin{array}{l}
       \bfu \\
       \bfv \\
\bfT \\
       \bfS \\
     \end{array}
\right)=\left(
\begin{array}{c}
  -\bfu \\
  -\bfv \\
  -\bfS \\
  -\bfT \\
\end{array}
\right)
$$
 Then it follows from (\ref{231}), (\ref{131}) that
$$
\Upsilon(\eta)=-\Ad_{\exp(\bfu)}\eta
 $$
If we accompany $\Upsilon$ with the action on the spectral
parameters $x_1\leftrightarrows x_2$
  then $\Upsilon$
interchanges the Lax operators (\ref{la2}) $\Upsilon(L_1(x_1))=-L_2(x_2)$,
$\Upsilon(L_2(x_2))=-L_1(x_1)$.

Consider the equations of motion (\ref{16}), (\ref{16a}),
(\ref{17}), (\ref{18}) in terms of the Lax operators (\ref{la2}).
Define $M_1$ as the introduced above the  $x_1$-independent operator  $M=\p_{\bfu}X$ (\ref{14}).
 Then on the first component we have
\beq{tl}
 \p_t
L_1(x_1)=[L_1(x_1),M_1]\,.
 \eq
 This equation implies
 $$
\underbrace{\frac{\p_t\bfT}{x_1}}_{\underline{1}}+
\underbrace{\frac{\p_t\bfT+\p_t\eta}{1-x_1}}_{\underline{2}}=
\underbrace{[\frac{\bfT}{x_1},M_1]}_{\underline{3}}+
\underbrace{[\frac{\bfT+\eta}{1-x_1},M_1]}_{\underline{4}}\,.
$$
 The equality $\underline{1}=\underline{3}$
   follows from
 (\ref{18}) and explicit form of $M_1=M=Y$ (\ref{14}). Then the equality
 $\underline{2}=\underline{4}$ follows from (\ref{15}).

 Define $M_2=\Upsilon(M_1)$. Then we come to lax equation on the second component\\
 $\p_t L_2(x_2)=[L_2(x_2),M_2]$ that is equivalent to (\ref{tl}).
%$M_\al$ operators on the both components  are $x_\al$-independent
%and assume the form
% \beq{mo}
%M_1=\p_{\bfu}X=M\,\,(\ref{14})\,,~~M_2=Ad_{e(\bfu)}M_1 \,. \eq
%Consider two Lax equations
% It
%is sufficient to consider the first one. We prove that it is
%equivalent to the equations of motion. In more details

%%%%%%%%%%%%%%%%%%%%%%%%%%%%%%%%%%%%%%%%%%%%%%%%%%%%%%%%%%%%%%%%%%%%%%%%%%%

\subsection{Classical $r$-matrix}

Let us change the variable $x_1$ in (\ref{la2}) to $y$:
 $$
 \coth(y)=(1-x_1)/x_1
 $$
Consider the following Lax operator:
  \beq{q032}
\ti L(y)=(1-x_1)L_1(x_1)\stackrel{(\ref{la2})}{=}\bfT\coth(y)-\clQ\,.
  \eq
  with
  $$
  \clQ=\sum_{\al\in R}\clQ_\al E_\al\,, ~~\clQ_{\pm\al}=\pm T_\al+X_{\pm \al}\,,
 $$
 where $X_{\pm \al}$ are defined in (\ref{131}).
 Since   $M$ is independent on the spectral parameter,
 we preserve its form (\ref{14}).
 % $$
% M=\sum_{\al\in R^+} Y_{\al}(E_\al-E_{\al})\,,~~
%Y_{\al}  =\frac{T_{\al}-S_{\al}\cosh(\bfu_{\al})}{\sinh^2(\bfu_{\al})}\,.
%   $$
The Lax equation
   \beq{q03}
 {\dot L}(x):=\{H,L(x)\}=[L(x),M]
   \eq
with the Lax pair (\ref{q032}), (\ref{14}) provides equations of
motion generated by the Hamiltonian (\ref{ha2}) and the Poisson
structure (\ref{pls}). Indeed, the Lax equation
(\ref{q03}) is equivalent to the already proven equations
  $$
 {\dot {\eta}}=[\eta,M],
  $$
where $\eta$ is defined by (\ref{231}), and
  $$
 {\dot \bfT}=[\bfT,M]\,.
  $$
The  $r$-matrix structure is given by
   \beq{q15}
 \displaystyle{
 \{L_1(x),L_2(y)\}=[L_1(x),r_{12}(x,y)]-[L_2(y),r_{21}(y,x)]
 }
   \eq
where
  \beq{q16}
  \begin{array}{c}
 \displaystyle{
 r_{12}(x,y)=\frac12\,(\coth(x-y)+\coth(x+y))\sum\limits_{i=1}^le_{i}\otimes
 e_{i}+
 }
 \\ \ \\
 \displaystyle{
 +\frac12\sum\limits_{\al\in R^+}\frac{(\al,\al)}{2}\, E_{\al}\otimes E_{-\al}
 (\coth(x-y)+\coth(\bfu_{\al}))+
  }
 \\ \ \\
 \displaystyle{
 +\frac12\,\sum\limits_{\al\in R^+}\frac{(\al,\al)}{2}\, E_{\al}\otimes E_{\al} (\coth(x+y)+\coth(\bfu_{\al}))\,.
 }
  \end{array}
  \eq
 Analogous result but without spectral parameter was obtained in
 \cite{Feher2},
 % ZAMENA - SSYLKA BYLA NA Feher - DOLZHNA BYT' NA Feher2
 %
 and for the algebra $\gu=so(N)$ the $r$-matrix was previously found in \cite{BAB}.
 Expression (\ref{q16}) is obtained in a similar way.
% The proof of (\ref{q15}) is also straightforward.
% \mo{It seems that some arguments in favor of (\ref{q15}) and (\ref{q16}) are relevant.
% Also a few words about the classical YB equation are needed.}
 Being written in the form (\ref{q15}) the Poisson brackets  provide
 the involutivity of the integrals of motion (\ref{il2}).

 One can also verify
that the $r$-matrix (\ref{q16}) provides $M$-matrix (\ref{14}) in
the following way:
   \beq{q17}
 \displaystyle{
 \tr_2(r_{12}(x,y)L_2(y))=\frac12\,(\coth(x-y)+\coth(x+y))\,L(x)-M\,.
 }
   \eq
%\mo{$\tr_2(r_{12}(x,y)L_2^{d_j}(y))=$?}

%%%%%%%%%%%%%%%%%%%%%%%%%%%%%%%%%%%%%%%%%%%%%%%%%%%%%%%%%%%%%%%%%%%%%%%%%%%%%%%%%%%%%%%%%%%

%%%%%%%%%%%%%%%%%%%%%%%%%%%%%%%%%%%%%%%%%%%%%%%%%%%%%%%%%%%%%%%%%%%%%%%%%%%%%%%%%%
\section{Higgs bundles. Model II}

\setcounter{equation}{0}

\subsection{Holomorphic bundles}
\paragraph{The base spectral curve.}
The base spectral curve is a singular curve $\Si^{II}$. It is defined by
the normalization of the rational curve
 \beq{sc2}
\pi\,:\,\mC P^1 \to\Si^{II}\,,~\pi(z=0)=\pi(z=\infty)
 \eq
In addition, we assume that $\Si^{II}$ has a marked point $z=1$ (see
Fig\,2.).
%
%
 %In particular, for $G^\mC=\SLN$
%and $K=$SU$(N)$ the quasi-compact structure $\clX$ is the space of
%unimodular  positive definite hermitian matrices.
%
%%%%%%%%%%%%%%%%%%%%%%%%%%%%%%%%%%%%%%%%%%%%%%%%%%%%%%%%%%%%%%%%%%%%%%%%%%%%%%
%
\paragraph{Vector bundles with quasi-compact structure.}
Let $\clP(G^\mC)$ be the principle $G^\mC$-bundle over $\Si^{II}$,
$V$ is a $G^\mC$ module  and $E^{II}(G^\mC)=\clP\times_{G^\mC}V$  is
the associated vector bundle.
 At the glued points we define the maps between the corresponding
 sections:
 \beq{hd2}
r\,:s|_{z=\infty}\in\G(E)\to s|_{z=0}\in\G(E)\,,~~r\in G^\mC\,.
 \eq
% To the point $z=1$ we attach the Riemannian  symmetric space $\clX=G^\mC/K$.
 Let
 $\bp+\bA$ $(\bp=\p_{\bz})$ be the antiholomorphic connections
acting on the sections $\G(E)$. The data
 \beq{dat}
\clD=\{(\bp+\bA)\,,~~r\in G^\mC\}
 \eq
 define the  vector bundle $E^{II}(G^\mC)$ over  $\Si^{II}$.
The group of  automorphisms $\clG$ of the  bundle $E^{II}(G^\mC)$ is
given by the  smooth maps
 \beq{32a}
\clG=\{f\,:\, C^\infty(\Si^{II})\to G^\mC\}\,.
 \eq
As above  at the marked point
 $z=1$  we replace $G^\mC$ with the maximal compact subgroup $K$:
 \beq{3.22}
f(z,\bz)|_{z=1}\in K\,.
 \eq
It means that we deal with
the quasi-compact vector bundle.
The gauge group action on the data (\ref{dat}) is of the form:
 \beq{ggt2a}
\bp+\bA\to f(\bp+\bA) f^{-1}\,,~~f(1)\in K\,,
 \eq
 \beq{ggt1a}
r\to f(\infty)r f^{-1}(0)\,.
 \eq

%\vspace{0.5cm}

%TeXCAD Picture [k11.tex]. Options:
%\grade{\on}
%\emlines{\off}
%\epic{\off}
%\beziermacro{\on}
%\reduce{\on}
%\snapping{\off}
%\pvinsert{% Your \input, \def, etc. here}
%\quality{8.000}
%\graddiff{0.005}
%\snapasp{1}
%\zoom{4.0000}
\unitlength 1mm % = 2.845pt
\linethickness{0.4pt}
\ifx\plotpoint\undefined\newsavebox{\plotpoint}\fi % GNUPLOT compatibility
\begin{picture}(116.75,103.75)(0,0)
%\circle(80.75,70.75){66}
\put(113.75,70.75){\line(0,1){1.2796}}
\put(113.725,72.03){\line(0,1){1.2777}}
\multiput(113.651,73.307)(-.03098,.31845){4}{\line(0,1){.31845}}
\multiput(113.527,74.581)(-.028867,.211342){6}{\line(0,1){.211342}}
\multiput(113.354,75.849)(-.031749,.180055){7}{\line(0,1){.180055}}
\multiput(113.131,77.11)(-.030105,.13898){9}{\line(0,1){.13898}}
\multiput(112.86,78.36)(-.031924,.123937){10}{\line(0,1){.123937}}
\multiput(112.541,79.6)(-.033369,.11146){11}{\line(0,1){.11146}}
\multiput(112.174,80.826)(-.0318711,.0931467){13}{\line(0,1){.0931467}}
\multiput(111.76,82.037)(-.0329261,.0852808){14}{\line(0,1){.0852808}}
\multiput(111.299,83.231)(-.0316821,.0734474){16}{\line(0,1){.0734474}}
\multiput(110.792,84.406)(-.0324765,.0679188){17}{\line(0,1){.0679188}}
\multiput(110.24,85.56)(-.0331364,.0629079){18}{\line(0,1){.0629079}}
\multiput(109.643,86.693)(-.0336797,.0583349){19}{\line(0,1){.0583349}}
\multiput(109.004,87.801)(-.0324957,.051558){21}{\line(0,1){.051558}}
\multiput(108.321,88.884)(-.0329036,.0479746){22}{\line(0,1){.0479746}}
\multiput(107.597,89.939)(-.0332287,.0446339){23}{\line(0,1){.0446339}}
\multiput(106.833,90.966)(-.0334788,.0415072){24}{\line(0,1){.0415072}}
\multiput(106.029,91.962)(-.0336606,.0385707){25}{\line(0,1){.0385707}}
\multiput(105.188,92.926)(-.0325286,.0344783){27}{\line(0,1){.0344783}}
\multiput(104.31,93.857)(-.033841,.033191){27}{\line(-1,0){.033841}}
\multiput(103.396,94.753)(-.0364526,.033079){26}{\line(-1,0){.0364526}}
\multiput(102.448,95.613)(-.0392162,.0329063){25}{\line(-1,0){.0392162}}
\multiput(101.468,96.436)(-.0421486,.0326676){24}{\line(-1,0){.0421486}}
\multiput(100.456,97.22)(-.0452699,.0323569){23}{\line(-1,0){.0452699}}
\multiput(99.415,97.964)(-.0509181,.0334894){21}{\line(-1,0){.0509181}}
\multiput(98.346,98.668)(-.0547873,.0330643){20}{\line(-1,0){.0547873}}
\multiput(97.25,99.329)(-.058977,.0325422){19}{\line(-1,0){.058977}}
\multiput(96.129,99.947)(-.0635387,.0319103){18}{\line(-1,0){.0635387}}
\multiput(94.986,100.522)(-.0728192,.0331004){16}{\line(-1,0){.0728192}}
\multiput(93.821,101.051)(-.0789845,.0322687){15}{\line(-1,0){.0789845}}
\multiput(92.636,101.535)(-.0925111,.0336713){13}{\line(-1,0){.0925111}}
\multiput(91.433,101.973)(-.101559,.032564){12}{\line(-1,0){.101559}}
\multiput(90.215,102.364)(-.112086,.031201){11}{\line(-1,0){.112086}}
\multiput(88.982,102.707)(-.13837,.032794){9}{\line(-1,0){.13837}}
\multiput(87.736,103.002)(-.15698,.03083){8}{\line(-1,0){.15698}}
\multiput(86.48,103.249)(-.210743,.03296){6}{\line(-1,0){.210743}}
\multiput(85.216,103.446)(-.254235,.029716){5}{\line(-1,0){.254235}}
\put(83.945,103.595){\line(-1,0){1.276}}
\put(82.669,103.694){\line(-1,0){1.2789}}
\put(81.39,103.744){\line(-1,0){1.2798}}
\put(80.11,103.744){\line(-1,0){1.2789}}
\put(78.831,103.694){\line(-1,0){1.276}}
\multiput(77.555,103.595)(-.254235,-.029716){5}{\line(-1,0){.254235}}
\multiput(76.284,103.446)(-.210743,-.03296){6}{\line(-1,0){.210743}}
\multiput(75.02,103.249)(-.15698,-.03083){8}{\line(-1,0){.15698}}
\multiput(73.764,103.002)(-.13837,-.032794){9}{\line(-1,0){.13837}}
\multiput(72.518,102.707)(-.112086,-.031201){11}{\line(-1,0){.112086}}
\multiput(71.285,102.364)(-.101559,-.032564){12}{\line(-1,0){.101559}}
\multiput(70.067,101.973)(-.0925111,-.0336713){13}{\line(-1,0){.0925111}}
\multiput(68.864,101.535)(-.0789845,-.0322687){15}{\line(-1,0){.0789845}}
\multiput(67.679,101.051)(-.0728192,-.0331004){16}{\line(-1,0){.0728192}}
\multiput(66.514,100.522)(-.0635387,-.0319103){18}{\line(-1,0){.0635387}}
\multiput(65.371,99.947)(-.058977,-.0325422){19}{\line(-1,0){.058977}}
\multiput(64.25,99.329)(-.0547873,-.0330643){20}{\line(-1,0){.0547873}}
\multiput(63.154,98.668)(-.0509181,-.0334894){21}{\line(-1,0){.0509181}}
\multiput(62.085,97.964)(-.0452699,-.0323569){23}{\line(-1,0){.0452699}}
\multiput(61.044,97.22)(-.0421486,-.0326676){24}{\line(-1,0){.0421486}}
\multiput(60.032,96.436)(-.0392162,-.0329063){25}{\line(-1,0){.0392162}}
\multiput(59.052,95.613)(-.0364526,-.033079){26}{\line(-1,0){.0364526}}
\multiput(58.104,94.753)(-.033841,-.033191){27}{\line(-1,0){.033841}}
\multiput(57.19,93.857)(-.0325286,-.0344783){27}{\line(0,-1){.0344783}}
\multiput(56.312,92.926)(-.0336606,-.0385707){25}{\line(0,-1){.0385707}}
\multiput(55.471,91.962)(-.0334788,-.0415072){24}{\line(0,-1){.0415072}}
\multiput(54.667,90.966)(-.0332287,-.0446339){23}{\line(0,-1){.0446339}}
\multiput(53.903,89.939)(-.0329036,-.0479746){22}{\line(0,-1){.0479746}}
\multiput(53.179,88.884)(-.0324957,-.051558){21}{\line(0,-1){.051558}}
\multiput(52.496,87.801)(-.0336797,-.0583349){19}{\line(0,-1){.0583349}}
\multiput(51.857,86.693)(-.0331364,-.0629079){18}{\line(0,-1){.0629079}}
\multiput(51.26,85.56)(-.0324765,-.0679188){17}{\line(0,-1){.0679188}}
\multiput(50.708,84.406)(-.0316821,-.0734474){16}{\line(0,-1){.0734474}}
\multiput(50.201,83.231)(-.0329261,-.0852808){14}{\line(0,-1){.0852808}}
\multiput(49.74,82.037)(-.0318711,-.0931467){13}{\line(0,-1){.0931467}}
\multiput(49.326,80.826)(-.033369,-.11146){11}{\line(0,-1){.11146}}
\multiput(48.959,79.6)(-.031924,-.123937){10}{\line(0,-1){.123937}}
\multiput(48.64,78.36)(-.030105,-.13898){9}{\line(0,-1){.13898}}
\multiput(48.369,77.11)(-.031749,-.180055){7}{\line(0,-1){.180055}}
\multiput(48.146,75.849)(-.028867,-.211342){6}{\line(0,-1){.211342}}
\multiput(47.973,74.581)(-.03098,-.31845){4}{\line(0,-1){.31845}}
\put(47.849,73.307){\line(0,-1){1.2777}}
\put(47.775,72.03){\line(0,-1){3.8368}}
\multiput(47.849,68.193)(.03098,-.31845){4}{\line(0,-1){.31845}}
\multiput(47.973,66.919)(.028867,-.211342){6}{\line(0,-1){.211342}}
\multiput(48.146,65.651)(.031749,-.180055){7}{\line(0,-1){.180055}}
\multiput(48.369,64.39)(.030105,-.13898){9}{\line(0,-1){.13898}}
\multiput(48.64,63.14)(.031924,-.123937){10}{\line(0,-1){.123937}}
\multiput(48.959,61.9)(.033369,-.11146){11}{\line(0,-1){.11146}}
\multiput(49.326,60.674)(.0318711,-.0931467){13}{\line(0,-1){.0931467}}
\multiput(49.74,59.463)(.0329261,-.0852808){14}{\line(0,-1){.0852808}}
\multiput(50.201,58.269)(.0316821,-.0734474){16}{\line(0,-1){.0734474}}
\multiput(50.708,57.094)(.0324765,-.0679188){17}{\line(0,-1){.0679188}}
\multiput(51.26,55.94)(.0331364,-.0629079){18}{\line(0,-1){.0629079}}
\multiput(51.857,54.807)(.0336797,-.0583349){19}{\line(0,-1){.0583349}}
\multiput(52.496,53.699)(.0324957,-.051558){21}{\line(0,-1){.051558}}
\multiput(53.179,52.616)(.0329036,-.0479746){22}{\line(0,-1){.0479746}}
\multiput(53.903,51.561)(.0332287,-.0446339){23}{\line(0,-1){.0446339}}
\multiput(54.667,50.534)(.0334788,-.0415072){24}{\line(0,-1){.0415072}}
\multiput(55.471,49.538)(.0336606,-.0385707){25}{\line(0,-1){.0385707}}
\multiput(56.312,48.574)(.0325286,-.0344783){27}{\line(0,-1){.0344783}}
\multiput(57.19,47.643)(.033841,-.033191){27}{\line(1,0){.033841}}
\multiput(58.104,46.747)(.0364526,-.033079){26}{\line(1,0){.0364526}}
\multiput(59.052,45.887)(.0392162,-.0329063){25}{\line(1,0){.0392162}}
\multiput(60.032,45.064)(.0421486,-.0326676){24}{\line(1,0){.0421486}}
\multiput(61.044,44.28)(.0452699,-.0323569){23}{\line(1,0){.0452699}}
\multiput(62.085,43.536)(.0509181,-.0334894){21}{\line(1,0){.0509181}}
\multiput(63.154,42.832)(.0547873,-.0330643){20}{\line(1,0){.0547873}}
\multiput(64.25,42.171)(.058977,-.0325422){19}{\line(1,0){.058977}}
\multiput(65.371,41.553)(.0635387,-.0319103){18}{\line(1,0){.0635387}}
\multiput(66.514,40.978)(.0728192,-.0331004){16}{\line(1,0){.0728192}}
\multiput(67.679,40.449)(.0789845,-.0322687){15}{\line(1,0){.0789845}}
\multiput(68.864,39.965)(.0925111,-.0336713){13}{\line(1,0){.0925111}}
\multiput(70.067,39.527)(.101559,-.032564){12}{\line(1,0){.101559}}
\multiput(71.285,39.136)(.112086,-.031201){11}{\line(1,0){.112086}}
\multiput(72.518,38.793)(.13837,-.032794){9}{\line(1,0){.13837}}
\multiput(73.764,38.498)(.15698,-.03083){8}{\line(1,0){.15698}}
\multiput(75.02,38.251)(.210743,-.03296){6}{\line(1,0){.210743}}
\multiput(76.284,38.054)(.254235,-.029716){5}{\line(1,0){.254235}}
\put(77.555,37.905){\line(1,0){1.276}}
\put(78.831,37.806){\line(1,0){1.2789}}
\put(80.11,37.756){\line(1,0){1.2798}}
\put(81.39,37.756){\line(1,0){1.2789}}
\put(82.669,37.806){\line(1,0){1.276}}
\multiput(83.945,37.905)(.254235,.029716){5}{\line(1,0){.254235}}
\multiput(85.216,38.054)(.210743,.03296){6}{\line(1,0){.210743}}
\multiput(86.48,38.251)(.15698,.03083){8}{\line(1,0){.15698}}
\multiput(87.736,38.498)(.13837,.032794){9}{\line(1,0){.13837}}
\multiput(88.982,38.793)(.112086,.031201){11}{\line(1,0){.112086}}
\multiput(90.215,39.136)(.101559,.032564){12}{\line(1,0){.101559}}
\multiput(91.433,39.527)(.0925111,.0336713){13}{\line(1,0){.0925111}}
\multiput(92.636,39.965)(.0789845,.0322687){15}{\line(1,0){.0789845}}
\multiput(93.821,40.449)(.0728192,.0331004){16}{\line(1,0){.0728192}}
\multiput(94.986,40.978)(.0635387,.0319103){18}{\line(1,0){.0635387}}
\multiput(96.129,41.553)(.058977,.0325422){19}{\line(1,0){.058977}}
\multiput(97.25,42.171)(.0547873,.0330643){20}{\line(1,0){.0547873}}
\multiput(98.346,42.832)(.0509181,.0334894){21}{\line(1,0){.0509181}}
\multiput(99.415,43.536)(.0452699,.0323569){23}{\line(1,0){.0452699}}
\multiput(100.456,44.28)(.0421486,.0326676){24}{\line(1,0){.0421486}}
\multiput(101.468,45.064)(.0392162,.0329063){25}{\line(1,0){.0392162}}
\multiput(102.448,45.887)(.0364526,.033079){26}{\line(1,0){.0364526}}
\multiput(103.396,46.747)(.033841,.033191){27}{\line(1,0){.033841}}
\multiput(104.31,47.643)(.0325286,.0344783){27}{\line(0,1){.0344783}}
\multiput(105.188,48.574)(.0336606,.0385707){25}{\line(0,1){.0385707}}
\multiput(106.029,49.538)(.0334788,.0415072){24}{\line(0,1){.0415072}}
\multiput(106.833,50.534)(.0332287,.0446339){23}{\line(0,1){.0446339}}
\multiput(107.597,51.561)(.0329036,.0479746){22}{\line(0,1){.0479746}}
\multiput(108.321,52.616)(.0324957,.051558){21}{\line(0,1){.051558}}
\multiput(109.004,53.699)(.0336797,.0583349){19}{\line(0,1){.0583349}}
\multiput(109.643,54.807)(.0331364,.0629079){18}{\line(0,1){.0629079}}
\multiput(110.24,55.94)(.0324765,.0679188){17}{\line(0,1){.0679188}}
\multiput(110.792,57.094)(.0316821,.0734474){16}{\line(0,1){.0734474}}
\multiput(111.299,58.269)(.0329261,.0852808){14}{\line(0,1){.0852808}}
\multiput(111.76,59.463)(.0318711,.0931467){13}{\line(0,1){.0931467}}
\multiput(112.174,60.674)(.033369,.11146){11}{\line(0,1){.11146}}
\multiput(112.541,61.9)(.031924,.123937){10}{\line(0,1){.123937}}
\multiput(112.86,63.14)(.030105,.13898){9}{\line(0,1){.13898}}
\multiput(113.131,64.39)(.031749,.180055){7}{\line(0,1){.180055}}
\multiput(113.354,65.651)(.028867,.211342){6}{\line(0,1){.211342}}
\multiput(113.527,66.919)(.03098,.31845){4}{\line(0,1){.31845}}
\put(113.651,68.193){\line(0,1){1.2777}}
\put(113.725,69.47){\line(0,1){1.2796}}
%\end
%\circle(91.75,70.25){44.011}
\put(113.756,70.25){\line(0,1){.9872}}
\put(113.734,71.237){\line(0,1){.9852}}
\put(113.667,72.222){\line(0,1){.9812}}
\multiput(113.557,73.204)(-.030892,.195059){5}{\line(0,1){.195059}}
\multiput(113.402,74.179)(-.033009,.161231){6}{\line(0,1){.161231}}
\multiput(113.204,75.146)(-.030157,.119691){8}{\line(0,1){.119691}}
\multiput(112.963,76.104)(-.031552,.105082){9}{\line(0,1){.105082}}
\multiput(112.679,77.05)(-.032611,.093205){10}{\line(0,1){.093205}}
\multiput(112.353,77.982)(-.033418,.083316){11}{\line(0,1){.083316}}
\multiput(111.985,78.898)(-.0314105,.0691588){13}{\line(0,1){.0691588}}
\multiput(111.577,79.797)(-.0320185,.0628458){14}{\line(0,1){.0628458}}
\multiput(111.129,80.677)(-.0324852,.0572564){15}{\line(0,1){.0572564}}
\multiput(110.641,81.536)(-.0328323,.0522576){16}{\line(0,1){.0522576}}
\multiput(110.116,82.372)(-.0330763,.0477478){17}{\line(0,1){.0477478}}
\multiput(109.554,83.184)(-.0332303,.0436484){18}{\line(0,1){.0436484}}
\multiput(108.955,83.969)(-.0333047,.0398972){19}{\line(0,1){.0398972}}
\multiput(108.323,84.727)(-.033308,.0364448){20}{\line(0,1){.0364448}}
\multiput(107.657,85.456)(-.033247,.0332513){21}{\line(0,1){.0332513}}
\multiput(106.958,86.155)(-.0364405,.0333126){20}{\line(-1,0){.0364405}}
\multiput(106.23,86.821)(-.0398929,.0333098){19}{\line(-1,0){.0398929}}
\multiput(105.472,87.454)(-.0436441,.0332359){18}{\line(-1,0){.0436441}}
\multiput(104.686,88.052)(-.0477436,.0330824){17}{\line(-1,0){.0477436}}
\multiput(103.874,88.614)(-.0522534,.032839){16}{\line(-1,0){.0522534}}
\multiput(103.038,89.14)(-.0572523,.0324925){15}{\line(-1,0){.0572523}}
\multiput(102.179,89.627)(-.0628417,.0320265){14}{\line(-1,0){.0628417}}
\multiput(101.3,90.076)(-.0691548,.0314194){13}{\line(-1,0){.0691548}}
\multiput(100.401,90.484)(-.083312,.033428){11}{\line(-1,0){.083312}}
\multiput(99.484,90.852)(-.0932,.032623){10}{\line(-1,0){.0932}}
\multiput(98.552,91.178)(-.105078,.031565){9}{\line(-1,0){.105078}}
\multiput(97.607,91.462)(-.119687,.030172){8}{\line(-1,0){.119687}}
\multiput(96.649,91.703)(-.161226,.03303){6}{\line(-1,0){.161226}}
\multiput(95.682,91.902)(-.195055,.030917){5}{\line(-1,0){.195055}}
\put(94.706,92.056){\line(-1,0){.9812}}
\put(93.725,92.167){\line(-1,0){.9852}}
\put(92.74,92.233){\line(-1,0){1.9744}}
\put(90.766,92.234){\line(-1,0){.9852}}
\put(89.78,92.167){\line(-1,0){.9813}}
\multiput(88.799,92.057)(-.195063,-.030867){5}{\line(-1,0){.195063}}
\multiput(87.824,91.903)(-.161235,-.032989){6}{\line(-1,0){.161235}}
\multiput(86.856,91.705)(-.119694,-.030142){8}{\line(-1,0){.119694}}
\multiput(85.899,91.464)(-.105086,-.031539){9}{\line(-1,0){.105086}}
\multiput(84.953,91.18)(-.093209,-.032599){10}{\line(-1,0){.093209}}
\multiput(84.021,90.854)(-.08332,-.033407){11}{\line(-1,0){.08332}}
\multiput(83.104,90.486)(-.0691628,-.0314017){13}{\line(-1,0){.0691628}}
\multiput(82.205,90.078)(-.0628499,-.0320105){14}{\line(-1,0){.0628499}}
\multiput(81.325,89.63)(-.0572605,-.0324779){15}{\line(-1,0){.0572605}}
\multiput(80.467,89.143)(-.0522618,-.0328256){16}{\line(-1,0){.0522618}}
\multiput(79.63,88.617)(-.0477521,-.0330702){17}{\line(-1,0){.0477521}}
\multiput(78.819,88.055)(-.0436526,-.0332248){18}{\line(-1,0){.0436526}}
\multiput(78.033,87.457)(-.0399014,-.0332996){19}{\line(-1,0){.0399014}}
\multiput(77.275,86.825)(-.036449,-.0333033){20}{\line(-1,0){.036449}}
\multiput(76.546,86.158)(-.0332555,-.0332428){21}{\line(-1,0){.0332555}}
\multiput(75.847,85.46)(-.0333173,-.0364363){20}{\line(0,-1){.0364363}}
\multiput(75.181,84.732)(-.0333149,-.0398887){19}{\line(0,-1){.0398887}}
\multiput(74.548,83.974)(-.0332414,-.0436399){18}{\line(0,-1){.0436399}}
\multiput(73.95,83.188)(-.0330885,-.0477394){17}{\line(0,-1){.0477394}}
\multiput(73.387,82.377)(-.0328456,-.0522492){16}{\line(0,-1){.0522492}}
\multiput(72.862,81.541)(-.0324998,-.0572481){15}{\line(0,-1){.0572481}}
\multiput(72.374,80.682)(-.0320345,-.0628376){14}{\line(0,-1){.0628376}}
\multiput(71.926,79.802)(-.0314282,-.0691508){13}{\line(0,-1){.0691508}}
\multiput(71.517,78.903)(-.033439,-.083308){11}{\line(0,-1){.083308}}
\multiput(71.149,77.987)(-.032635,-.093196){10}{\line(0,-1){.093196}}
\multiput(70.823,77.055)(-.031579,-.105074){9}{\line(0,-1){.105074}}
\multiput(70.539,76.109)(-.030187,-.119683){8}{\line(0,-1){.119683}}
\multiput(70.297,75.152)(-.03305,-.161222){6}{\line(0,-1){.161222}}
\multiput(70.099,74.184)(-.030941,-.195051){5}{\line(0,-1){.195051}}
\put(69.944,73.209){\line(0,-1){.9812}}
\put(69.833,72.228){\line(0,-1){.9852}}
\put(69.767,71.243){\line(0,-1){2.9596}}
\put(69.832,68.283){\line(0,-1){.9813}}
\multiput(69.943,67.302)(.030842,-.195067){5}{\line(0,-1){.195067}}
\multiput(70.097,66.327)(.032968,-.161239){6}{\line(0,-1){.161239}}
\multiput(70.295,65.359)(.030126,-.119698){8}{\line(0,-1){.119698}}
\multiput(70.536,64.402)(.031525,-.10509){9}{\line(0,-1){.10509}}
\multiput(70.819,63.456)(.032587,-.093213){10}{\line(0,-1){.093213}}
\multiput(71.145,62.524)(.033396,-.083325){11}{\line(0,-1){.083325}}
\multiput(71.513,61.607)(.0313929,-.0691668){13}{\line(0,-1){.0691668}}
\multiput(71.921,60.708)(.0320025,-.062854){14}{\line(0,-1){.062854}}
\multiput(72.369,59.828)(.0324706,-.0572647){15}{\line(0,-1){.0572647}}
\multiput(72.856,58.969)(.032819,-.052266){16}{\line(0,-1){.052266}}
\multiput(73.381,58.133)(.0330641,-.0477563){17}{\line(0,-1){.0477563}}
\multiput(73.943,57.321)(.0332192,-.0436568){18}{\line(0,-1){.0436568}}
\multiput(74.541,56.535)(.0332945,-.0399057){19}{\line(0,-1){.0399057}}
\multiput(75.174,55.777)(.0332987,-.0364533){20}{\line(0,-1){.0364533}}
\multiput(75.84,55.048)(.0332386,-.0332598){21}{\line(0,-1){.0332598}}
\multiput(76.538,54.349)(.036432,-.0333219){20}{\line(1,0){.036432}}
\multiput(77.266,53.683)(.0398844,-.03332){19}{\line(1,0){.0398844}}
\multiput(78.024,53.05)(.0436357,-.033247){18}{\line(1,0){.0436357}}
\multiput(78.809,52.451)(.0477352,-.0330946){17}{\line(1,0){.0477352}}
\multiput(79.621,51.889)(.052245,-.0328523){16}{\line(1,0){.052245}}
\multiput(80.457,51.363)(.057244,-.0325071){15}{\line(1,0){.057244}}
\multiput(81.316,50.875)(.0628336,-.0320425){14}{\line(1,0){.0628336}}
\multiput(82.195,50.427)(.0691468,-.031437){13}{\line(1,0){.0691468}}
\multiput(83.094,50.018)(.083303,-.033449){11}{\line(1,0){.083303}}
\multiput(84.01,49.65)(.093192,-.032646){10}{\line(1,0){.093192}}
\multiput(84.942,49.324)(.10507,-.031592){9}{\line(1,0){.10507}}
\multiput(85.888,49.039)(.119679,-.030203){8}{\line(1,0){.119679}}
\multiput(86.845,48.798)(.161218,-.033071){6}{\line(1,0){.161218}}
\multiput(87.813,48.599)(.195047,-.030966){5}{\line(1,0){.195047}}
\put(88.788,48.445){\line(1,0){.9812}}
\put(89.769,48.334){\line(1,0){.9852}}
\put(90.754,48.267){\line(1,0){1.9744}}
\put(92.729,48.266){\line(1,0){.9852}}
\put(93.714,48.332){\line(1,0){.9813}}
\multiput(94.695,48.442)(.195071,.030817){5}{\line(1,0){.195071}}
\multiput(95.671,48.596)(.161243,.032948){6}{\line(1,0){.161243}}
\multiput(96.638,48.794)(.119702,.030111){8}{\line(1,0){.119702}}
\multiput(97.596,49.035)(.105094,.031512){9}{\line(1,0){.105094}}
\multiput(98.542,49.319)(.093217,.032575){10}{\line(1,0){.093217}}
\multiput(99.474,49.644)(.083329,.033386){11}{\line(1,0){.083329}}
\multiput(100.39,50.012)(.0691708,.0313841){13}{\line(1,0){.0691708}}
\multiput(101.29,50.42)(.062858,.0319945){14}{\line(1,0){.062858}}
\multiput(102.17,50.867)(.0572688,.0324633){15}{\line(1,0){.0572688}}
\multiput(103.029,51.354)(.0522701,.0328123){16}{\line(1,0){.0522701}}
\multiput(103.865,51.879)(.0477605,.0330581){17}{\line(1,0){.0477605}}
\multiput(104.677,52.441)(.0436611,.0332136){18}{\line(1,0){.0436611}}
\multiput(105.463,53.039)(.0399099,.0332895){19}{\line(1,0){.0399099}}
\multiput(106.221,53.672)(.0364575,.033294){20}{\line(1,0){.0364575}}
\multiput(106.95,54.338)(.033264,.0332343){21}{\line(1,0){.033264}}
\multiput(107.649,55.036)(.0333265,.0364278){20}{\line(0,1){.0364278}}
\multiput(108.315,55.764)(.0333251,.0398802){19}{\line(0,1){.0398802}}
\multiput(108.948,56.522)(.0332526,.0436314){18}{\line(0,1){.0436314}}
\multiput(109.547,57.307)(.0331007,.047731){17}{\line(0,1){.047731}}
\multiput(110.11,58.119)(.0328589,.0522408){16}{\line(0,1){.0522408}}
\multiput(110.635,58.954)(.0325144,.0572398){15}{\line(0,1){.0572398}}
\multiput(111.123,59.813)(.0320505,.0628295){14}{\line(0,1){.0628295}}
\multiput(111.572,60.693)(.0314458,.0691428){13}{\line(0,1){.0691428}}
\multiput(111.981,61.592)(.03346,.083299){11}{\line(0,1){.083299}}
\multiput(112.349,62.508)(.032658,.093188){10}{\line(0,1){.093188}}
\multiput(112.675,63.44)(.031606,.105066){9}{\line(0,1){.105066}}
\multiput(112.96,64.385)(.030218,.119675){8}{\line(0,1){.119675}}
\multiput(113.202,65.343)(.033092,.161214){6}{\line(0,1){.161214}}
\multiput(113.4,66.31)(.030991,.195043){5}{\line(0,1){.195043}}
\put(113.555,67.285){\line(0,1){.9812}}
\put(113.666,68.266){\line(0,1){.9852}}
\put(113.733,69.252){\line(0,1){.9984}}
%\end
\put(113.75,71.5){\circle*{1.581}} \put(59,71){\circle*{1.414}}
\put(54.75,75.5){$z=1$} \put(116.75,75.25){$z=\infty$}
\put(116,66.25){$z=0$} \put(60,60.25){$\Sigma^{II}$}
\end{picture}
\vspace{-3.5cm}
\begin{center}
%\centerline{\texttt{Fig.1 The base spectral curve }}
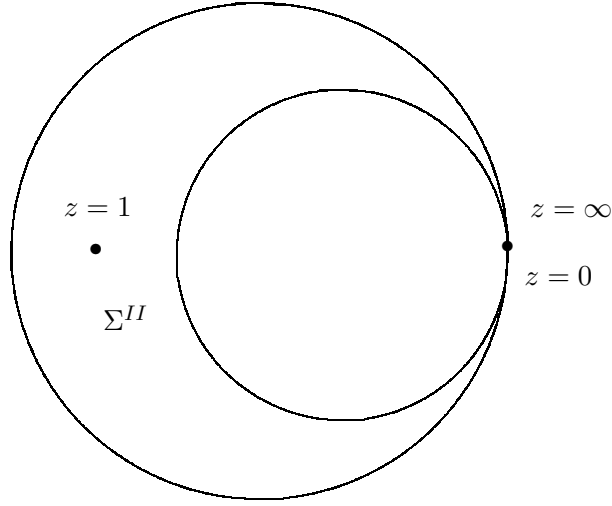
\texttt{Fig.2 The base spectral curve $\Si^{II}$ }
\end{center}

%%%%%%%%%%%%%%%%%%%%%%%%%%%%%%%%%%%%%%%%%%%%%%%%%%%%%%%%%%%%%%%%%%%%%%%%%%%%%%%%%%%

%%%%%%%%%%%%%%%%%%%%%%%%%%%%%%%%%%%%%%%%%%%%%%%%%%%%%%%

%\vspace{10mm}

% For $y\in\clX$
% \beq{xt}
% y\to yf^{-1}|_{z=1}\,.
% \eq
% \beq{ggt3}
%h\to f_2(0) h f_1^{-1}(0)\,,~f_\al(0)=f_\al(z_\al,\bz_\al)|_{z_\al=0} \,,
% \eq
%where, according to (\ref{3.2}) $f_1,f_2\in K$ in (\ref{ggt1}) and
%(\ref{ggt3}) .

\paragraph{The moduli space $\Bun_{II}(G^\mC)$.}
The coset space $\Bun_{II}(G^\mC)=\clD/\clG$  is the moduli of
quasi-compact $G^\mC$-bundles over the singular curve $\Si^{II}$.

 Let us prove that  a big cell $\Bun^0_{II}(G^\mC)$ in the moduli space is
 \beq{msh1}
\Bun^0_{II}(G^\mC)= (\clH^\mR,\ti K)=\{h^{-1} \exp(\bfu)\,|\,h\in\ti
K\,,\,\exp(\bfu)\in \clH^\mR\}\,,
 \eq
 compare with (\ref{msh}).
It has dimension (see (\ref{dk}))
 \beq{dms1}
\dim_\mR\,
\Bun^0_{qc}(\Si^{II},G^\mC)=\rank(G^\mR)+\dim\,K-\rank\,K=\dim\,K=
2\sum_{j=1}^ld_j-l\,.
 \eq
First, we prove that by the gauge group action (\ref{ggt2a}) the
generic configuration of variables  $\bA$ can be chosen to be
trivial
 \beq{cs2}
\bA=0\,,
 \eq
 and after this gauge fixing  we stay with  the residual constant gauge
transformations taking values in the subgroup $K$ as in (\ref{cgt}).
 The condition (\ref{3.22}) prevents the connection
from trivialization. To go around we consider the symmetric space
$\clX^\mC=K\setminus G^\mC$. In these terms (\ref{cs2})  means that
$f(z)|_{z=1}$ preserves the point $x_0$ corresponding to the coset
$K$ in  $\clX^\mC$. If we omit (\ref{3.22}) then the gauge
transformations (\ref{ggt2a}) of the bundles over $\mC P^1$ allows
one to choose the gauge (\ref{cs2}). The residual gauge
transformations $\clG^{res}$ are the constant maps $\Si^{II}$ to
$G^\mC$. We choose $f\in\clG^{res}$ as follows. Let $g$ be the value
at $z=1$ of the corresponding transformation $\bA\to 0$. It acts on
$x_0$ as $x_0\to y=x_0g$. The stationary subgroup corresponding to
$y$  is  $g^{-1}Kg$. Then  acting by $g^{-1}$ on $y$ we reduce to
the residual gauge transformations to $K$. After this procedure we
stay with the condition (\ref{cs2}) and the residual constant gauge
transformations which take values in the subgroup $K$.

The action of these  transformations on $r$ (\ref{ggt1a}) takes the
form $r\to frf^{-1}\,,~~f\in K$. As in (\ref{fg3}) we have:
 \beq{fg4}
 frf^{-1}= h^{-1} \exp(\bfu)\,,~~~h\in
K\,,~~\bfu\in\gh^\mR\,.
\eq
%where $f$ is the transformation "diagonalizing" $p$.
%The transformation (\ref{ggt3}) describe the action of the group
%$K\times K$ on  $K$. The group $K=(K\times K)/K$  can be considered
%as the compact symmetric space.
 The element $h$ in (\ref{fg4})
is defined up to the Cartan torus $\clT$-action (\ref{rg7}):
 $$
h\to shs^{-1}\,,~~s\in \clG^{res}_2=\clT\subset K\,.
 $$
 Then $h\in\ti K$ (\ref{or}) and we come to (\ref{msh1}).

%%%%%%%%%%%%%%%%%%%%%%%%%%%%%%%%%%%%%%%%%%%%%%%%%%%%%%%%%%%%%%%%%%%

\subsection{Higgs bundles}

The  Higgs bundle $\clH(G^\mC)$ over $\Si^{II}$ is  described by
the pairs
 \beq{1.2a}
\{(\p_{\bA}=\bp+\bA,\Phi)\,,~(r,\xi)\in T^*G^\mC\}\,.
 \eq
Here $\Phi=\Phi(z,\bz)dz\in {\rm End}(E)\otimes\Om^{(1,0)}(\Si^{II})$ is the
Higgs field, and
 $$
(r,\xi)\in T^*(\Hom(E|_{z=\infty}\to E|_{z=0}))\sim T^*G^\mC\,.
 $$
%
% $$
%(h,\nu)\in T^*(\Hom(E|_{z_1=0}\to E|_{z_2=0})) \sim T^*K\,.
% $$
%
The Higgs bundle $\clH_G$ is a symplectic manifold equipped with the
form
 $$
\om=\int_{\Si^{II}}\left(( D\Phi,D\bA )+ (D(\xi, r^{-1}Dr)\de(0) \right)\,.
 $$
 The group of the automorphisms $\clG$ of the bundle $E^{II}$ (\ref{ggt2a}), (\ref{ggt1a})
 is lifted to the symplectic automorphisms of the Higgs bundle $\clG$:
 \beq{pk}
\Phi\to \Ad_f\Phi\,,~~\xi\to \Ad_{f(0)}\xi\,.
 \eq
These transformations are generated by the moment map
 \beq{mu30}
 \mu=\p_{\bz}\Phi+\de(z,\bz)|_{z=0}\xi+\de(z,\bz)|_{z=\infty}\Ad_r\xi \,.
 \eq
% \beq{m20}
%\mu_2=\p_{\bz_2}\Phi_2+\de(z_2,\bz_2)|_{z_2=\infty}(g\eta g^{-1})|_{\gk^*}+
%\de(z_2,\bz_2)|_{z_2=0}(h\nu h^{-1})\,.
% \eq
Impose the moment map constraints $\mu=0$.
They mean that the Higgs field $\Phi (z,\bz)$ is meromorphic on
$\Si^{II}$, have simple poles at $z=0,\infty$ with the residues
 $$
 \left\{
 \begin{array}{ll}
z=0: & \Res\,\Phi=\xi\,, \\
 z_\al=\infty: & \Res\,\Phi=-\Ad_r\xi
 \end{array}
 \right.
 $$
 and
  \beq{res}
\Res\,\Phi(z=1)|_{\gk}=0\,,~~~({\rm quasi-compactness})\,.
 \eq
 Let $\Res\,\Phi(z)|_{z=1}=\clP$.
 Then the Higgs field (on the surface $\mu=0$) assumes the form:
 \beq{mo2}
 \Phi=\frac{\xi}{z}-\frac{\clP}{z-1}\,,~~\clP|_\gk=0\,,\,\,\Res\,\Phi_{z=\infty}=-\Ad_r\xi\,.
 \eq
The sum of residues vanishes $\xi+\clP-\Ad_r\xi=0$. It means that
 \beq{lm}
  (\xi-\Ad_r\xi )|_{\gk}=0\,.
   \eq
%    or
%  \beq{mo5}
%     p=(\xi-r\xi r^{-1})|_{\gp^\mC}\,.
% \eq
Thus,
 $$
\Phi=\frac{\xi}{z}-\frac{\xi-\Ad_r\xi}{z-1}\,.
 $$
Now fix the gauge as (\ref{cs2}). It means that $r=h^{-1}\exp(\bfu)$.
 The Lax operator $L(z)$ is identified with the gauge transformed Higgs field.
 From (\ref{mo2}) we find
 \beq{la5}
 L(z)=\frac{\xi}{z}-\frac{\clP}{z-1}\,,
 ~~\clP=\xi-\Ad_{h^{-1}\exp(\bfu)}\xi\,.
 \eq
 Compare this Lax operator with the Lax operator $L_2(z_2)$ (\ref{la}) in the Model I.
 From (\ref{TS}c) we find that
 \beq{c12}
 \Ad_h( L(z))=L_2(z_2)|_{z_2=z}\,.
 \eq
In these terms the moduli space of the Higgs bundle $\ti\clM_{ II}(G^\mC)=\mu^{-1}(0)/\clG$
 are defined by the pair
  \beq{kr}
\ti\clM_{II}(G^\mC)=T^*\clH^\mR\times T^*K=
\{  (\xi\in\gg^\mC\,,r= h^{-1}\exp(\bfu))\cup(\ref{lm})\,.
  \eq
 It coincides with $\ti\clR^{red}_{II}$ (\ref{ifo}).
As above we pass from $T^*K$ to the orbits $(\clO_K\times\clO_K)$
with the torus action.
After reduction we obtain
\beq{moda}
\clM_{II}(G^\mC)=T^*\clH^\mR\times((\clO_K\times\clO_K)//\clT)\sim\clR^{red}_{II}(G^\mC)\,,
\eq
where $\bfT$ is defined by (\ref{me}a), $\eta$ satisfies (\ref{me}b), and
$\dim\,\clM_{II}(G^\mC)=4\sum_{j=1}^l(d_j-1)$ (see (\ref{dre1})).
Thus after the symplectic reduction we come to the real symplectic space.
The integrals of motion in this models are constructed by means of the differentials
(\ref{dbe}) on the curve $\Si^{II}$. In this way we come to the following expressions
\beq{imo2}
I_{jk}=(\xi^{d_j-k}\clP^k)\,,~~k=0,\dots,d_j\,.
\eq
Due to arguments presented at the end of Section 2.3.2.  the number of
integrals is equal to $\sum_{j=1}^ld_j$. We still have the deficit of the integrals.

%%%%%%%%%%%%%%%%%%%%%%%%%%%%%%%%%%%%%%%%%%%%%%%%%%%%%%%%%%%%%%%%%%%%%
\subsection{Real forms}
As in Section 3.4 we pass to the real form of this construction.
We consider the antiholomorphic automorphism $z\to\bz$ of $\Si^{II}$ and the involutive
automorphism $\si$ (\ref{gr}).
The involution (\ref{01}) acts on $\clM_{II}(G^\mC)$ (\ref{moda}) as
\beq{in2}
i(L(z))=\si (L(\bz))\,.
\eq
The fixed point set is described by the Lax operator
\beq{la10}
 L^{II}(x)=\frac{\eta}{x}-\frac{\eta-\Ad_{h^{-1}\exp(\bfu)}\eta}{x-1}\,.
 \eq
where $x\in\mR P^1$, $h\in U$ and $\eta$ is defined in (\ref{231}), (\ref{131}).
Thus, from (\ref{inte}) and (\ref{kr})
\beq{tm2}
\ti\clM_{II}(G^\mR)=\{T^*\clH^\mR\times T^*U)\}\sim\ti\clR^{red}_{II}(G^\mR)\,.
\eq
Passing to the orbits as (\ref{grs})   we obtain
$$
\clM_{II}(G^\mR)=\{T^*\clH^\mR\times(\clO_U\times\clO_U)\}\sim\clR^{red}_{II}(G^\mR)\,.
$$
In this way we come to integrable system.

%%%%%%%%%%%%%%%%%%%%%%%%%%%%%%%%%%%%%%%%%%%%%%%%%%%%%%%%%%%%%%%%%%%%%%%%%%%%%%%%%%%%%%%%%%%

%%%%%%%%%%%%%%%%%%%%%%%%%%%%%%%%%%%%%%%%%%%%%%%%%%%%%%%%%%%%%%%%%%%%%%%%%%%%%%%%

\section{Universal  bundle}

\setcounter{equation}{0}

\subsection{Vector bundle}

%\emph{The base spectral curve:}\\
Consider a $G^\mC$-bundle over the
spectral curve $\Si^I$  for the model I (Fig 1, Section 3.1).
As in that case we define two vector bundles $E_\al(G^\mC)=\clP_\al\times_{G^\mC}V$
 over the components $\Si_\al$.

% At the glued points we define the maps of the corresponding fibers
%\beq{hd}
%\varrho_\infty\,:s|_{z_1=\infty}\in\G(E_1)\to s|_{z_2=\infty}\in\G(E_2)\,,~~
%\varrho_0\,:s|_{z_1=0}\in\G(E_1)\to s|_{z_2=0}\in\G(E_2)\,,
%\eq
%but now  $\varrho_0\,,\,\varrho_\infty\in G^\mC$.
%
%At the point $z_1=1$ we fix the symmetric space $\clX_1=G^\mC/K$. Here
%$K$ are the maximal compact subgroups $K\subset G^\mC$.
%%\footnote{Of course $K_\al$ are conjugated.}
%
% At the point $z_2=1$ on $\Si_2$ we fix a trivialization $s^0$ of the fiber $s_{z_2}$.

% It means that there are the set of isomorphisms $\clI_2=\{Isom(V_{z_2}\to V^0\}$ attached to $z_2$.
% the flag variety $Fl_2=G^\mC/B_2$, where $B$ is a
% Borel subgroup $B\subset G^\mC$.

Let
 $\bp_\al+\bA_\al$ $(\bp_\al=\p_{\bz_\al})$ be the antiholomorphic connections
acting on the sections $\G(E_\al)$. The gluing maps $\varrho_0\in G^\mC$, $\varrho_\infty\in G^\mC$
play the role of $r_0$ and $r_\infty$ (\ref{hd1}).
The data
\beq{dat11}
\clD=\{(\bp_\al+\bA_\al)\,,~\al=1,2\,,~(\varrho_0\,,\,\varrho_\infty)\in G^\mC\}
\eq
 define the vector bundle $E^{univ}(G^\mC)$  over
 the singular curve $\Si$ with two marked points
  ($z_1=1$, $z_2=1$).
%The main difference with \cite{Ne} is that we fix a real  $K$-structure at the fiber
%over  the point $z_1=1$. For $K=$SU$(N)\subset\SLN$ it is an Hermitian structure.

\noindent
\paragraph{Gauge group.}
The gauge group $\clG$ of the  bundle $E^{univ}(G^\mC)$
is  the pair of smooth maps $\clG_\al$
\beq{gub}
\clG_\al=\{f_\al\in\Si_\al\in C^\infty(\Si_\al)\to G^\mC\,,~f_1(z_1=1)\in K\,,~
f_2(z_2=1)=Id\}\,.
\eq
The restriction at the point $z_1=1$ means that the component $E_1(G^\mC)$ has quasi-compact
structure, while the second restriction implies that
we fix a trivialization at the point $z_2=1$ of the bundle $E_2(G^\mC)$.

%Let $\si$ be the involutive automorphism of $G^\mC$, such that its fixed point set $K$
%is the maximal compact subgroup $K\subset G^\mC$.
% At the marked points
% $z_\al=0,\infty$ and $z_\al=1$ we replace $G^\mC$ on the maximal compact subgroup $K$
%\beq{3.2}
%f_\al(z_\al,\bz_\al)|_{z_\al=0,1,\infty}\in K\,.
%\eq
%We call such holomorphic bundles the bundles with \emph{the quasi-compact structure}.
%In particular, for $G^\mC=\SLN$ and $K=$SU$(N)$ the quasi-compact structure
%means that the gauge group preserves a positive definite hermitian structures
%at the marked points.
%Since at $z_2=1$ the trivialization is fixed, $f(z_2)|_{z_2=1}=1$.

The gauge group action on the data (\ref{dat11}) is as follows:
\begin{subequations}\label{gtr}
  \begin{align}
 & \bp_\al+\bA_\al\to f_\al(\bp_{\al}+\bA_\al) f^{-1}_\al\,,\\
 &\varrho_\infty\to f_2(\infty)\varrho_\infty f_1^{-1}(\infty)\,,~
 f_\al(\infty)=f_\al(z_\al,\bz_\al)|_{z_\al=\infty}\,,\,,\\
 &\varrho_0\to f_2(0)\varrho_0  f_1^{-1}(0)\,,~f_\al(0)=f_\al(z_\al,\bz_\al)|_{z_\al=0} \,.
 \end{align}
  \end{subequations}
% For $y_1\in\clX_1$ and $y_2\in Isom(s(z_2=1),s^0)$
% \beq{xt1}
% y_1\to f_1(1)y_1\rho^{-1}(f_1)(1))\,,~~
% y_2\to f_2(1)(y_2)\,.
% \eq

\noindent
\paragraph{The moduli space $\Bun^{univ}(\Si,G^\mC)$.}

The quotient $\Bun^{univ}(\Si,G^\mC)=\clD/\clG$ is the moduli of
holomorphic $G^\mC$-bundles over the singular curve $\Si$ with the
quasi-compact structure (at the point $z_1=1$).

Due to the restrictions (\ref{gub}) of the gauge group, the
Universal bundle is an analog of the bundle for the Model II.
%There is analog of type III Universal bundle equivalent
%to the type II. We don't construct
%it here, but
We use this fact to trivialize the connections $\bA_\al=0$.
After that we stay with the constant gauge transformations
\beq{ostg}
\clG^{res}_1 =K\,,~~\clG^{res}_2= G^\mC\,.
\eq
Let $f_2(0)=f\in G^\mC$ and $f_1(0)=k\in K$. Taking $f=k\varrho_0^{-1} $ (see (\ref{gtr}c))
we transform
 $$
\varrho_0\to Id\,.
 $$
Then (\ref{gtr}b) assumes the form $\varrho_\infty \to k\varrho_0^{-1}\varrho_\infty k^{-1}$.
 Using (\ref{fg3}) we find that $\varrho_\infty$
is transformed to the form
 $$
\varrho_\infty \to h^{-1}\exp(\bfu)\,,~~h\in \ti K\,,\, (\ref{or})\,;~~\exp(\bfu)\in\clH^\mR\,.
 $$
Let  $\clV_{2}$ be a fiber over $z_2=1$ and  $g_2:$ $\clV_{2}\to V$
is a trivialization
 ($\Isom(\clV_2,V)=\{g_2\}$). This space is the principal homogeneous space over $G^\mC$.
In these notations
 $$
\Bun^{univ}(\Si,G^\mC)=\ti K\times\clH^\mR\times \Isom(\clV_{2},V) \,.
 $$

%%%%%%%%%%%%%%%%%%%%%%%%%%%%%%%%%%%%%%%%%%%%%%%%%%%%%%%%%%%%%%%%%%%%%%%%%%%%%%%%%%%%%%

\subsection{Higgs bundle}

The Higgs bundle $\clH^{univ}(G^\mC)$ corresponding to the bundle $E^{univ}(G^\mC)$ is defined by the pairs
 $$
\{(\bp_\al+\bA_\al,\Phi_\al)\,,~(\varrho_0,\la_0)\,,~(\varrho_\infty,\la_\infty)\}\,.
%~(p_1,y_1)\}
 $$
Here
 $$
\Phi_\al=\Phi_\al(z_\al,\bz_\al)dz_\al\in End(E_\al)\otimes\Om^{(1,0)}(\Si_\al)
 $$
are the components of the Higgs fields.
%As above at $z_1=1$ attach the cotangent bundle
% $$
%\{(p_1,y_1)\,,~p_1\in\gp^\mC\}\in T^*\clX_1
% $$
%and at $z_2=1$ the cotangent bundle
% $$
%T^*Isom(s_{2},s^0)\sim T^*G^\mC=\{(p_2,y_2)\,,~p_2\in\gg^\mC\,,~y_2\in G^\mC\}\,.
% $$
 At $z_1=0$ and $z_1=\infty$
we have
 $$
(\varrho_\infty,\la_\infty)\in T^*(\Hom(V|_{z_1=\infty}\to V|_{z_2=\infty}))\sim T^*G^\mC\,,
 $$
 $$
(\varrho_0,\la_0)\in T^*(\Hom(V|_{z_1=0}\to V|_{z_2=0})) \sim T^*G^\mC\,.
 $$
The Higgs bundle  $\clH^{univ}(G^\mC)$ is a symplectic manifold equipped with the form
 $$
 \begin{array}{c}
\om=\sum_{\al=1,2}\int_{\Si_\al}( D\Phi_\al,D\bA_\al)
-\int_{\Si_1}\sum_{i=0,\infty}\de(z_i,\bz_i)|_{z_i}D(\la_i, \varrho_i^{-1}D\varrho_i)+
 \\ \ \\
+\int_{\Si_2}\sum_{i=0,\infty}\de(z_i,\bz_i)|_{z_i}D(\la_i,
\varrho_i^{-1}D\varrho_i)\,.
 \end{array}
 $$
%\beq{2.2}
%\om=\sum_{\al=1,2}\int_{\Si_\al}( D\Phi_\al,D\bA_\al)+
%\int_{\Si_1}(\de(z_1,\bz_1)|_{z_1=\infty}D(\varrho_\infty,\varrho_\infty^{-1}D \varrho_\infty)+
%\eq
% $$
%\int_{\Si_1}\de(z_1,\bz_1)|_{z_1=0}D(\varrho_0,\varrho_0^{-1}D \varrho_0)
%%+\de(z_1,\bz_1)|_{z_1=1}\om^{ss}(p_1,y_1)))
%%+\int_{\Si_2}\de(z_2,\bz_2)|_{z_2=1}\om^{cb}\,.
% $$
%where $\om^{cb}(p_2,y_2)=D(p_2,y_2^{-1}Dy_2)$ is the canonical form on the cotangent bundle
% $T^*Isom(s_{2},s^0)$.
%
%By passing to the variables corresponding to the curve $\Si_2$
%  $$
% (\varrho_0\,,\,\varrho_\infty)\to(\varrho^{-1}_0\,,\,\varrho^{-1}_\infty)\,,~~
% (\la_0\,,\,\la_\infty)\to(-\Ad^{-1}_{\varrho_0}\la_0\,,\,-\Ad^{-1}_{\varrho_\infty}\la_\infty)\,.
%  $$
%one can rewrite the boundary terms in (\ref{2.2}) as the integral over the second component
%\beq{222}
%\int_{\Si_2}(\de(z_2,\bz_2)|_{z_2=\infty}D(\la_\infty,D \varrho_\infty\varrho_\infty^{-1})+
%\int_{\Si_2}\de(z_2,\bz_2)|_{z_2=0}D(\la_0,D \varrho_0\varrho_0^{-1})\,.
%\eq
%
 The gauge group  $\clG$ of the bundle $E^{univ}(G^\mC)$ (\ref{gtr})
  can be lifted to the
 symplectic automorphisms  of the Higgs bundle  $\clH^{univ}(G^\mC)$ as
 \beq{48a}
 \Phi_\al\to\Ad^{-1}_{f_\al}\Phi_\al\,,
 \eq
\beq{48b}
\la_0\to  \Ad^{-1}_{f_1(0)} \la_0
 \,,~~\la_\infty         \Ad^{-1}_{f_1(\infty)} \la_\infty  \,.
\eq
%\beq{48c}
%p_1\to p_1\,,~~p_2\to \Ad_{f_2(1)} p_2 \,.
%\eq
%
The moment maps corresponding to this actions are
 $$
\mu_1=-\bp\Phi_1+\de(z_1,\bz_1)|_{z_1=0}\la_0+\de(z_1,\bz_1)|_{z_1=\infty}\la_\infty\,,
 $$
 $$
\mu_2=-\bp\Phi_2+\de(z_2,\bz_2)|_{z_2=0}\Ad^{-1}_{\varrho_0}(\la_0)+
\de(z_2,\bz_2)|_{z_2=\infty}\Ad^{-1}_{\varrho_\infty}\la_\infty\,.
 $$
Here we have fixed the gauge by the conditions $\bA_\al=0$. The
moment map constraints $\mu_1=0$ and $\mu_2=0$ mean that $\Phi_\al$
are meromorphic with poles at $z_\al=0,\infty$.
 In addition, due to (\ref{gub}) $\Phi_1(z_1)$ has a pole at $z_1=1$,
such that $\Res\,\Phi_1(z_1)=-\clK_1\in\gk$. Similarly,
$\Phi_2(z_2)$ has a pole at $z_2=1$ with
$\Res\,\Phi_2(z_2)=-\clX_2\in\gg^\mC$. Therefore,
 $$
\Phi_1=\frac{\la_0}{z_1}-\frac{\clK_1}{z_1-1}\,,~~
\Phi_2=\frac{\Ad_{\varrho_0}\la_0}{z_2}-
\frac{\clX_2}{z_2-1}\,.
 $$
Since the sum of residues vanishes,
 $$
\la_0+\la_\infty+\clK_1=0\,,
 $$
 \beq{2po}
\Ad_{\varrho_0}\la_0+\Ad_{\varrho_\infty}\la_\infty+\clX_2=0\,.
 \eq
 In this way
the gauge transformed Higgs fields take the form:
 \beq{hf1}
\Phi_1=\frac{\la_0}{z_1}-\frac{\la_0+\la_\infty}{z_1-1}\,,~~
\Phi_2=\frac{\Ad_{\varrho_0}\la_0}{z_2}-
\frac{\Ad_{\varrho_0}\la_0+\Ad_{\varrho_\infty}\la_\infty}{z_2-1}\,.
 \eq

%The residual gauge transformations are
%the constant maps $\Si_\al\to G^\mC$. As it was explained by acting on the symmetric space
%$\clX$ we  reduce the gauge group to
%\beq{rgu}
%\clG^{res}=K_1\times G_2^\mC
%\eq
%acting on $\Si_1\times\Si_2$.

\paragraph{Reduction to model I.}
Let $\varrho_0=pk$ ($\varrho_0\in G^\mC$) be the polar decomposition (\ref{dp3}). Here $k\in K$.
It is
exactly the same subgroup that defines $\clG^{res}_1$ (\ref{ostg}).
%Use the gauge fixing (\ref) and the residual gauge symmetry (\ref).
 Present $f_2\in G^\mC$  in the form
 \beq{gfi}
f_2=k_2p^{-1}\,,
 \eq
where $k_2$ is an arbitrary element of $K$ and $p=p(\varrho_0)$ is fixed.
In this way we partly fix the gauge.
 Due to (\ref{gtr}c) the gauge transformed $\varrho_0$ belong to $K$
 \beq{gfi2}
\varrho_0\in K\,.
 \eq
  To come to the model I we impose the second
class constraints, transversal to the gauge fixing (\ref{gfi}):
 \beq{scc} \la_0|_{\gp^*}=0\,.
  \eq
  The transversality follows from
the orthogonality of the Cartan decomposition (\ref{cd1}). This
condition is an analog of the moment maps constraints, but it does
not generate vector fields on the phase space.
At this step we have the residual gauge group $K_1\times K_2$ acting
on  $\Si_1\times\Si_2$. Acting by this group we can transform
$\varrho_\infty$ to the Cartan subgroup $\clH^\mR$:
 $$
\varrho_\infty=e(\bfu)\,.
 $$
The action $f_\al\in K_\al$ leads to the moment map constraints
 \beq{m4} \mu_1=(\la_0+\la_\infty)|_{\gk^*}=0\,, \eq
  \beq{m5}
\mu_2=(\Ad_{\varrho_0}\la_0+\Ad_{\varrho_\infty}\la_\infty)|_{\gk^*}=0\,.
 \eq
 Due to (\ref{gfi2}) and (\ref{scc}) we can assume the notations (\ref{ov})
  $$
 \la_0=\bfT\,,~~\varrho_0=h\,.
  $$
 From (\ref{m4}), (\ref{m5}) we have
  $$
 \la_\infty=\eta\,.
  $$
Then Lax operators of the model I (\ref{la}) coincides with  (\ref{hf1}).

%%%%%%%%%%%%%%%%%%%%%%%%%%%%%%%%%%%%%%%%%%%%%%%%%%%%%%%%%%%%%%%%%%%%%%%%%%%
\bigskip
\paragraph{Reduction to model II.}
Consider the symplectic action of $f_2\in G^\mC$  (\ref{48b}). By  (\ref{gtr}c)
$\varrho_0$ can be transformed to the unity element $\varrho_0=1$.
 The moment map constraints coming from (\ref{48b}) take the form
 $$
\Ad_{\varrho_0}\la_0+\Ad_{\varrho_\infty}\la_\infty=0\,.
 $$
It means that there are no poles of the Higgs fields $\Phi_2$ on the
$\Si_2$ component and the singular curve $\Si_1\cup\Si_2$ is
shrinked  to the base spectral (\ref{sc2}) Fig.\,2. and
$\varrho_0=\varrho_\infty=r$. The residual gauge symmetries is the
compact subgroup $K$ (\ref{ostg}). In this way we come to the model
II.

In this way we realize the diagram (\ref{0}).

%%%%%%%%%%%%%%%%%%%%%%%%%%%%%%%%%%%%%%%%%%%%%%%%%%%%%%%%%%%%%%%%%%%%%%%%%%%%%%%%%%
\section{Higgs bundles. Model III}

\setcounter{equation}{0}

\subsection{Description of model}
 Here we define alternative quasi-compact bundle corresponding to the Model III.
 Consider the spectral curve $\Si^{II}$ and
 attach the Riemannian  symmetric space $\clX^\mC=K\setminus G^\mC$ to the point $z=1$.
 We add it to the data (\ref{dat}):
 \beq{dat1}
\clD^{III}=\{(\bp+\bA)\,,~\rho\in G^\mC\,,~g\in\clX^\mC\}\,.
 \eq
 Then  $g\in G^\mC$ is defined up to the multiplications $g\sim kg$, $k\in K$.
The gauge group is the smooth map
 \beq{3200}
f\,:\, C^\infty(\Si)\to G^\mC\,.
 \eq
 It acts on $\clX^\mC$ as
 \beq{xt}
 g\to gf^{-1}|_{z=1}\,.
 \eq
After fixing the gauge  $\bA=0$ by the residual constant gauge transformations one can
transform $\rho$ to the Cartan subgroup
 $$
\rho\to\exp{\bfu^\mC}\in\clH^\mC\,.
 $$
Since $f\in\clH^\mC$ does not change the gauge we have
 \beq{mbu}
\Bun_{III}(G^\mC)=\{\clH^\mC\setminus\clX^\mC\,,\,\clH^\mC\}\,.
 \eq
From (\ref{dgx}) we have
 $$
\dim_{\mR}\,\Bun_{III}(G^\mC)=\sum_{j=1}^l (2 d_j-1)\,.
 $$
The corresponding Higgs bundle $\clH(G^\mC)$ is described by
the pairs
 \beq{1.2}
\{(\p_{\bA}=\bp+\bA,\Phi^{III})\,,~(\rho,\varsigma)\in T^*G^\mC\,,~(\zeta,g)\in T^*\clX\}\,,
 \eq
 where $\zeta\in\gp^\mC$ (\ref{cd1}).
The Higgs bundle is equipped with the form
 \beq{2.2}
\om=\int_{\Si}\left(( D\Phi^{III},D\bA )+ (D(\varsigma, \rho^{-1}D\rho)\de(0) +
(\zeta,Dgg^{-1})\de(z,\bz)|_{z=1}\right)\,.
 \eq
 The symplectic transformations of $\Phi^{III}$ and $\varsigma$ are standard
 (\ref{pk}), while $\zeta\to \zeta$ is given by (\ref{301}).
 The moment map generating these transformations is (compare with (\ref
 {mu30})) as follows:
 $$
 \mu=\p_{\bz}\Phi^{III}-\de(z,\bz)|_{z=0}\varsigma+\de(z,\bz)|_{z=\infty}\Ad_r(\varsigma)
 -\de(z,\bz)|_{z=1}\Ad^{-1}_g\zeta \,.
 $$
Taking $\mu=0$ we obtain
  \beq{f3}
 \Phi^{III}(z)=\frac{\varsigma}z-\frac{\bfP}{z-1}\,,\,\,
 ~~(\bfP=\Ad^{-1}_g\zeta \,,~(\ref{css}))\,,
 \eq
 and $Res\,\Phi^{III}(z)|_{z=\infty}=-\Ad_r(\varsigma)$.
 The sum of residues vanishes. It is the condition $\mu_{III}=0$ (\ref{m31})
\beq{re3}
 \varsigma-\Ad_{\exp\bfu^\mC}(\varsigma)-\bfP=0\,.
 \eq
 The solutions of this equation take the form (\ref{kc3}).
%
%
%This relations allows one to find $\xi$. Let
% $$
% \bfP=\sum_j^l\bfP_j+\sum_{\al\in R}\bfP_\al E_\al\,,~~
% \bfP_\al\in\mC\,.
% $$
% Then from (\ref{re3})
% $$
% \xi=\bfv_\mC+\sum_j^l\bfP_j+\sum_{\al\in R}\frac{\bfP_\al}{1-\exp(\bfu_\al^\mC)} E_\al\,.
% %+\frac{\bar\bfP_\al}{1-\exp(\bfu_{-\al}^\mC)} E_{-\al}\,.
% $$
 Then taking into account the symplectic action of the Cartan group $\clH^\mC$ we come to
 the expression (\ref{33}).
 %\beq{kc3}
% \xi=\bfv_\mC+\sum_{\al\in R}\frac{\bfP_\al}{1-\exp(\bfu_\al^\mC)} E_\al\,,
% \eq
% where $\bfP_\al$ is defined up to the multiplications $\bfP_\al\sim w\bfP_\al$,
% $w\in\mC^*$.
%
 One can see that the moduli space of the Higgs bundle in the Model III coincides
 with the phase space   $\clR^{red}_{III}(G^\mC)$      (\ref{34}):
 \beq{md3}
\clM_{III}(G^\mC)=T^*\clH^\mC\times (T^*\clX//\clH^\mC)=
\{(\bfv_\mC,\exp(\bfu^\mC)\,,\,(\bfP_\al,\al\in R)\}\sim\clR^{red}_{III}(G^\mC) \,,
 \eq
$$
\dim\,\clM_{III}(G^\mC)=\dim_\mR(T^*\clX)=2\sum_{j=1}^l(2d_j-1)\,.
$$

%%%%%%%%%%%%%%%%%%%%%%%%%%%%%%%%%%%%%%%%%%%%%%%%%%%%%%%%%%%%%%%%%%%%%%

\paragraph{Comparison of the Higgs bundles for Model III and Model II.}

Similarly to the finite-dimensional case
 the Higgs bundle for the Model III (\ref{1.2})
with the automorphism group (\ref{3200}) is isomorphic to
 the Higgs bundle of the Model II (\ref{1.2a}) with the automorphism group (\ref{3.22}).
It means that the reduction of the gauge group at the point $z=1$ to
the subgroup $K\subset G^\mC$ is equivalent to the attachment  the
symmetric space $\clX^\mC=K\setminus G^\mC$ to the same point.

As in Section 2.3.2 take the gauge transformation $f(z)|_{z=1}=p$ in
(\ref{3200}), where $p$ is defined by the polar decomposition of
$g$: $(g=kp)$. It acts on the symmetric space $\clX^\mC=K\setminus
G^\mC$ attached to $z=1$ in such a way that  $y\in\clX^\mC\to Id$.
Then we stay with the Higgs bundle of the Model II (\ref{1.2a}) with
the automorphism group (\ref{3.22}). We will prove the
symplectomorphism $\clM_{II}(G^\mC)\sim\clM_{III}(G^\mC)$
  for arbitrary Higgs bundles in a separate publication.

 \subsection{Real form.}
 As above we pass to the fixed point set of the involution
(\ref{02}) acting on $\clM_{III}(G^\mC)$.
The fixed point set is described by the Lax operator
\beq{la30}
 L^{III}(x)=\frac{\varsigma}{x}-\frac{\bfP}{x-1}\,,
 \eq
where $x\in\mR P^1$ and
$$
\varsigma=\bfv+\sum_{\al\in R}\frac{\bfP_\al}{1-\exp\bfu_\al} E_\al\,,~~\bfP_\al\in\mR\,.
$$
 In this way we obtain
 $$
\clM_{III}(G^\mR)=T^*\clH^\mR\times (T^*\clX^\mR//\clH^\mR)=
\{(\bfv,\exp(\bfu)(\ti\bfP_\al,\al\in R\}\,,
 $$
 $$
\dim\,\clM_{III}(G^\mR)=\dim_\mR(T^*\clX^\mR)=2\sum_{j=1}^ld_j\,.
$$
From (\ref{i230}), (\ref{is23}) we find
$\clM_{III}(G^\mR)\sim\clR^{red}_{III}(G^\mR)$ (\ref{clr3}).
Similarly to (\ref{is23}) we find from (\ref{tm2}) that
$$
\clM_{III}(G^\mR)\sim\ti\clM_{II}(G^\mR)\,.
$$

%%%%%%%%%%%%%%%%%%%%%%%%%%%%%%%%%%%%%%%%%%%%%%%%%%%%%%%%%%%%%%%%%%%%%%%%%%%%%%%%%%%%%
%%%%%%%%%%%%%%%%%%%%%%%%%%%%%%%%%%%%%%%%%%%%%%%%%%%%%%%%%%%%%%%%%%%%%%%%%%%%%%%%%%%%%%
  \section{Calogero-Sutherland system}

%%%%%%%%%%%%%%%%%%%%%%%%%%%%%%%%%%%%%%%%%%%%%%%%%%%%%%%%%%%%%%%%%%%%%%%%%%%%%

\subsection{CS and Model III}

Consider a bundle over the curve $\Si^{II}$ (Fig.\,2).
We change the data (\ref{dat1}) relating to the Model III in the following way.
Let $B$ be the Borel subgroup of $G^\mC$.
Replace the symmetric space $\clX^\mC$ attached to the point $z=1$
 with the flag variety $Fl^\mC=B\setminus G^\mC$. It means that $g$ is defined up to the left shift
 $g\sim bg$, $b\in B$.
 The gauge is fixed by the conditions $\bA=0$. Using the residual constant
gauge transformations $G^\mC$ we can put $r$ in the form  $r=\exp(\bfu^\mC)\in \clH^\mC$.
 In this way we come to the moduli space of the quasi-parabolic
 bundles over  $\Si^{II}$ (instead of (\ref{mbu}))
 $$
\Bun^{qp}_{III}(G^\mC)=\{Fl^\mC/\clH^\mC,\clH^\mC\}\,.
$$
%\beq{cs3}
%f(z,\bz)|_{z=1}\in G^\mC.
%\eq
%
In the corresponding Higgs bundles we replace the cotangent bundle $T^*\clX^\mC$
 with the coadjoint orbit
 $$
 \clO_\nu=\{\Ad_g\nu\,|\,g\in G^\mC\,,\,\nu\in\gh^\mC\}\,.
 $$
Here we assume that $\al(\nu)>0$ for $\al\in R^+$.
 It can be proved that  as in Section 2.3.1 it can be equivalently defined as
a result of the symplectic reduction $\clO_\nu\sim B\setminus\setminus T^*G^\mC$.
 It means that in (\ref{oct}) and (\ref{kak}) the maximal compact subgroup $K$ is replaced
 by
 the Borel subgroup $B$. Then $\clO_\nu$ can be identified with
 the set of pairs $(\zeta,g)$, where $g$ is an element of the coset space $B\setminus G^\mC$
 and $\zeta$ satisfies the moment map constraint $\zeta|_{\gb^*}=\nu$, where $\gb^*=$Lie$^*(B)$ is the
 Lie coalgebra, and $\nu\in\gh^*$.
 % In other words the orbits are the set
%  of pairs with equivalence relations
%  $$
%  (g,\zeta)\sim (\,bg\,,\,\Ad^*_b\,(\zeta)\,)\,,
%  $$
%  where $\Ad^*$ is the coadjoint action on the Lie coalgebra $\gb^*$.
 As in (\ref{css}) one can represent the orbit in terms of $B$ invariant expressions
$$
\clO_\nu=\{\clS=\Ad^*_{g^{-1}}\zeta\,|\,g\sim bg\,,\,\zeta|_{\gb^*}=\nu\,,\,
 b\in B\,,\,\nu\in\gh^\mC\}\,.
$$
Here the coadjoint action $\Ad^*$ is defined on the coalgebra $(\gg^\mC)^*$.
For  the positive nilpotent subalgebra $\gn^+$ and $x\in\gn^+$ we have  $x|_{\gb^*}=0$.
The subalgebra $\gn^+$ is a cotangent space to the flag variety $B\setminus G^\mC$ at
the point corresponding to $B$.
 Since $\zeta|_{\gb^*}=\nu$ it can be represented as
$$
\zeta=\nu+x\,,~x\in\gn^+\,,~\nu\in\gh^\mC\,.
$$
Certainly, the coadjoint action $\Ad_B^*$ on $\zeta$ does not change
$\nu$, but for $\nu\neq 0$ it acts freely on $\gn^+$. Therefore,
 $\clO_\nu$ is a principle homogeneous space  over the cotangent bundle $T^*Fl$
 to the flag variety attached to $z=1$. The cotangent bundle $T^*Fl$ corresponds
  to the case $\nu=0$.

The Higgs bundle is defined by the data (see (\ref{1.2})):
 $$
\{(\p_{\bA}=\bp+\bA\,,\,\Phi_{III}^{CS})\,,~(r,\varsigma)\in T^*G^\mC\,,~(\zeta,g)\in \clO_\nu\}\,,
 $$
where $\Phi_{III}^{CS}$ has the poles at $z=0,1,\infty$.
% Higgs field $\Phi$ and $\xi\in\gg^\mC$ dual.
%we define
%the Lax operator $L^{CS}(z)$ of the spin CS system.
% It has the following form.
Similarly to (\ref{f3})
 $$
 \Phi_{III}^{CS}=\frac{\varsigma}{z}-\frac{\clS}{z-1}\,,~~\clS\in\clO_\nu\,.
 %\,\,\Res\,\Phi^{CS}_{z=\infty}=-\Ad_r\xi\,.
 $$
Since the sum of the residues of $\Phi_{III}^{CS}(z)$ vanishes $
\varsigma-\Ad_{\exp(r)}(\varsigma)=\clS$. Solutions of this equation
is the  Lax operator without spectral parameter.
 $\varsigma=\eta^{CS}$ for the Calogero-Sutherland system.
  Let
 $$
 \clS=(\clS_{\gh^\mC}+\sum_{\al\in R}\clS_\al E_\al)\in\clO_{G^\mC}\subset\gg^\mC\,.
 $$
Then
 \beq{ecs}
 \eta^{CS}=\bfv+\clS_{\gh^\mC}+\sum_{\al\in R}\frac{\clS_\al}{1-\exp(\bfu_\al)}E_\al\,.
 \eq
 The spectral dependent Lax operator takes the form:
 \beq{la6}
 L_{III}^{CS}(z)=\frac{\eta^{CS}}{z}-\frac{\clS}{z-1}\,.
 \eq
 In fact, one should pass to the symplectic quotient $\ti\clS\in\clS_{G^\mC}//\clH^\mC$.
 The moment constraints corresponding to this action means
 that the Cartan part $\clS_{\gh^\mC}$ on the quotient space vanishes and $\eta^{CS}$ (\ref{ecs})
 takes the form:
 $$
 \eta^{CS}=\bfv+\sum_{\al\in R}\frac{\ti\clS_\al}{1-\exp(\bfu_\al)}E_\al\,.
 $$
From (\ref{ecs}) and (\ref{are6})  we find the quadratic Hamiltonian of the CS system
$$
H=\oh(\eta^{CS},\eta^{CS})=\oh(\bfv,\bfv)+
\oh\sum_{\al\in R^+}\frac{\ti\clS_\al\ti\clS_{-\al}}{(\al,\al)\sinh^2(\frac{\bfu_\al}2)}\,.
$$
The phase space of CS system is
 \beq{c1}
\clR_{III}^{CS}= T^*H^\mC\times\clS//\clH^\mC\,,~~\dim\,\clR_{III}^{CS}=\dim\,\clO_\nu\,.
 \eq
 %In our case we replace the orbit $\clO_{G^\mC}=\{\clS\}$ by the
% cotangent space $\gp^\mC=\{\bfP\}$ to $G^\mC/K$ (\ref{la5}) and then by the
% residual gauge transformations "kill" these degrees of freedom.
% At this stage we stayed with the gauge group $K$. As a result we come to
% the phase space (\ref{mod}).

%%%%%%%%%%%%%%%%%%%%%%%%%%%%%%%%%%%%%%%%%%%%%%%%%%%%%%%%%%%%%%%%%%%%%%%%%%%%%%%%%%

\subsection{CS and Model II}
  Define the Model II type description of the Calogero-Sutherland system.
We start with the same  spectral curve  $\Si^{II}$.
  The holomorphic bundle is defined by the data  (\ref{dat}))
$\clD=\{(\bp+\bA)\}$,  $r\in G^\mC$,
  where $r$ is defined in (\ref{hd2}).
The gauge group is (\ref{32a}), but the restriction (\ref{3.22}) is
replaced by
 \beq{cs7} f(z,\bz)|_{z=1}\in B\,, ~~(f(z,\bz)\in\clG)\,,
 \eq
 where $B$ is a Borel subgroup of $G^\mC$.
It means that the holomorphic bundle has \emph{the quasi-parabolic structure}.
As for the Model II we transform $\bA$ to $\bA=0$ and stay with the constant maps
$f\,:\,\Si_{II}\to B$.
%To compare with  the Model II case  let us use (\ref{cs7}).
The residual gauge transformation means that $r$ becomes the element of the quotient space
 \beq{slf}
\clF =\{r\sim  frf^{-1}\,,~~r\in G^\mC\,,~f\in B\}\,.
 \eq
The poles of the Higgs field $\Phi_{II}^{CS}$ allows one to write it in the standard way.
Let $\ti r$ be an element of $\clF$. Then
$$
 \Phi_{II}^{CS}=\frac{\varsigma}{z}-\frac{\varsigma-\Ad_{r}\varsigma}{z-1}\,.
$$
%Let $\gb=Lie(B)$ and $\gb^*$ is the Lie coalgebra.
The analog of the moment map equation is
\beq{c4}
(\varsigma-\Ad_{r}\varsigma))|_{\gb^*}=0\,,~r\in \clF\,.
 \eq
 In this construction
the phase space of the CS system is described as
\beq{c2}
\clR_{II}^{CS}=\{\clF\times {\rm solutions~of~}(\ref{c4})\}\,.
 \eq
 Notice, that $\dim\,\clR_{II}^{CS}=\dim\,\clR_{III}^{CS}$ (\ref{c1}).
 In fact, these phase spaces are symplectomorphic as in (\ref{i23}).

%%%%%%%%%%%%%%%%%%%%%%%%%%%%%%%%%%%%%%%%%%%%%%%%%%%%%%%%%%%%%%%%%%%%%%%%%%%%%%%

%%%%%%%%%%%%%%%%%%%%%%%%%%%%%%%%%%%%%%%%%%%
%%%%%%%%%%%%%%%%%%%%%%%%%%%%%%%%%%%%%%%%%%%%
\appendix

\section{Simple Lie groups: Notations and decompositions \cite{He}}
Let $G^\mC$ be a simple  complex Lie group, $\gg^\mC$ its Lie
algebra of rank $l$ and $\gh^\mC$ is a Cartan subalgebra.

\paragraph{Chevalley basis in $\gg$.}
Let $\{\al\in R\}$ be the root system.
The algebra $\gg^\mC$ has the root decomposition
 \beq{CD}
\gg^\mC=\gh^\mC+\gL\,,~~\gL=\sum_{\be\in R}\gR_\be\,,  ~~\dim_{\mC}\,\gR_\be=1\,.
 \eq
Let $\Pi\subset R$ be a subsystem of simple roots. The Chevalley basis in $\gg^\mC$ is generated by
 \beq{CBA}
\{E_{\be_j}\in\gR_{\be_j}\,,~\be_j\in R\,,~~H_{\al_k}\in\gh^\mC\,,~\al_k\in\Pi\}\,,
 \eq
where $H_{\al_k}$ are defined by the relation    $\al(H_\al)=2$.
The commutation relations in this basis assume the form
 \begin{subequations}\label{cbcr}
 \begin{align}
&[E_{\al_k},E_{-\al_k}]=H_{\al_k}\,,\\
&[H_{\al_k},E_{\pm\al_j}]= \pm a_{kj}E_{\pm\al_k}\,,~~\al_k\,,\al_j\in\Pi\,,\\
&[E_\al,E_{\be}]=C_{\al,\be}E_{\al+\be}~{\rm if~} \al+\be\in R\,,~~C_{\al,\be}=0\,,~{\rm if~} \al+\be\notin R\,,
 \end{align}
 \end{subequations}
 where $a_{kj}$ is the Cartan matrix and $C_{\al,\be}$ are the structure constants of $\gg$.

 Let $(~,~)$ be an invariant scalar product in $\gh^\mC$. The condition $\al(H_\al)=2$
 allows one to identify $H_\al$, $(\al\in\Pi)$ (the basis simple coroots) with the roots as
 \beq{hal}
H_\al=\al^\vee=\frac{2\al}{(\al,\al)}\,.
 \eq
Then the scalar product on the  Chevalley basis assumes the form
 \beq{kk}
(H_\al,H_\al)=\frac{4}{(\al,\al)}\,,
 \eq
and, from (\ref{cbcr}a),
 \beq{are6}
(E_\al,E_\be)=2\de_{\al,-\be}/{(\al,\al)}\,.
 \eq
For the canonical basis $\{e_j\}$, $(j=1\ldots,l)$ in $\gh^\mC$ we
have the expansion $H_\al=\sum_{j=1}^l\frac{2\al(j)}{(\al,\al)}e_j$
$\,(\al(j)=\al(e_j))$. Then from the first relation in (\ref{cbcr})
for any $\al\in R$:
  \beq{ee}
 [E_{\al},E_{-\al}]=\sum_{j=1}^l\frac{2\al(j)}{(\al,\al)}e_j\,.
  \eq
And
 \beq{ee99}
 [e_j,E_{\al}]=\al(j)E_\al\,.
  \eq
 The structure constants are real $C_{\alpha,\beta}\in \mR$\,, antisymmetric
 $C_{\alpha,\beta}=-C_{\beta,\alpha}$ and
  \begin{subequations}\label{sc1}
%  \beq{sc1}
 \begin{align}
&\,C_{-\alpha,-\beta}=-C_{\alpha,\beta}\,,\\
&\,C_{\alpha+\be,-\al}=-\frac{(\be,\be)}{(\al+\be,\al+\be)} C_{\alpha,\beta}\,.
 \end{align}
% \eq
 \end{subequations}

\paragraph{Real forms.} Let $\rho$ be the involutive automorphism $\rho$
$(\rho^2={\rm Id})$ acting on the root basis and the Cartan
subalgebra as
 \beq{rho}
\rho\,:\,(x\to -\bar x\,,~{\rm for~}x\in\gh^\mC\,,~~c_\al
E_\al\to-\bar c_\al E_{-\al})\,.
 \eq
The fixed points is the maximal compact subalgebra
 \beq{b2}
\gk=\{x\in \gg^\mC\,|\,\rho(x)=x\}\subset\gg^\mC\,.
 \eq
 We denote by the same letter the involution acting on $G^\mC$
 \beq{b2a}
K=\{g\in G^\mC\,|\,\rho(g)=g\}\subset G^\mC\,.
 \eq
The corresponding group $K\subset G^\mC $ is the maximal compact
subgroup. Let
$\gp^\mC$ be a subspace in the Lie algebra $\gg^\mC$ such that
\beq{b10}
 \gp^\mC=\{x\in \gg^\mC\,|\,\rho(x)=-x\}\subset\gg^\mC\,.
 \eq
It is the tangent space to the Riemannian non-compact symmetric
space
   $G^\mC/K$ at the point corresponding to $K$.
The orthogonal decomposition of the complex Lie algebra $\gg^\mC$
\beq{cd1}
 \gg^\mC=\gk\oplus\gp^\mC\,,~~\,~~((\gk,\gp^\mC)=0)\,.
  \eq
  is the \emph{Cartan decomposition} of $\gg^\mC$.

 Denote by $\clH^\mC$  a Cartan subgroup of $G^\mC$ and by
$\clT\subset K$ is the corresponding Cartan torus of $K$\,.
Define the corresponding Lie algebras
 $$
\gh^\mC={\rm Lie}(\clH^\mC)\,,~~\gt={\rm Lie}(\clT)\,.
 $$
Let $\{e_j\}$, be a canonical basis in $\gh^\mC$.
The basis in $\gk$ has the form
 \beq{bk}
\{\imath e_j\,,~ j=1,\ldots,l\}\,,~~\{\si_\al^1=(E_\al-E_{-\al})\,,~\si_\al^2=\imath (E_\al+E_{-\al})\,,
\al\in R^+\}\,,
 \eq
where $R^+$ is a set of positive roots.
% The coset space  $G^\mC/K$ is a  Riemannian non-compact symmetric space.

Let $\si$ be an involutive automorphism of $\gg^\mC$ which fixed points is \emph{the normal form}
$\gg^\mR$
 \beq{gr}
\gg^\mR=\{x\in \gg^\mC\,|\,\si(x)=x\}\,.
 \eq
The normal real form $\gg^\mR$ has the same basis (\ref{CBA}) as
$\gg^\mC$. Let $\gh^\mR\subset \gh^\mC$ be the Cartan subalgebra of
$\gg^\mR$. We have (compare with (\ref{cd1}))
 \beq{ck}
\gh^\mC=\gt\oplus\gh^\mR\,.
 \eq
By the action of the group $K\times K$  generic elements $g\in G^\mC$ can be transformed
 to the Cartan form
 \beq{dp}
g=k_1rk_2^{-1}\,,~~k_1\,,k_2\in K\,, ~~r=\bfe(\bfu)\in\clH^\mR\,,~~(\bfe(x)\doteq\exp\,( x))\,,
 \eq
where $\clH^\mR$ is a Cartan subgroup of $G^\mR$, (Lie\,$(\clH^\mR=\gh^\mR$),
$\bfu\in\gh^\mR$ (see (\ref{ck})). In fact $\bfu$ can be chosen from a Weyl chamber.
For example, one can take $\bfu_\al\doteq\al(\bfu)>0$.

%The involutive automorphisms $\si$ and $\rho$ commutes.
The automorphism $\te=\si\circ\rho$ acts on $\gg^\mC$.
Define subalgebra
 \beq{gu}
\gu=\{x\in \gk\,|\,\si(x)=x\}=\{x\in \gg^\mR\,|\,\rho(x)=x\}\,.
 \eq
and let $\te(\gp^\mR)=-\gp^\mR$ for $\gp^\mR\subset\gg^\mR$.
 Then
  \beq{cd}
 \gg^\mR=\gu\oplus\gp^\mR~~~((\gu,\gp^\mR)=0)
  \eq
  is the \emph{Cartan decomposition} of $\gg^\mR$.
 The real form $\gg^\mR$ is normal
if $\gp^\mR$ contains $\gh^\mR$.
The Killing form is non-degenerate on $\gg^\mR$ and the subspaces in (\ref{cd})
are orthogonal $(\gu,\gp^\mR)=0$.

Define the basis in the normal form $\gg^\mR$ according with the Cartan decomposition (\ref{cd}).
The basis in the space $\gp^\mR$ is
 $$
\gp^\mR \to \{\gh^\mR\to(e_1,\ldots,e_l)\,,~ P_\al=(E_\al+E_{-\al})\,,\,\al\in R^+\}
 $$
The basis in the Lie algebra $\gu$ is
 \beq{bu}
\gu\to \{U_\al=(E_\al-E_{-\al})\,,~~\al\in R^+\}
 \eq
with the commutation relations
 \beq{cur}
[U_\al,U_\be]=C_{\al,\be}U_{\al+\be}-C_{\al,-\be}U_{\al-\be}\,.
 \eq
and the norm (see (\ref{are6})
 \beq{nu}
(U_\al,U_\be)=-\frac{4}{(\al,\al)}\de_{\al\be}\,.
 \eq
%while
% \beq{nu1}
%(P_\al,P_\be)=\frac{4}{(\al,\al)}\de_{\al\be}\,,
% \eq
Thus, the Killing form (\ref{are6})
 is negative define on the subalgebra $\gu$.

Let $G^\mR$ be the normal subgroup ($\Lie\,G^\mR=\gg^\mR$) and $U=K\cap G^\mR$
is its maximal compact subgroup  ($\Lie\,U=\gu$). Similar to (\ref{dp}) for $g\in G^\mR$
\beq{dp1}
g=k_1rk_2^{-1}\,,~~k_1\,,k_2\in U\,, ~~r=\bfe(\bfu)\in\clH^\mR\,,~~\bfu\in\gh^\mR\,.
 \eq
It can be rewritten as \emph{the polar decomposition} of $g\in G^\mR$
\beq{dp2}
g=yk\,,~~(y=k_1rk_1^{-1}\,,~k=k_1^{-1}k_2\in U)\,.
 \eq
The similar formula holds for $G^\mC$
\beq{dp3}
g=k_1rk_2^{-1}=yk\,,~~k_1\,,k_2\in K\,,~~(y=k_1rk_1^{-1}\,,~r=\bfe(\bfu)\in\clH^\mR)\,.
 \eq

The subgroups $G^\mR$, $K$ and $U$ are
presented in Table 1.
 $$
 \begin{tabular}{|c|c|c|c|c|c|}
  \hline
  \hline
  &&&&&\\
  $G^\mC$&$G^\mR$ & $K$&$U$& $d_j$ & rank$\,U$ \\
  &&&&&\\
  \hline
  \hline
 $\SLN$  &SL$(N,\mR)$ &SU$(N)$& SO$(N)$ & $2,3\ldots,N$&$[N/2]$ \\
  SO$(2N\!+\!1,\mC)$& SO$(N\!+\!1,N)$ &SO$(2N\!+\!1)$ & SO$(N\!+\!1)\times$SO$(N)$  & $2,4,\ldots,2N$&$N$ \\
  Sp$(N,\mC)$&Sp$(N,\mR)$ &Sp$(N)$ & U$(N)$ & $2,4,\ldots,2N$&$N$ \\
   SO$(2N,\mC)$&SO$(N,N)$&SO$(2N)$& SO$(N)\times$SO$(N)$ & $2,4,\ldots,$& $2[N/2]$ \\
 & & & & $2N\!-\!2,N$ &  \\
  G$^\mC_2$& G$^\mR_2$& G$_2$& SU$(2)\times$SU$(2)$ &2, 6&2 \\
  F$_4^\mC$ &F$_4^\mR$&  F$_4$ & Sp$(3)\times$SU$(2)$&2,6,8,12 & 4\\
  E$_6^\mC$& E$_6^\mR$& E$_6$& Sp$(4)$ &$2,5,6,8,9,12$& 4 \\
  E$_7^\mC$& E$_7^\mR$ & E$_7$& SU$(8)$ &$2,6,8,10,14,18 $& 7 \\
    E$_8^\mC$&E$_8^\mR$ & E$_8$& SO$(16)$ &$2,8,12,14,18,\!$& 8 \\
%   &  &  \\
  & & & & 20,24,30 & \\
  \hline
 \end{tabular}
 $$
 \begin{center}
\texttt{Table 1}\\
The groups $G^\mC$, $G^\mR$, $K$, $U$
 \end{center}

\paragraph{Dimensions of algebras.}
Let $d_j$ be the order of the invariants of algebra $\gg^\mC$ and
$\rank(\gg^\mC)=l$
 \beq{dj}
d_1=2,\ldots,d_l=h\,,~~j=1,\ldots,l\,,
 \eq
where $h$ is the Coxeter number.
The dimensions of the algebras are expressed in terms of $d_j$.
\beq{dgc}
\dim_{\mC}\,G^\mC=\sum_{j=1}^l(2d_j-1)\,.
 \eq
The real dimension of $K$ is
 \beq{dk}
\dim_{\mR}\,K=\sum_{j=1}^l(2d_j-1)\,,
 \eq
and
 \beq{dgr}
\dim_{\mR}\,G^\mR=\dim_{\mR}\,K=\sum_{j=1}^l (2 d_j-1)\,.
 \eq
 Let $\clX^\mC=G^\mC/K$ be the Riemannian symmetric space. Then from (\ref{dgc}) and (\ref{dk})
 \beq{dgx}
\dim_{\mR}\,\clX^\mC=\sum_{j=1}^l (2 d_j-1)\,.
 \eq
The dimension of the compact form $U$ is
 \beq{du}
\dim_\mR\,U=\oh(\dim_\mR\,K-l)=\sum_{j=1}^l(d_j-1)
 \eq
 The corresponding Riemannian symmetric space $\clX^\mR=G^\mR/U$ be the. Then from (\ref{dgr})
  and (\ref{du})
 \beq{dgxr}
\dim_{\mR}\,\clX^\mR=\sum_{j=1}^l  d_j\,.
 \eq

%
% From (\ref{cd}) we find
% \beq{dpr}
% \dim_\mR\gp^\mR=\sum_{j=1}^ld_j\,.
% \eq
%

  Let $R^+$ be a subset of positive roots generated by simple roots $\Pi$ and
  $B$ is the corresponding Borel subgroup of $G^\mC$. Its Lie algebra Lie$(B)=\gh^\mC+
 \sum_{\al\in R^+}\gR_\al$. The flag variety $Fl$ is the homogeneous space
  $Fl=G^\mC/B$.
 It has dimension
 \beq{df}
 \dim_{\mC}\,Fl=\sum_{j=1}^l(d_j-1)\,.
 \eq

\paragraph{Poisson brackets on co-algebra $\gg^*$.} The general form of the
Lie-Poisson brackets for the functions on $\gg^*$  is
 \beq{lpb}
\{F_1(x),F_2(x)\}=\lan x,[\nabla F_1,\nabla F_2]\ran\,,~~x\in\gg^*\,,
 \eq
where $\nabla F_j(x)\in\gg$ is the gradient defined as a dual element to any $y\in\gg^*$
 $$
\lan y,\nabla F_j(x)\ran:=\frac{d}{dt}F_j(x+ty)|_{t=0}\,.
 $$

Let $\clU(\gg)$ be the universal enveloping algebra and $P(R)\in\clU(\gg)$  symmetric
$ad$-invariant polynomial. These polynomial belong to the center $\clZ(\clU)$ of  $\clU(\gg)$.
The algebra $\clZ(\clU)$ is generated by the rank$(\gg)$ homogeneous polynomials of degrees $d_j$
 (\ref{dj}), invariant with respect to the adjoint action.
The Casimir functions with respect to the Lie-Poisson brackets (\ref{lpb}) are the
coefficients of $P(R)$.

\bigskip
Consider in details the Lie-Poisson brackets on the compact Lie algebra $\gu^*$.
An arbitrary element of $\gu$ is decomposed in the basis (\ref{bu}) as
 \beq{4.2e}
\bfT=\sum_{\al\in R^+}T_\al U_\al\,,
 \eq
with
 \beq{sp}
(\bfT,\bfT)=-4\sum_{\al\in R^+}\frac{T_\al^2}{(\al,\al)}\,.
 \eq
By means of (\ref{nu}) define the dual  basis to the  basis $U_\al$, $\al\in R^+$ ($(u^\al,U_\be)=\de^\al_\be$)
%in $\gk^*$ dual to (\ref{19a})
 $$
u^\al=-\frac{(\al,\al)}4U_{\al}\,.
 $$
In this way $T_\al$,  can be considered as  linear functions
on the Poisson algebra on $\gu^*$:
 $$
T_\al=\lan\bfT,u^{\al}\ran\,.
 $$
The Poisson brackets (\ref{lpb}) are defined by means of the form
$d\vartheta=d\lan\bfT,h^{-1}dh\ran$ (\ref{8}). Because $\nabla T_\al=u^{\al}$ we have
 $$
\{T_\al,T_\be\}=\lan\bfT,[u^{\al},u^{\be}]\ran\,,
 $$
Let
 \beq{na}
N_{\al,\be}=-C_{\al,\be}\ka_{\al,\be}\,,~\ka_{\al,\be}=
  \frac{(\al,\al)(\be,\be)}{4(\ga,\ga)}\,,~~\ga=\al+\be\in R\,.
 \eq
Then from (\ref{cur}) we find
 \beq{20a}
\{T_\al,T_\be\}=N_{\al,\be}T_{\al+\be}-N_{\al,-\be}T_{\al-\be}\,.
 \eq
It can be proved similarly that
 \beq{30a}
\{S_\al,S_\be\}=-N_{\al,\be}S_{\al+\be}+N_{\al,-\be}S_{\al-\be}\,.
 \eq
 The brackets
 \beq{30b}
\{T_\al,S_\be\}=0\,.
 \eq
 follow from commutativity of the left and right momenta.
From (\ref{sc1}b) and (\ref{na}) we also find
 \beq{22c}
N_{\al+\be,-\al}= \frac{(\al+\be,\al+\be)}{(\be,\be)}N_{\be,\al}=\frac{\al^2}4C_{\al\be}\,.
 \eq

%%%%%%%%%%%%%%%%%%%%%%%%%%%%%%%%%%%%%%%%%%%%%%%%%%%%%%%%%%%%%%%%%%%%%%%%%%%%%%%%%

\section{Equations of motion of gyrostat on simple compact group}

If the one of the spin variables (say $\bfS$) is fixed, then the
Hamiltonian (\ref{i1}) describe a gyrostat (a rigid top with
additional rotator moment.  So technically as a intermediate result
it is convenient to consider dynamic of a gyrostat on the group $U$.
It is defined by the Poisson brackets (\ref{20a}) on the coalgebra
$\gu^*$ for  the momenta $\bfT$ (\ref{4.2e}) and the  Hamiltonian:
 \beq{qh}
H^{top}=\sum_{\nu\in R^+}\left(\oh T_\nu^2f(\nu)+  T_\nu g(\nu)\right)=
\oh\sum_{\nu\in R}\left(\oh T_\nu^2f(\nu)+  T_\nu g(\nu)\right)\,.
 \eq
Here $f(\nu)$ are the components of "the inverse inertia tensor"
 of the top that maps $\gu^*\to\gu$. We assume that
 \beq{ii}
f(\nu)=f(-\nu)\,,~~g(\nu)=-g(-\nu)\,.
 \eq
The equations of motion assume the form
 \beq{fe2}
 \begin{array}{c}
  \dot{T}_\ga=\{H^{top},T_\ga\}=\oh\sum_{\nu\in R}
  ((N_{\nu,\ga}T_{\nu}T_{\nu+\ga}-N_{\nu,-\ga}T_{\nu}T_{\nu-\ga}))f(\nu)+ \\
  \\
  (N_{\nu,\ga}T_{\nu+\ga}-N_{\nu,-\ga}T_{\nu-\ga})g(\nu))=
 \end{array}
 \eq
 $$
\oh\sum_{\nu\in R}  \gS_{\nu,\ga}(\bfT)+(N_{\nu,\ga}T_{\nu+\ga}-N_{\nu,-\ga}T_{\nu-\ga})g(\nu))\,.
 $$
where $ \gS_{\nu,\ga}(\bfT)=T_{\nu}T_{\nu+\ga}(N_{\nu,\ga}-N_{\nu+\ga,-\ga}) f(\nu)$.
Using (\ref{22c}) we rewrite  it as
% \beq{e1}
%\dot{T}_\ga=\sum_{\nu\in R_+}\gS_{\nu,\ga}(\bfT)
%-\gS_{\nu,-\ga}(\bfT)
%+(N_{\nu,\ga}T_{\nu+\ga}-N_{\nu,-\ga}T_{\nu-\ga}) g(\nu)\,,
% \eq
 $$
\gS_{\nu,\ga}(\bfT)=N_{\nu,\ga}T_{\nu+\ga} T_{\nu}
\left(f(\nu)-\frac{(\nu+\ga)^2}{\nu^2}
f(\nu+\ga)\right)\,.
 $$
Let $\ga=\al+\be$. Then for $\nu=-\al$
 $$
\gS_{\nu,\ga}(\bfT)|_{\nu=-\al}=N_{-\al,\al+\be}T_{\al} T_{\be}
\left(f(\al)-\frac{\be^2}{\al^2}
f(\be)\right)\,.
 $$
%and for $\nu=\al$
% $$
%\gS_{\nu,-\ga}(\bfT)|_{\nu=\al}=N_{\al,-\al-\be}T_{\al} T_{\be}
%\left(f(\al)-\frac{\be^2}{\al^2}
%f(\be)\right)\,.
% $$
%Due to 1(\ref{sc1}) and (\ref{na}) $N_{-\al,\al+\be}=-N_{\al,-\al-\be}$.
Again from (\ref{22c}) we find that
$N_{-\al,\al+\be}=-\frac{\al^2}4 C_{\al\be}$.
Then
 \beq{s1}
\gS_{-\al,\ga}(\bfT)=-\f1{4}C_{\al,\be}T_\al T_\be(\al^2f(\al)-\be^2f(\be)).
 \eq
Also by means of (\ref{22c}) we analyze in (\ref{fe2}) the second term
$(N_{\nu,\ga}T_{\nu+\ga}-N_{\nu,-\ga}T_{\nu-\ga})g(\nu)$.
For $\ga=\al+\be$ we have
 $$
N_{\nu,\ga}T_{\nu+\ga}g(\nu)|_{\nu=-\be}=-C_{\al,\be}T_\al g(\be)\frac{\be^2}4\,.
 $$
Also
 $$
-N_{\nu,-\ga}T_{\nu-\ga}g(\nu)|_{\nu=\be}=-C_{\al,\be}T_\al g(\be)\frac{\be^2}4\,.
 $$
Thus
 $$
(N_{\nu,\ga}T_{\nu+\ga}-N_{\nu,-\ga}T_{\nu-\ga})g(\nu)=-\sum_{\al+\be=\ga}
C_{\al\be}(T_\al \be^2g(\be)-T_\be \al^2g(\al))\,.
 $$
 Substituting this expression and (\ref{s1}) in (\ref{fe2}) we come to the final expression
  \beq{fe}
\dot{T}_\ga=\f1{4}\sum_{\al+\be=\ga}
C_{\al\be}
\Bigl(T_\al T_\be(\al^2f(\al)
-\be^2 f(\be))-(T_\al \be^2g(\be)-T_\be \al^2g(\al)) \Bigr)\,.
 \eq
From (\ref{30a}) we have
 \beq{fe1}
\dot{S}_\ga=-\f1{4}\sum_{\al+\be=\ga}
C_{\al\be}
\Bigl(S_\al S_\be(\al^2f(\al)
- \be^2 f(\be))- (S_{\al}\be^2 g(\be)-S_{\be}\al^2 g(\al) )\Bigr)\,.
 \eq
%%%%%%%%%%%%%%%%%%%%%%%%%%%%%%%%%%%%%%%%%%%%%%%%%%%%%%%%%%%%%%%%%%%%%%%%%%%%%%%%%

\section{Proof of the equivalence of the  Lax equations
and the equations of motion}

 In terms of Lax operators (\ref{12}) and (\ref{14}) the
 Lax equation assumes the form
 $$
\p_tP+\p_tX=
[P,Y]+[X,Y]|_{\gh^\mR}+[X,Y]|_{\gL^\mR}\,.
 $$
It means for arbitraries $j=1,\dots,l$ and $\ga\in R^+$
 $$
\underbrace{\p_tv_j}_{\underline{1}}+
\underbrace{\p_tX_{\pm\ga}}_{\underline{2}}=
\underbrace{[P,Y]|_{E_{\pm\ga}}}_{\underline{3}}+
\underbrace{[X,Y]|_{e_j}}_{\underline{4}}+\underbrace{[X,Y]|_{E_{\pm\ga}}}_{\underline{5}}\,.
 $$
Then the equations $\underline{1}=\underline{4}$ are equivalent to
(\ref{16a}). In fact, due to (\ref{ee}) we have
 $$
[X,Y]|_{e_j}=\sum_{\al\in R^+}[X_\al E_\al+X_{-\al}
E_{-\al},Y_\al(E_\al-E_{-\al})]|_{e_j}= -\sum_{\al\in
R^+}\frac{2\al(j)Y_\al(X_\al +X_{-\al})}{(\al,\al)}e_j\,.
 $$
 Now consider
the equations  $\underline{2}=\underline{3}+\underline{5}$.
 $$
\underline{2}=\p_tX_{\pm\ga}E_{\pm\ga}=\left(\sum_jv_j\p_{u_j}\left(
\frac{T_\ga e(\mp\bfu_\ga)-S_\ga}{2\sinh(\bfu_\ga)}\right)+
\frac{\p_tT_\ga e(\mp\bfu_\ga)-\p_tS_\ga}{\sinh(\bfu_\ga)}\right)E_{\pm\ga}=
 $$
 $$
\stackrel{(\ref{14})}{=}\underbrace{\sum_jv_j\ga_jY_{\ga}E_{\pm\ga}}_{\underline{2a}}+
\underbrace{\frac{\p_tT_\ga
e(\mp\bfu_\ga)-\p_tS_\ga}{\sinh(\bfu_\ga)}}_{\underline{2b}}
E_{\pm\ga}\,.
 $$
Since
 $$
\underline{3}=[P,Y]|_{E_{\pm\ga}}=\sum_jv_j\ga_jY_{\pm\ga}
 $$
we have $\underline{2a}=\underline{3}$.
Compare $\underline{2b}$ and  $\underline{5}$:
 $$
\underline{5}=\sum_{\al+\be=\ga}C_{\al\be}(X_\al Y_\be -X_\be Y_\al)E_\ga=
\sum_{\al+\be=\ga}C_{\al\be}\clF(\al,\be)E_\ga\,,
 $$
 $$
\clF(\al,\be)=
\underline{S_\al S_\be}\left(\frac{\cosh(\bfu_\al)}{\sinh^2(\bfu_\al)\sinh(\bfu_\be)}-
\frac{\cosh(\bfu_\be)}{\sinh^2(\bfu_\be)\sinh(\bfu_\al)}\right)+
 $$
 $$
\underline{T_\al T_\be}\left(-\frac{e(\mp\bfu_\al)}{\sinh^2(\bfu_\al)\sinh(\bfu_\be)}+
\frac{e(\mp\bfu_\be)}{\sinh^2(\bfu_\be)\sinh(\bfu_\al)}\right)+
 $$
 $$
\underline{S_\al T_\be}\left(\frac{1}{\sinh(\bfu_\al)\sinh^2(\bfu_\be)}-
\frac{e(\mp\bfu_\be)\cosh(\bfu_\al)}{\sinh(\bfu_\be)\sinh^2(\bfu_\al)}\right)+
 $$
 $$
\underline{S_\be T_\al}\left(-\frac{1}{\sinh^2(\bfu_\al)\sinh(\bfu_\be)}+
\frac{e(\mp\bfu_\al)\cosh(\bfu_\be)}{\sinh^2(\bfu_\be)\sinh(\bfu_\al)}\right)\,.
 $$
On the other hand from the equations of motion for the spin variables (\ref{17}) and (\ref{18})
 $$
\underline{2b}=\frac{\p_tT_\ga e(\mp\bfu_\ga)-\p_tS_\ga}{\sinh(\bfu_\ga)}E_{\pm\ga}
=\f1{\sinh(\bfu_\ga)}\sum_{\al+\be=\ga}C_{\al,\be}\clR(\al,\be)E_{\pm\ga}\,,
 $$
 $$
\clR(\al,\be)=\left(\underline{S_\al S_\be}-\underline{T_\al T_\be}e(\mp\bfu_\ga)\right)
\left(\f1{\sinh^2(\bfu_\al)}-\f1{\sinh^2(\bfu_\be)}\right)+
 $$
 $$
\underline{S_\al
T_\be}\left(\frac{\cosh(\bfu_\be)}{\sinh^2(\bfu_\be)}-
\frac{\cosh(\bfu_\al)}{\sinh^2(\bfu_\al)}e(\mp\bfu_\ga)\right)+
\underline{S_\be
T_\al}\left(\frac{\cosh(\bfu_\be)}{\sinh^2(\bfu_\be)}e(\mp\bfu_\ga)-
\frac{\cosh(\bfu_\al)}{\sinh^2(\bfu_\al)}\right)\,.
 $$
The equality $\underline{2b}=\underline{5}$ follows from  the following addition relations
for trigonometric functions
 \beq{20}
\f1{\sinh(x+y)}(\f1{\sinh^2y}-\f1{\sinh^2x})=\frac{\cosh y}{\sinh x\sinh^2y}-
\frac{\cosh x}{\sinh y\sinh^2x}\,,
 \eq
 \beq{21}
\frac{\exp(-x-y)}{\sinh(x+y)}(\f1{\sinh^{2}y}-\f1{\sinh^{2}x})=-\frac{\exp (- y)}{\sinh^2 x\sinh y}+
\frac{\exp (-x)}{\sinh^2 y\sinh x}\,,
 \eq
 \beq{22}
\f1{\sinh (x+y)}\left(\frac{\exp(-x-y)\cosh(y)}{\sinh^{2}(y)}-\frac{\cosh(x)}
{\sinh^{2}(x)}\right)=\frac{\exp(-x)\cosh y}{\sinh x\sinh^2y}-
\f1{\sinh y\sinh^2x}\,.
 \eq

%%%%%%%%%%%%%%%%%%%%%%%%%%%%%%%%%%%%%%%%%%%%%%%%%%%%%%%%%%%%%%%%%%%%%%%%%%%%%%%%%%%

\end{document}